\documentclass[11pt,a4paper]{article}
\usepackage{graphicx}
\usepackage{epstopdf, epsfig}
\usepackage{amsmath,amssymb}
\usepackage{color}
\usepackage{tikz}
\usepackage{natbib}
\usepackage{url}

\newcommand\ppt[1]{\cfrac{\partial#1}{\partial t}} 
\newcommand\ppx[1]{\cfrac{\partial#1}{\partial x}} 
\newcommand\ppy[1]{\cfrac{\partial#1}{\partial y}} 
\newcommand\ppz[1]{\cfrac{\partial#1}{\partial z}} 
\newcommand\pppy[1]{\cfrac{\partial^2#1}{\partial y^2}} 
\newcommand\pppx[1]{\cfrac{\partial^2#1}{\partial x^2}} 
\newcommand\pppz[1]{\cfrac{\partial^2#1}{\partial z^2}} 
\newcommand\ppi[1]{\cfrac{\partial#1}{\partial x_i}} 
\newcommand\ppj[1]{\cfrac{\partial#1}{\partial x_j}} 
\newcommand\abs[1]{\lvert #1\rvert} 

\title{A numerical study of a variable-density \\ low-speed turbulent mixing layer} 
\author{A. Almagro,
M. Garc\'{\i}a-Villalba, 
O. Flores\\
Departamento de Bioingenier\'{\i}a e Ingenier\'{\i}a Aeroespacial \\
Universidad Carlos III de Madrid \\
28911 Legan\'es, SPAIN}
\date{}

\begin{document}

\maketitle 

\begin{abstract}
Direct numerical simulations of a temporally-developing, low-speed, variable-density, turbulent, plane mixing layer 
are performed. 
The Navier-Stokes equations in the low-Mach number approximation are
solved using a novel algorithm based on an extended version of the velocity-vorticity 
formulation used by \cite{Kim1987} for incompressible flows. 
Four cases with density ratios $s=1, 2, 4$ and 8 are considered. 
The simulations are run with a Prandtl number of 0.7 and achieve a $Re_\lambda$ up to 
150
during the self-similar evolution of the mixing layer. 
It is found that the growth rate of the mixing layer decreases with increasing density ratio,
in agreement with theoretical models of this phenomenon.
 Comparison with high-speed data shows that the reduction of the growth rates with increasing the
 density ratio has a weak dependence with the Mach number. 
In addition, the shifting of the mixing layer to the low-density stream has been characterized by analyzing
one point statistics within the self-similar interval. 
This shifting has been quantified, and 
related to the growth rate of the mixing layer 
under the assumption  that the shape of the mean velocity and density profiles do not change with the density ratio. This leads to a predictive model for the reduction of the growth rate of the momentum thickness, which agrees reasonably well with the available data. 
Finally, the effect of the density ratio on the turbulent structure has been analyzed using flow visualizations and spectra. It is found that  
with increasing density ratio the longest scales in the high density side are gradually inhibited. 
A gradual reduction of the energy in small scales with increasing density ratio is also observed.

\end{abstract}

\section{Introduction}

Variable density effects in turbulent flows are often encountered in the natural environment and in 
many engineering applications \citep{Chassaing2002, turner:1979}. 
In the oceans, density variations are 
due to temperature and salinity variations \citep{thorpe:2005}, while in the atmosphere they are due to both temperature and moisture changes \citep{wyngaard:2010}. In both situations, the buoyancy effects are mainly due to gravity.
In absence of gravity, density effects may still be important due to pressure and/or temperature fluctuations.
For example in aeronautical applications, density variations due to high speed in gas flows are very relevant 
\citep{lele:1994,gatski:2013}. In that case, the main effect is due to velocity induced pressure variations. 
In other applications, density variations due to dilatation effects are
important even at low speeds. This is for example the case in combustion applications 
\citep{williams:1985,peters:2000}, where the heat release by chemical reaction leads to the thermal
expansion of the fluid.
An additional kind of density effect is associated with the mixing of two non-reactive fluids of different density
or to the mixing of different temperature bodies of the same fluid \citep{Chassaing2002,Dimotakis2005}. 
In this work we are concerned with the latter since we study a variable-density low-speed temporal
turbulent mixing layer in the absence of gravity.

As reviewed by \cite{Dimotakis1986}, in spatially-developing turbulent shear layers
the density ratio influences the spreading rate of the layer, 
the entrainment rate and the convective velocity of the large-scale eddies.
The influence on the spreading rate was already observed in early experiments \citep{Brown1974}. 
However, the effect of increasing the Mach number, $M$, was found to be more drastic and that led to a main
focus on compressibility effects in subsequent works \citep{bogdanoff:1983,papamoschou:1988,clemens:1992,hall:1993,vreman:1996,mahle:2007,obrien:2014,jahanbakhshi:2016}.
A notable exception is the work of \cite{Pantano2002} who studied both compressibility effects and density ratio
effects in direct numerical simulations of turbulent compressible temporal mixing layers.  
They found that, with increasing density ratio, the shear layer growth rate decreases substantially 
and that the dividing streamline is shifted towards the low-density stream.
The variation of the density ratio by \cite{Pantano2002} was performed at high speed, with a convective Mach number $M_c=0.7$,
so that density variations due to both pressure effects and temperature effects were likely to affect the
flow. 
In this work we try to separate these two effects by considering a variable-density layer at low speed 
in the limit $M_c \rightarrow 0$, using the low Mach number approximation \citep{mcmurtry:1986,cook:1996,nicoud:2000}.

The current understanding of the effect of the density ratio on the structure of the turbulent mixing layer is
still unsatisfactory. 
Part of the problem is that it is difficult to perform experiments at low speeds with a large density ratio.
Numerical studies are also scarce and most of them deal with the initial stages of transition to turbulence,
and not with the turbulent regime itself.
Most numerical studies consider variable density effects in the limit of incompressible flow, i.e. the velocity
field is solenoidal, the density is given by an advection equation and the energy equation is therefore decoupled
from the momentum equation. 
For instance, \cite{knio:1992} reported calculations of a variable-density, incompressible, temporal mixing layer. 
They performed visualizations of the vorticity and scalar fields and of the motion of material surfaces, focusing
on the manifestation of three dimensional instabilities. 
They found an asymmetric entrainment pattern favouring the low-density stream.
Also in the incompressible regime, \cite{soteriou:1995} performed two-dimensional simulations
of spatially-developing variable-density mixing layers. 
They found that the speed of the unstable waves is biased toward that of the high-density stream and also that
the entrainment of the high-density stream is inhibited relative to the low-density stream.
The instability characteristics of variable-density incompressible mixing layers have been studied
by \cite{reinaud:2000} and \cite{fontane:2008}.
On the modelling side, \cite{Ramshaw2000} developed a simple model for predicting the thickness of 
a variable-density mixing layer. 
\cite{Ashurst2005} included variable-density effects in a one-dimensional turbulence approach. 
Using this approach they studied both temporally-developing and spatially-developing mixing layers,
and despite the limitations of the approach, their results provide information concerning
the expected behaviour of the mixing layers at high density ratios. 
In addition, there are also not so many studies in the literature using Large Eddy Simulation (LES) in variable density turbulent flows. Some examples are \citet{wang2008}, who analyzed spatially developing axisymmetric jets, and \citet{mcmullan2011}, who considered a spatially developing mixing.

In this work we address the following issues. How is the growth rate of the turbulent mixing layers affected by the
free-stream density ratio? What is the turbulent structure of variable-density mixing layers? What are the 
differences of the low speed case, $M\rightarrow0$, with respect to the high speed case, $M_c=0.7$ \citep{Pantano2002}?
The manuscript is organized as follows. In \S \ref{sec:setup} the computational setup is described including
the details of a novel algorithm developed to solve the low Mach number approximation of the Navier-Stokes equations.
This is followed by a description of the simulation parameters in \S \ref{sec:simulations}.
Results are presented in \S \ref{sec:results}. First, we analyze the self-similar evolution of the mixing layers.
Secondly, we characterize their growth rate and compare to a model proposed in the literature.
Third, we analyze the mean density and Favre averaged velocity, and propose a semi-empirical model for the observed shifting. After this, we complete the characterization of the vertical profiles with mean temperature. This is followed in \S\ref{sec:rms} by the analysis of the higher order statistics. Section \ref{sec:results} finalizes with the analysis of the flow structures, using flow visualizations and premultiplied spectra of temperature and velocity.   
Conclusions are provided in \S \ref{sec:conclusions}.

\section{Computational setup} \label{sec:setup}

The flow under consideration is a three-dimensional, temporally-evolving mixing layer developing
between two streams of different density, $\rho_t$ (upper stream) and $\rho_b$ (lower stream). 
The flow is assumed to be homogeneous in the horizontal directions, $x$ and $z$, 
while it is inhomogeneous in the vertical direction, $y$. 
The lower stream flows at a velocity $\Delta U/2$ in the positive $x$ direction, while the upper stream flows at a velocity $\Delta U/2$ in the oposite direction, so that the velocity difference between both streams is $\Delta U$. For the present work, $\rho_b > \rho_t$, although since we do not consider gravity effects, the case with $\rho_b < \rho_t$  can be obtained by changing the direction of the $y$-axis.

As explained in the introduction, for the present study we consider that temperature and density
fluctuations are much more significant than pressure fluctuations.
Therefore, the governing equations are the Navier-Stokes under the low Mach number approximation
\citep{mcmurtry:1986,cook:1996,nicoud:2000} together with the equation of state. These 
equations read (Einstein's summation convention is employed) 
\begin{align}
 \ppt{\rho}+\ppi{(\rho u_i)} & =0, \label{eq:continuity} \\
\ppt{(\rho u_i)}+\ppj{(\rho u_i u_j)} & =-\ppi{p^{(1)}}+\ppj{\tau_{ij}},
\label{eq:momentum} \\
\rho C_p \ppt{T}+\rho C_p u_i \ppi{T} & =\ppi{}\left( \kappa \ppi{T} \right),
\label{eq:energy}\\
 p^{(0)}& =\rho R T,
\label{eq:EOS1}
\end{align}
where $\rho$ is the fluid density, $u_i$ are the velocity components, $T$ is the temperature, 
$\tau_{ij}$ is the viscous stress tensor,  $\kappa$  is the thermal conductivity,
$C_p$ is the specific heat at constant pressure and $R$ is the specific gas constant.
Within the low Mach number approximation, the variables are expanded in a Taylor series 
where the Mach number is the small parameter. 
The first two terms of the pressure expansion appear in eqs. (\ref{eq:continuity}-\ref{eq:EOS1}), 
denoted $p^{(0)}$ and $p^{(1)}$.
The former, $p^{(0)}$, is usually called the thermodynamic
pressure, since it only appears in the equation of state. 
In the present case, $p^{(0)}$ can be considered to be constant, since the temporal mixing layer is an open system 
\citep{nicoud:2000}. 
The latter, $p^{(1)}$, plays the same role as in incompressible flow and it is usually called
the mechanical pressure.
The viscous stress tensor is given by $\tau_{ij}=\mu(\partial u_i/\partial x_j +\partial 
u_j/\partial x_i
-2/3\delta_{ij}(\partial u_k/\partial x_k))$, where $\mu$ is the dynamic viscosity and $\delta_{ij}$
is the Kronecker delta.
Note that in the low Mach number approximation the heating due to viscous dissipation 
in the energy equation is negligible, as discussed by \cite{cook:1996}.
In the present study, the fluid properties ($\mu,\kappa,C_p$) are assumed to be constant, 
independent of the temperature. 

The equations (\ref{eq:continuity}-\ref{eq:EOS1}) can be made non-dimensional using 
a reference density $\rho_0=(\rho_b+\rho_t)/2$, a reference temperature $T_0=p^{(0)}/\rho_0R$,
a characteristic velocity $\Delta U$ (the velocity difference between the
two streams) and a characteristic length $\delta_m^0$ (the initial momentum thickness of the mixing
layer, further discussed below). The resulting non-dimensional numbers that govern the problem
are the Reynolds number, $Re=\rho_0 \Delta U \delta_m^0/\mu$, the Prandtl number 
$Pr = \mu C_p/\kappa$ and the density ratio, $s=\rho_b/\rho_t$. 

Considering the role played by the mechanical pressure, we  solve the governing
equations using an algorithm analogous to the algorithm for incompressible flow of \cite{Kim1987}.
In that work, the momentum equation is recast in terms of two evolution equations, 
the first one for the vertical component of the vorticity, $\omega_y$, and the second one for the 
laplacian of the vertical component of the velocity, $\nabla^2v$. 
In that way, pressure is removed from the equations and continuity is enforced by construction.
In order to employ a similar formulation, we decompose the momentum vector   
\begin{equation}
\rho \vec{u}= \vec{m}+\nabla{\psi},
\label{eq:helmholtz}
\end{equation}
where $\vec{m}$ is a divergence-free component  and $\nabla{\psi}$ is a curl-free component.
We define $\Omega_y$ as the vertical component of the vector $\nabla \times \rho \vec{u}= 
\nabla \times \vec{m}$ and $\phi$ as the laplacian of the vertical component of $\vec{m}$,
$\phi=\nabla^2 m_y$. 
Hence, as described in the appendix \ref{sec:appendix}, equations  (\ref{eq:continuity}-\ref{eq:EOS1}) can be recast as 
evolution equations
for $\rho$, $\Omega_y$, $\phi$ and $T$, which together with the equation of state $\rho T = \rho_0 T_0$, 
results in a system of five equations and five unknowns.

The details of the algorithm used to integrate in time this coupled system of equations is described in 
detail in appendix \ref{sec:appendix}.  For completeness, we provide here a brief description. 
The time integration is performed using a three-stage low-storage Runge-Kutta scheme. 
At each stage, the evolution equations for $\Omega_y$, $\phi$ and $T$ (namely, momentum and energy 
equations) are used to update
explicitly these variables. 
Then, $\rho$ is computed using the equation of state.
Once $\rho$ is known, we estimate $\partial \rho/\partial t$ and use the continuity equation 
to solve for $\psi$.

 The spatial discretisation is based on a Fourier decomposition for the homogeneous 
directions $x$ and $z$, with 7th and 5th order compact finite differences for first and second 
derivatives in the vertical direction, as in \cite{Hoyas:2006}. 
The computation of the non-linear terms in the evolution equations for $\rho$, $\Omega_y$, 
$\phi$ and $T$ (equations \ref{eq:phieq}-\ref{eq:cont3}) is pseudo-spectral, using the 
2/3 rule to remove the aliasing error associated with quadratic terms. 
Note that due to the non-linearity appearing in the equation of state, it is not possible to 
completely remove aliasing errors in the present formulation. 
The solution of the Poisson equation for $\psi$ (see appendix \ref{sec:appendix}) is done in Fourier space, 
solving a penta-diagonal linear system for each Fourier mode with an LU decomposition.  
No explicit filtering or smoothing is used in the present formulation.

Concerning the boundary conditions, from a physical point of view the velocity and density 
fluctuations should tend to zero as $y\rightarrow \pm \infty$,
with an additional constraint that relates the entrainment and the ambient pressure.
From a computational point of view, we impose free-slip
boundary conditions for the fluctuations of $\vec{m}$,  homogeneous Dirichlet boundary conditions
for the density fluctuations and homogeneous Neumann boundary conditions for the  $\psi$. 
In terms of entrainment, the global mass balance in the system leads to one equation with two 
unknowns, 
namely the mass flux through the upper and lower boundaries of the system.
A second equation is obtained imposing that the ratio of these two mass fluxes
should be equal to the square root of the density ratio \citep{Dimotakis1986,dimotakis:1991}.
This condition is equivalent to the one imposed by \cite{Higuera1994},
matching the mass fluxes to an outer wave region where acoustic effects are important. Further
details are provided in appendix \ref{sec:appendix}. 

Finally, initial conditions are provided specifying the mean streamwise velocity and density 
profiles
\begin{align}
\overline{u} (y) & = \frac{\Delta U}{2} \tanh \left(-\frac{y}{2\delta_m^0} \right),\\
\overline{  \rho} (y) &   = \rho_0 \left( 1 + \lambda(s) \tanh \left(-\frac{y}{2\delta_m^0} \right) 
\right),
\end{align}
where $\lambda(s)=(\rho_b-\rho_t)/(\rho_b+\rho_t) = (s-1)/(s+1)$.
The mean spanwise and vertical velocity components are set to zero.
In order to promote a quick transition to turbulence, random velocity fluctuations are added.
This is done in a manner similar to \citet{Pantano2002}, \citet{da2008invariants} and others: a random 
solenoidal velocity fluctuation field with a 10\% turbulence intensity and
a peak wavenumber of $k_0 \delta_m^0 \approx 0.84$.  
The region in space were the fluctuating velocity field is defined is limited by a gaussian filter,   $e^{-\left( y/\delta_m^0 \right)^2}$. Also, no fluctuations are imposed on wavenumbers smaller than $k_x\delta_m \approx 0.05$, so that the initial transient of the mixing layer is as {\em natural} as possible, as discussed by \citet{da2008invariants}.  

It should be noted that in the previous paragraphs we have been using $\delta_m^0$ to denote the 
initial value of the momentum thickness $\delta_m$. For a variable density boundary layer the 
momentum thickness is defined as
\begin{equation}
\delta_m(t) = \frac{1}{\rho_0 \Delta U^2} \int_{-\infty}^\infty {\overline{ \rho } \left( 
\frac{1}{2}  \Delta U - \tilde{u} \right) \left( \frac{1}{2}  \Delta U + \tilde{u} \right) dy},
\label{eq:momthickness}
\end{equation}
where $\tilde{u} = {\overline{ \rho u }}/{\overline{ \rho }}$ denotes the Favre average of $u$, and 
$\overline{u}$ is the standard Reynolds average (i.e.,  averaged over the 
homogeneous directions and over the different runs performed for each density ratio). The Favre perturbations are defined as $u^{\prime \prime}=u - \tilde{u}$, so 
that the turbulent stress tensor, $R_{ij}$, is defined as
\begin{equation}
 R_{ij} = \frac{\overline{ \rho u_i^{\prime \prime} u_j^{\prime \prime} }}{\overline{ \rho }}.
 \label{eq:Rij}
\end{equation}
Note that in the following, if not stated otherwise, the mean velocities and perturbations are Favre-averaged. On the other hand,
 density and temperature quantities are always Reynolds-averaged.
For completeness, we also provide here the definition of the vorticity thickness 
\begin{equation} 
\delta_w(t) = \frac{\Delta U}{ \abs{\partial\tilde{u}/\partial y }_{max}},
\label{eq:vorthickness}
\end{equation}
which is similar to the {\em visual} thickness of the mixing layer (see \citealp{Ramshaw2000, Brown1974, dimotakis1991turbulent} and experimental works in general), and it  
will be used in the discussion of the results in the following sections.

\section{Simulation parameters} \label{sec:simulations}

As mentioned above, the set-up of the simulations consists of a three-dimensional 
temporally-evolving mixing layer with two streams with different density. 
A total of four density ratio cases have been studied in this work, namely $s=\rho_b/\rho_t = 1$, 
2, 4 and 8. 
Four different realizations have been run for each density ratio (with different random initial conditions, discussed below), in order to perform ensemble averaging. 
For the case with $s=1$, the temperature is treated as a passive scalar: density is constant in 
time and space, and the energy equation is solved for the temperature disregarding the equation of 
state.   
The Reynolds and Prandtl numbers are fixed for all cases, with $Re=160$ and $Pr=0.7$.
The value of other relevant parameters are presented in Table \ref{tab:SimParams}. 
For instance, the Reynolds number based on the Taylor micro-scale, $Re_\lambda$, is moderately 
large 
for the $s=1$ case ($Re_\lambda = 150$), although it decreases with 
the density ratio ($Re_\lambda = 95$ for $s=8$). 

In terms of temporal resolution, all simulations presented here are run with a $CFL=0.5$. 
The computational domain is $L_x \times L_y \times L_z = 
461\delta_m^0 \times 368 \delta_m^0 \times 173\delta_m^0$, roughly twice larger in every direction than that employed by \citet{Pantano2002}. The plane $y=0$ is at the center of the computational domain, so that the upper and lower vertical boundaries are at $y=\pm L_y/2=184\delta_m^0$. 

The computational domain is discretized using $1536 \times 851 \times576$ collocation grid points, resulting in a spatial resolution in the homogeneous directions of $\Delta x = \Delta z=0.30 \delta_m^0$ before dealiasing (collocation points). 
In the vertical direction, the grid points are equispaced in the central 
part of the domain ($\abs{y} \le 20\delta_m^0$), with a resolution  $\Delta y = 0.2 \delta_m^0$. In the region $20\delta_m^0 \le \abs{y} \le 150\delta_m^0$ the resolution decreases with a maximum stretching of 1\%, up to a maximum grid spacing of $\Delta y = 0.85\delta_m^0$. Finally, in order to avoid numerical issues in the calculation of the vertical derivatives at the boundaries, the grid spacing is reduced again in the region $150\delta_m^0 \le \abs{y} \le 184\delta_m^0$ with a maximum stretching of $3\%$, resulting in a resolution of $\Delta y = 0.3\delta_m^0$ at the top and bottom boundaries of the computational domain.

As shown in Table \ref{tab:SimParams}, the resolution of the simulations is 
very good in terms of the local Kolmogorov lengthscale $\eta$ (i.e., averaged in horizontal planes only). 
The horizontal grid spacing is smaller than $1.8\eta$ during the 
self-similar evolution of the mixing layer.  
The vertical resolution is slightly better, to account for the worse 
resolution properties of compact finite differences compared to Fourier expansions \citep{lele:1992}. 
For reference, the resolution in the compressible simulations of \citet{Pantano2002} is $\Delta x /\eta \approx 3-4$. 
 Compared to typical resolution of DNS of incompressible flows, 
 the values of the resolution reported in Table \ref{tab:SimParams} would  indicate that our simulations are slightly  
 over-resolved (e.g,  \citealp{moin1998direct} recomends $\Delta x = 8\eta$ in the streamwise direction,  and $\Delta y = 4\eta$  in the shear-wise direction for homogeneous shear turbulence). 
 However,  it should be noted that the non-linear terms of equations (\ref{eq:phieq})-(\ref{eq:cont3}) are not quadratic, resulting in stronger aliasing and stricter limitations in the resolution than typically encountered in incompressible flows.

\begin{figure}
\begin{center}
\begin{minipage}{0.49\linewidth}
\centerline{$(a)$}
\includegraphics[width=\linewidth]{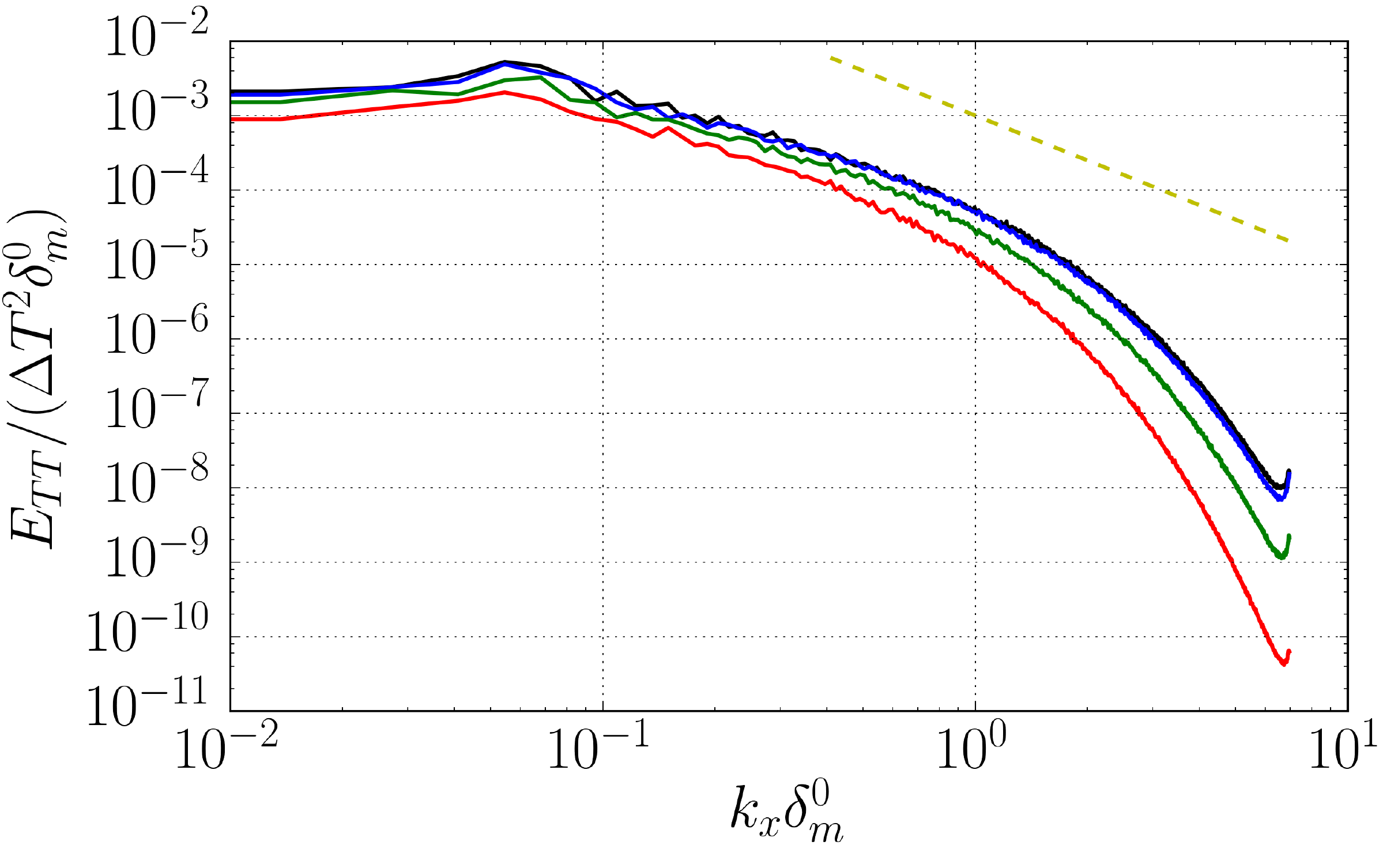}
\end{minipage}
\begin{minipage}{0.49\linewidth}
\centerline{$(b)$}
\includegraphics[width=\linewidth]{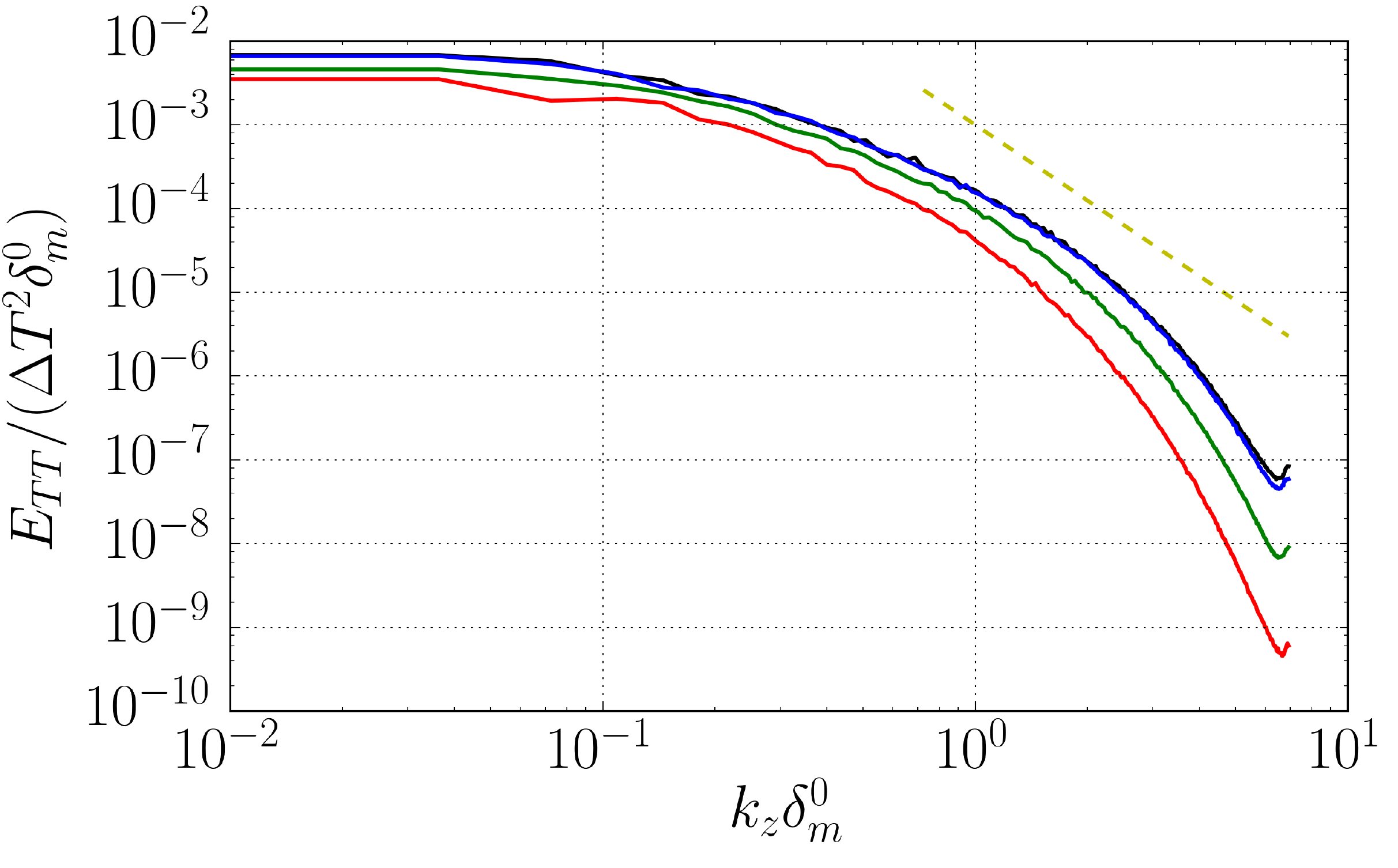}
\end{minipage}
\begin{minipage}{0.49\linewidth}
\centerline{$(c)$}
\includegraphics[width=\linewidth]{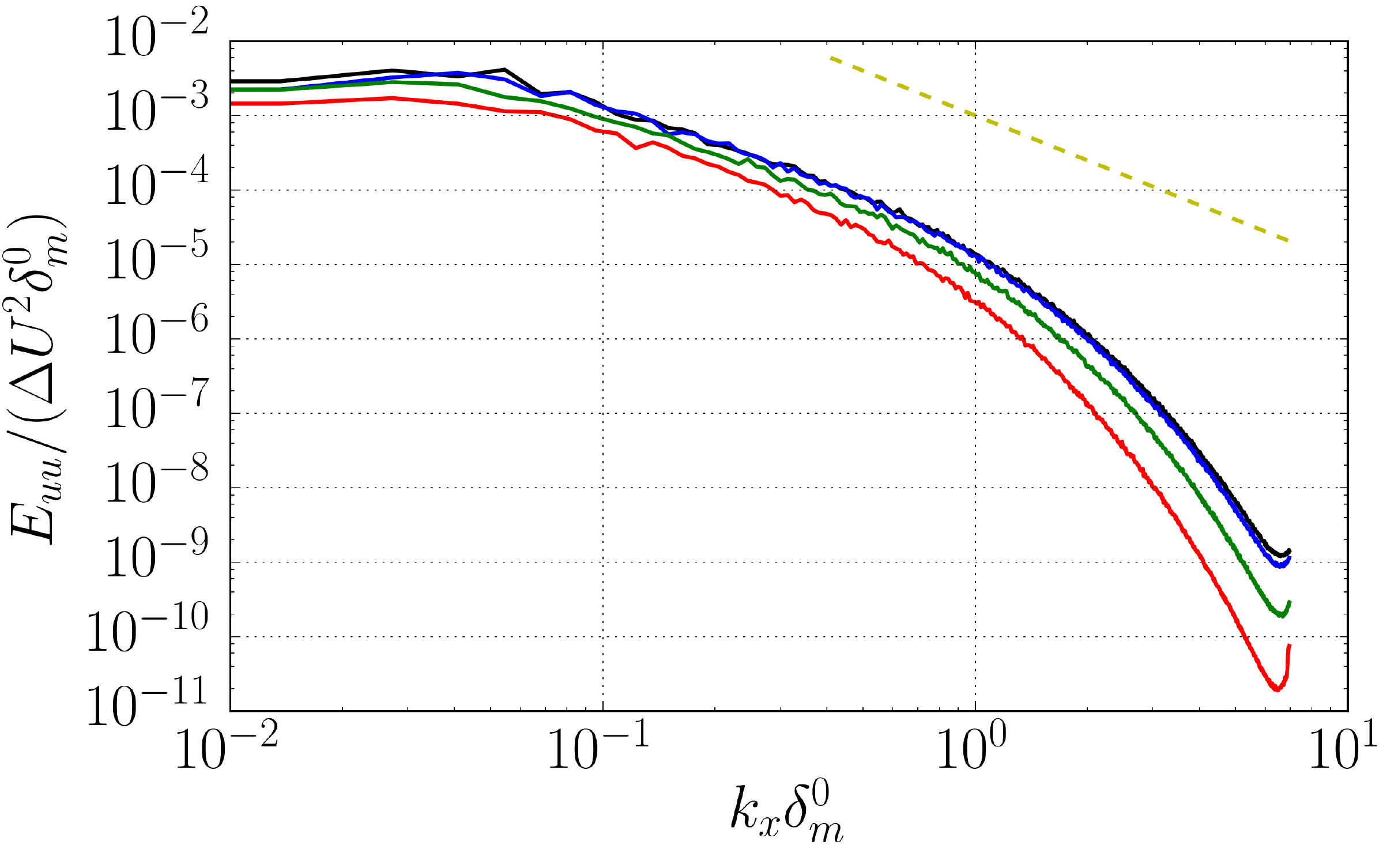}
\end{minipage}
\begin{minipage}{0.49\linewidth}
\centerline{$(d)$}
\includegraphics[width=\linewidth]{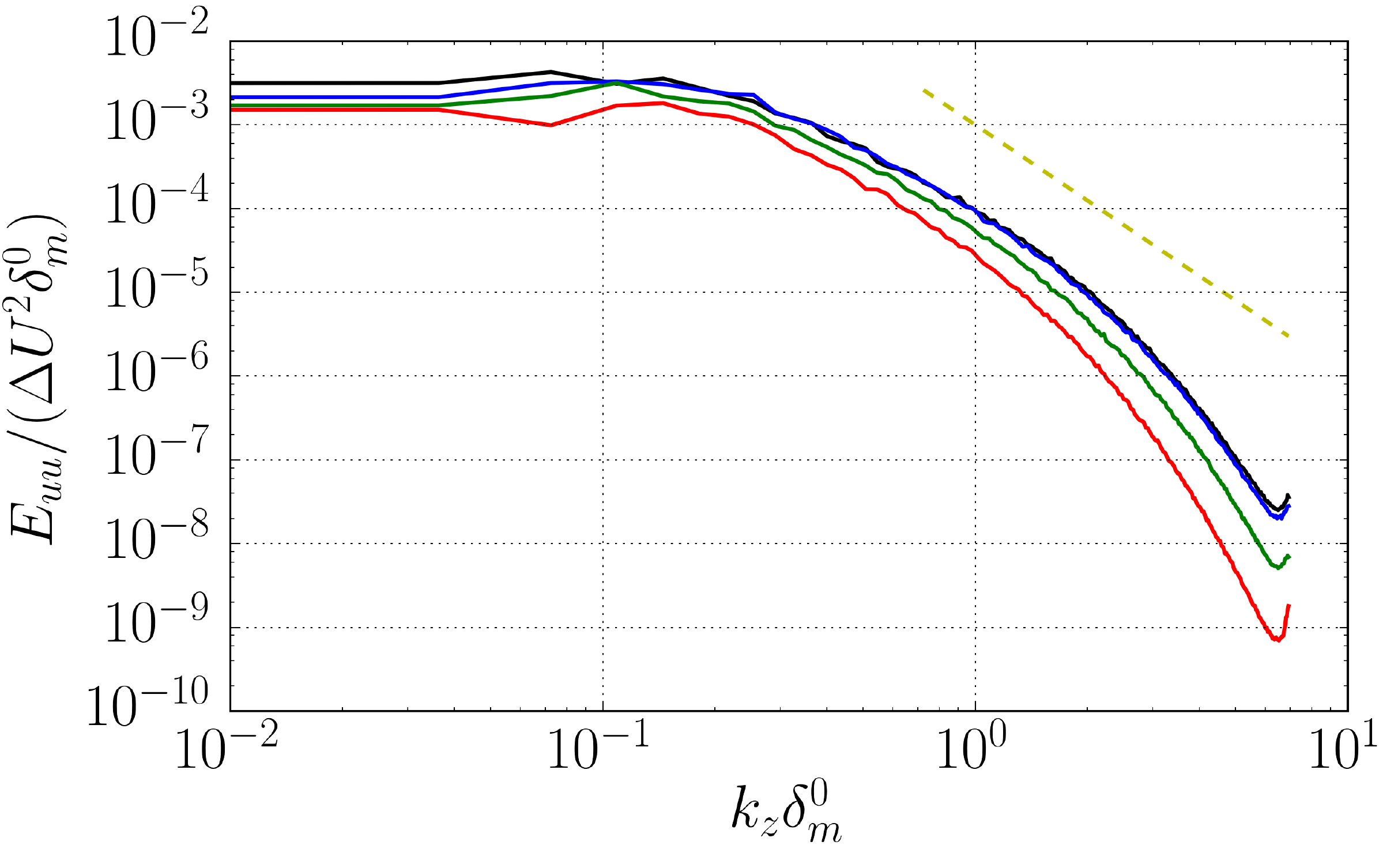}
\end{minipage}
\end{center}
\caption{
1D spectra at mid plane of the computational domain ($y=0$), at the beginning of the self-similar range. 
({\em a}) $E_{TT}(k_x \delta_m^0)$. 
({\em b}) $E_{TT}(k_z \delta_m^0)$. 
({\em c}) $E_{uu}(k_x \delta_m^0)$. 
({\em d}) $E_{uu}(k_z \delta_m^0)$. 
Different colours correspond to different density ratios: black, $s=1$; blue, $s=2$; green, $s=4$ and red, $s=8$. 
}
\label{fig:1Dspectra}
\end{figure}

The extent of the aliasing errors can be examined in figure \ref{fig:1Dspectra}, which shows the one-dimensional spectra of the streamwise velocity and temperature ($E_{uu}$ and $E_{TT}$) as functions of the streamwise and spanwise wavenumbers ($k_x$ and $k_z$). The spectra is computed at center of the computational domain ($y=0$) and at the beginning of the self-similar range discussed in section \ref{sec:SelfsimilarEvolution}. 
For those times, the spatial resolution and the aliasing errors are more critical, since the Kolmogorov length scale slowly grows during the self-similar evolution of the mixing layer (not shown). 
Besides that, the one-dimensional spectra in figure \ref{fig:1Dspectra} only shows a slight energy pile-up at the largest wavenumbers as a consequence of the aliasing errors, similar to those observed in DNS of homogeneous isotropic turbulence for incompresible flows (see for instance \citealp{kaneda2006high}). 
Note that in the present case the aliasing errors do not preclude the existence of a viscous range (where the energy decays faster that the -5/3 law, indicated in figure \ref{fig:1Dspectra} by dashed lines), and that the energy levels associated to the energy pile up are up to five orders of magnitude smaller than those of the  energy containing scales.

Finally, it should be noted that the use of relatively large computational domains is motivated by 
two reasons, 
fidelity of the turbulent structures in the mixing layer and statistical convergence. 
First, a large domain in the $y$-direction allows the mixing layer to grow for longer 
times before confinement effects develop, resulting in a longer self-similar range. 
In the present simulations,  the visual thickness of the mixing layer at the end of the self-similar range 
is smaller than 30\% of the vertical size of the computational domain.
 
Second, the horizontal size of the domain 
also needs to be large enough to capture the largest structures of the flow. 
For reference, in our simulations less than 6\% of the turbulent kinetic energy is contained in infinitely large modes in the streamwise ($k_x =0$) and spanwise ($k_z=0$) directions at the end of the self-similar range, when the turbulence structures are largest. As discussed later in section \ref{sec:spectra}, this percentage is a little bit larger  for the temperature variance ($\approx 15\%$), which tends to have a stronger signature in $k_z=0$ modes than the turbulent kinetic energy. 
Also, in order to improve the statistical convergence, the horizontal averaging is complemented with an ensemble average over the four independent runs (i.e., with different initial conditions) performed for each density ratio. 
%

\begin{table}
\begin{center}
\begin{tabular}{ccccccc}
$s$ & 
$\tau_0$ - $\tau_f$ & $Re_w$ & $Re_\lambda$ & $(\Delta x /\eta)_{\max}$ & $(\Delta y/\eta)_{\max}$  
& $D_w$  \\
\hline
 1    & 380-520  & 4200-6300    & 140-150       & 1.7-1.6      & 1.1   - 1.05    &   4.8           \\
 2    & 400-520  & 4500-5800    & 130-140       & 1.6-1.5      & 1.05 - 0.95    &   5.2              \\
 4    & 440-620  & 4500-6500    & 110-120       & 1.4-1.3      &  0.9  - 0.8      &   6.1             \\
 8    & 550-730  & 4900-7000    &    85-95        & 1.2-0.9     &  0.7  - 0.6       &   7.7             \\

 \hline
 
\end{tabular}
\caption{ Relevant parameters of the simulations within self-similar period. All the ranges 
correspond to the values of the parameter at the beginning ($\tau=\tau_0$) and end ($\tau=\tau_f$) 
of the self-similar evolution, discussed in section \ref{sec:SelfsimilarEvolution}. $Re_w = \rho_0 
\Delta U \delta_w /\mu$, 
where $\delta_w$ is the vorticity thickness. $Re_\lambda = q \lambda/\nu$, where $\lambda$ is the 
Taylor microscale and $q^2$ is twice the turbulent kinetic energy. 
$\Delta x$ and $\Delta y$ are the streamwise and vertical grid spacings in collocation points, respectively.
$\eta$ is the Kolmogorov lengthscale.
$D_w = \delta_w/\delta_m$, where
$\delta_m$ is the momentum 
thickness and $\delta_w$ is the vorticity thickness.} 
\label{tab:SimParams}
\end{center}
\end{table}

\section{Results} \label{sec:results}

\subsection{Self-similar evolution}
\label{sec:SelfsimilarEvolution}

It is well known that temporal mixing layers reach a self-similar evolution after an initial 
transient, in which the initial perturbations evolve into the structure of the fully developed 
turbulent mixing layer 
\citep{Rogers1994, Pantano2002}.
In the self-similar evolution, the mixing layer thickness grows linearly with time, and large-scale 
quantities scaled with the variation across the mixing layer (i.e., $\Delta U$, $\rho_b-\rho_t$, 
etc.)
collapse into a single profile when plotted as a function of $y/\delta_m(t)$ or $y/\delta_w(t)$.  

\begin{figure}
\begin{center}
\begin{minipage}{0.49\linewidth}
\centerline{$(a)$}
\includegraphics[width=\linewidth]{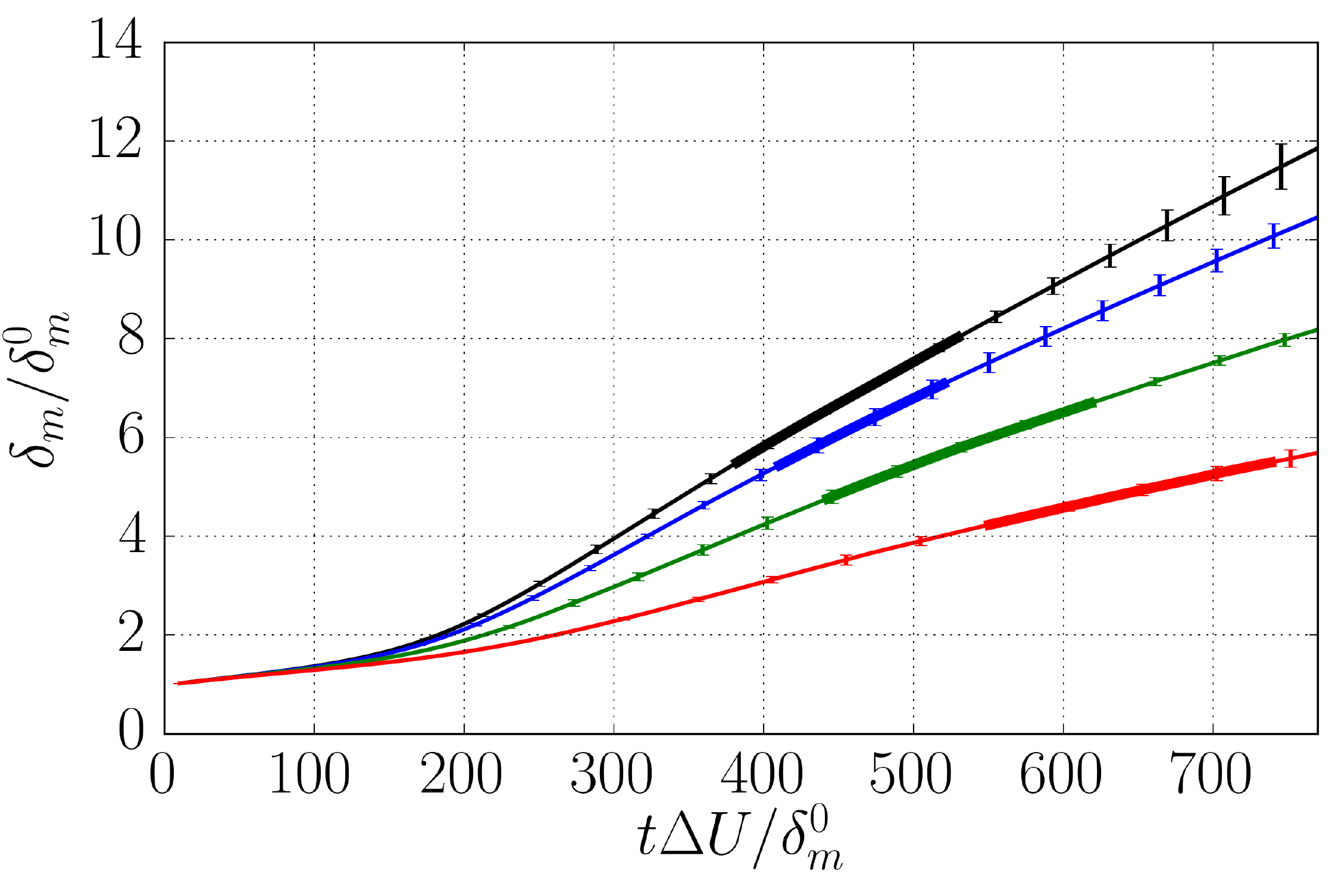}
\end{minipage}
\begin{minipage}{0.49\linewidth}
\centerline{$(b)$}
\includegraphics[width=\linewidth]{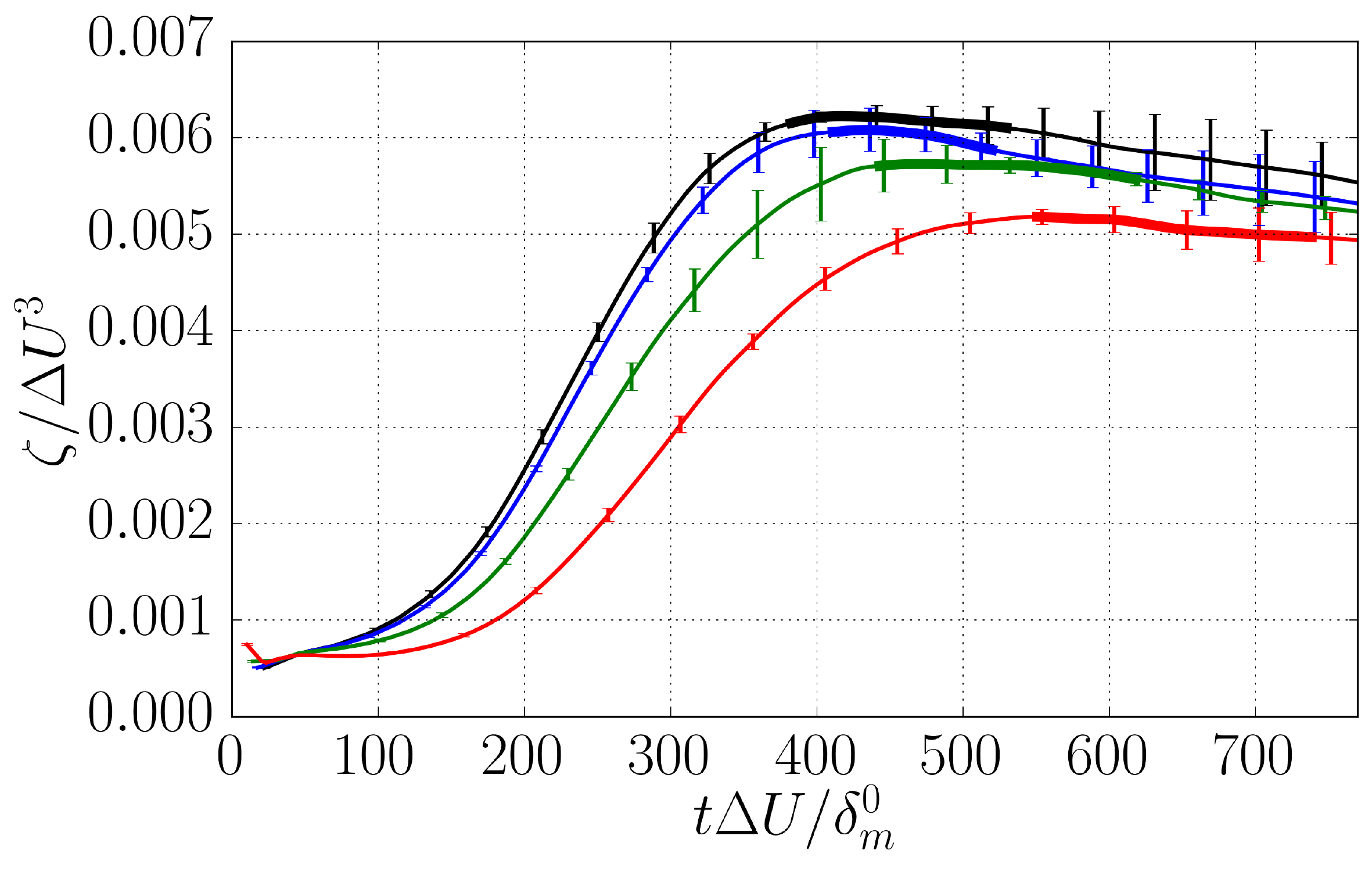}
\end{minipage}
\end{center}
\caption{ 
Temporal evolution of ({\em a}) the momentum thickness $\delta_m$ divided by the initial momentum 
thickness $\delta_m^0$, and ({\em b}) the non-dimensional integrated turbulent energy 
dissipation rate, $\zeta/\Delta U^3$. Line types are black for $s=1.0$, blue for $s=2.0$, green for 
$s=4.0$ and red for $s=8$. 
The values correspond to the ensemble average of the 4 runs for each density ratio, and the error bars are the corresponding standard deviation.   
The thick line shows the ranges of self-similar evolution for each density ratio.
\label{fig:selfsim}
}
\end{figure}

In order to evaluate the self-similar evolution of the present DNS results, figure \ref{fig:selfsim}({\em a}) 
shows the evolution of $\delta_m(t)$ for the four cases considered here. The variablity in $\delta_m$ is estimated using the standard deviation of the momentum thicknesses over the four runs, and is indicated with error-bars in the figure. Also, figure \ref{fig:selfsim}({\em b}) shows 
the time evolution of 
the integrated dissipation rate of turbulent kinetic energy
\begin{equation} 
\zeta=\int_{-\infty}^\infty \overline{\varepsilon} dy. 
\end{equation}
The quantity $\zeta$ scales with $\Delta U^3$ and, therefore, should be constant with time,
once self-similarity has been achieved.
The expression for the dissipation rate of turbulent kinetic energy for variable density flows can
be found in  \cite{Chassaing2002}, and is reproduced here for completeness
\begin{equation}
\overline{ \rho } \, \overline{ \varepsilon } =  \frac{4}{3} \mu \overline{\theta'^2} +
\mu \overline{\omega'_{i} \omega'_{i}} +
2\mu\left( \frac{\partial^2 \overline{ u'_i u'_j}}{\partial x_i \partial x_j} - 
2\frac{\partial\overline{\theta' u'_j} }{\partial x_j} \right),
\label{eq:epsilon}
\end{equation}
where primed variables denote fluctuations with respect to the mean,
$\theta=\partial{u_i}/\partial{x_i}$ is the divergence of the velocity,
and $\omega_i$ are the components of the vorticity.
The results presented in figure \ref{fig:selfsim} show that self-similarity is achieved after an initial transient, with $\delta_m(t)$ growing linearly with time and $\zeta(t)$ becoming approximately constant (at least within the errors in $\zeta$).  However, comparing  
figures \ref{fig:selfsim}({\em a}) and \ref{fig:selfsim}({\em b}) it can be observed
that 
the linear growth of $\delta_m$ starts at $\tau = t \Delta U/\delta_m^0\approx 200$, a time at which 
$\zeta$ is still growing. 
This behavior was also observed by \cite{Rogers1994}, and it 
 indicates that the  determination of the time interval where self-similarity is achieved needs a careful consideration, 
 and should not be determined exclusively from a linear evolution of $\delta_m(t)$.

\begin{figure}
\begin{center}
\begin{minipage}{0.49\linewidth}
\centerline{$(a)$}
\includegraphics[width=\linewidth]{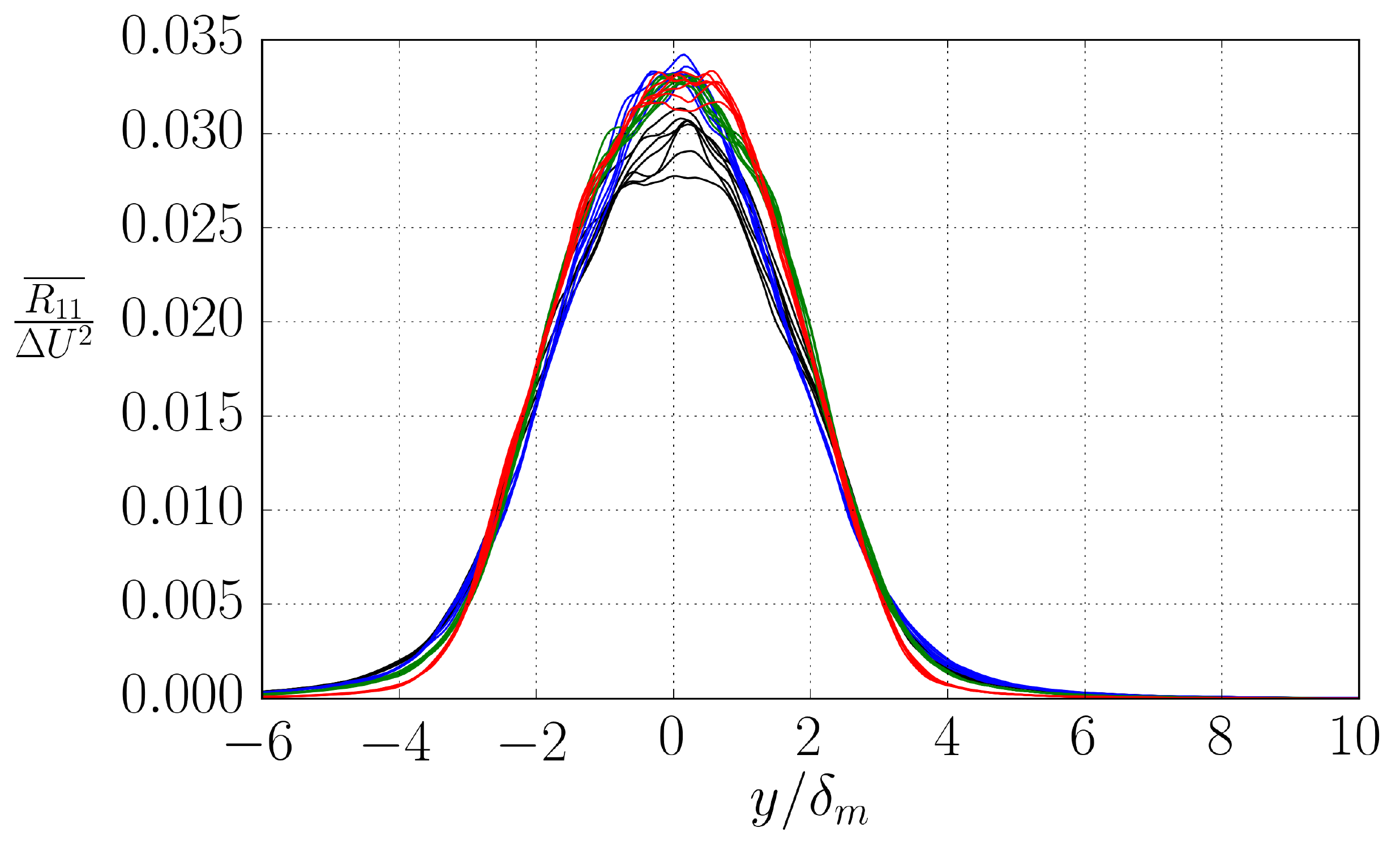}
\end{minipage}
\begin{minipage}{0.49\linewidth}
\centerline{$(b)$}
\includegraphics[width=\linewidth]{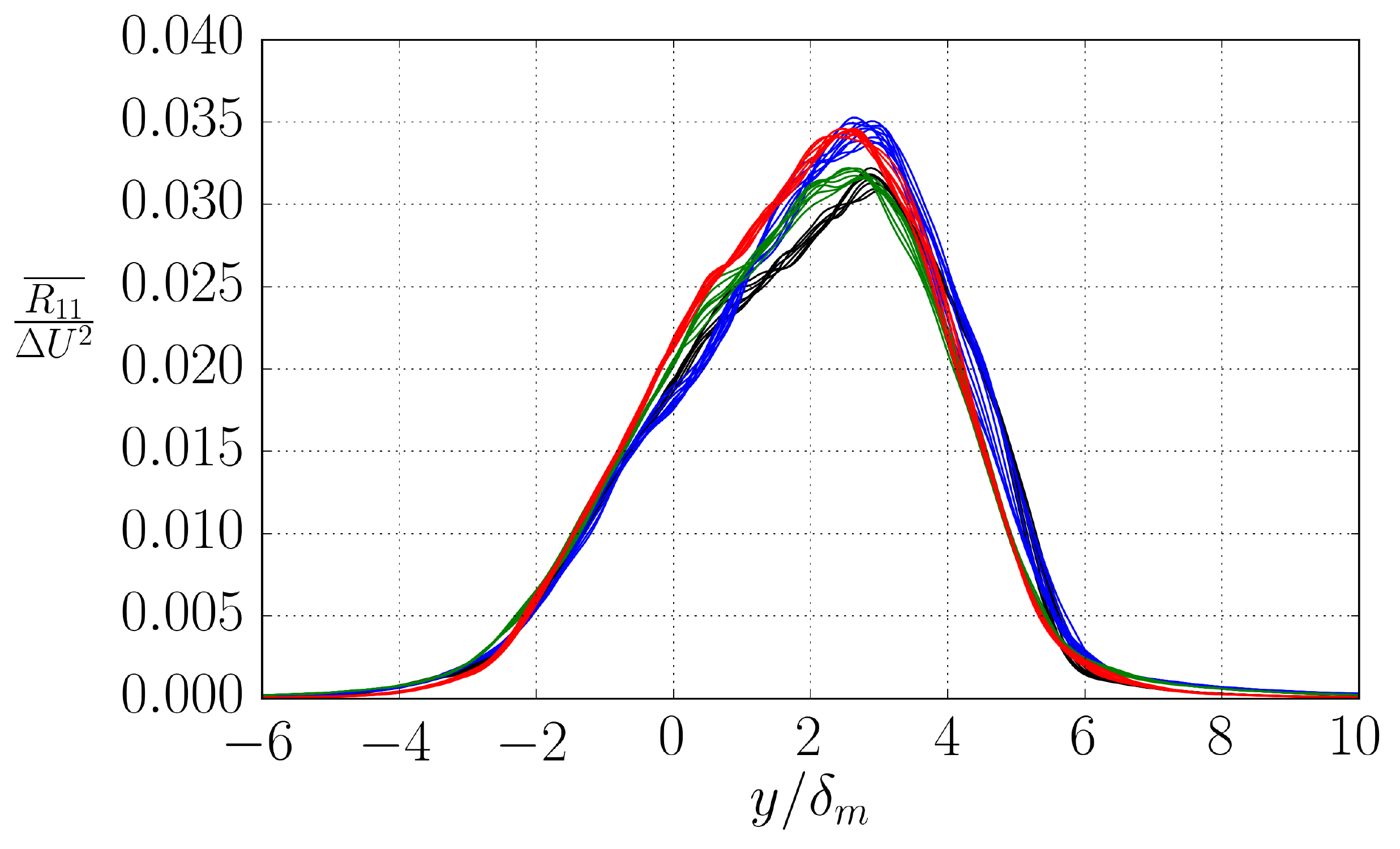}
\end{minipage}
\end{center}
\caption{ 
Reynolds stress $R_{11}$ profiles within the self-similar range for ({\em a}) case $s=1$, all runs with a total of 26 profiles,  and ({\em b}) case $s=4$, all runs, with a total of 31 profiles.  Colors are used to differentiate between runs.
\label{fig:selfsim2}
}
\end{figure}

In the present study, and for the purpose of collecting statistics, 
we have defined the time interval $[\tau_0,\tau_f]$ where the mixing layer is self-similar 
by analyzing the collapse of the 
instantaneous (i.e., averaged in the horizontal directions only) profiles of the normal Reynolds stresses,  $R_{11}(y/\delta_m,\tau)$, $R_{22}(y/\delta_m,\tau)$,
and $R_{33}(y/\delta_m,\tau)$.
We have computed the temporal mean and standard deviation of these Reynolds stresses for several 
time intervals, 
selecting for each run the longest time interval 
 in which the standard deviation of the normal Reynolds stresses is smaller than 5\% of the maximum. 
The resulting time intervals (more explicitly, the maximal time interval over the four runs for each density ratio)  are shown in figure \ref{fig:selfsim} and reported in table \ref{tab:SimParams}, yielding 
a total self-similar range of at least 10 eddy-turnover times per density ratio.
For illustration, figure \ref{fig:selfsim2} 
 shows all the $R_{11}$ profiles 
within the self-similar range for the cases $s=1$ and $s=4$, 
using different color for each run.  
The agreement of the profiles is good, especially taking into account that there are 26 and 31 curves on each plot, respectively.  The differences are more apparent near the maximum of the Reynolds stresses. It is interesting to note that 
the variability of the profiles within each run is small, similar to that reported by \cite{Pantano2002}. On the other hand, the variability between different runs is a bit larger, and it is probably linked to differences between the largest structures developed in each run (i.e., by different realizations of the initial conditions), emphasizing the importance of running several  realizations of each density ratio to accumulate statistics for the largest structures in the mixing layer.

\subsection{Effects of the density ratio on the growth rate}
\label{sec:growth} 

Once the self-similar time interval has been defined, we analyse the effect that the density ratio 
has on the growth rate of the temporal mixing layer, comparing the results of the present zero Mach cases 
with those obtained by \cite{Pantano2002} for convective Mach number $M_c=0.7$. First, consider the growth rate 
of the momentum thickness, $\dot\delta_m$, which 
is evaluated here following the expression derived in \cite{vreman:1996}, 
\begin{equation}
\dot\delta_m \approx  -  \frac{2}{\rho_0 \Delta U^2} \int_{-\infty}^{\infty}\overline{\rho} R_{12} \ppy{\tilde{u}} dy.
\label{eq:dotdeltam}
\end{equation}
This expression is obtained differentiating (\ref{eq:momthickness}) with respect to time, and neglecting viscous terms. An alternative method to compute $\dot\delta_m$ is to fit a linear law to the data shown in figure \ref{fig:selfsim}({\em a}). The differences in the mean and standard deviation of the growth rate of the  momentum thickness obtained from both methods are small: for $s=1$, the first method yields $\dot\delta_m/\Delta U = 0.0168 \pm 0.0003$, while the second method yields $\dot\delta_m/\Delta U = 0.0170\pm 0.0002$. 

The value of the growth rate of the momentum thickness for $s=1$ is in good agreement with previous works, especially taking into account the scatter of the available data. For instance,  in the ``unforced" experiments quoted by \citep{dimotakis1991turbulent} the growth rate of the momentum thickness varies from 0.014 to 0.022. 
Also, \cite{Rogers1994} report $\dot\delta_m /\Delta U = 
0.014$ in simulations of incompressible temporal mixing layers, and the experimental data of \cite{Bell1990} yield a value of 0.016. 
For $M_c = 0.3$ and $s=1$, \cite{Pantano2002} report $\dot\delta_m/\Delta U \approx 0.0184$, a value that decreases to $0.0108$ when the Mach number is increased to $M_c = 0.7$.  

\begin{figure}
\begin{center}
\begin{minipage}{0.49\linewidth}
\centerline{$(a)$}
\includegraphics[width=\linewidth]{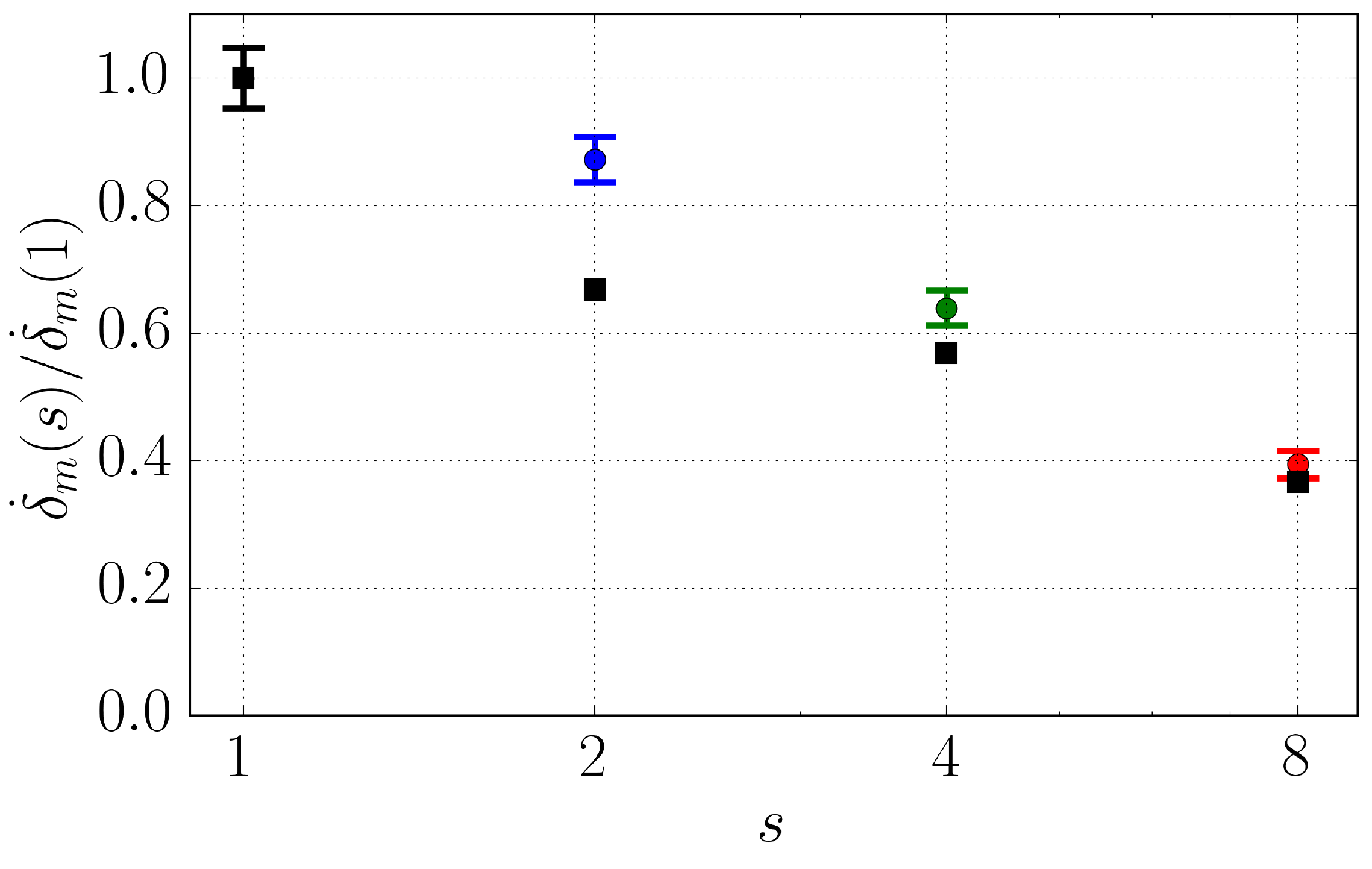}
\end{minipage}
\begin{minipage}{0.49\linewidth}
\centerline{$(b)$}
\includegraphics[width=\linewidth]{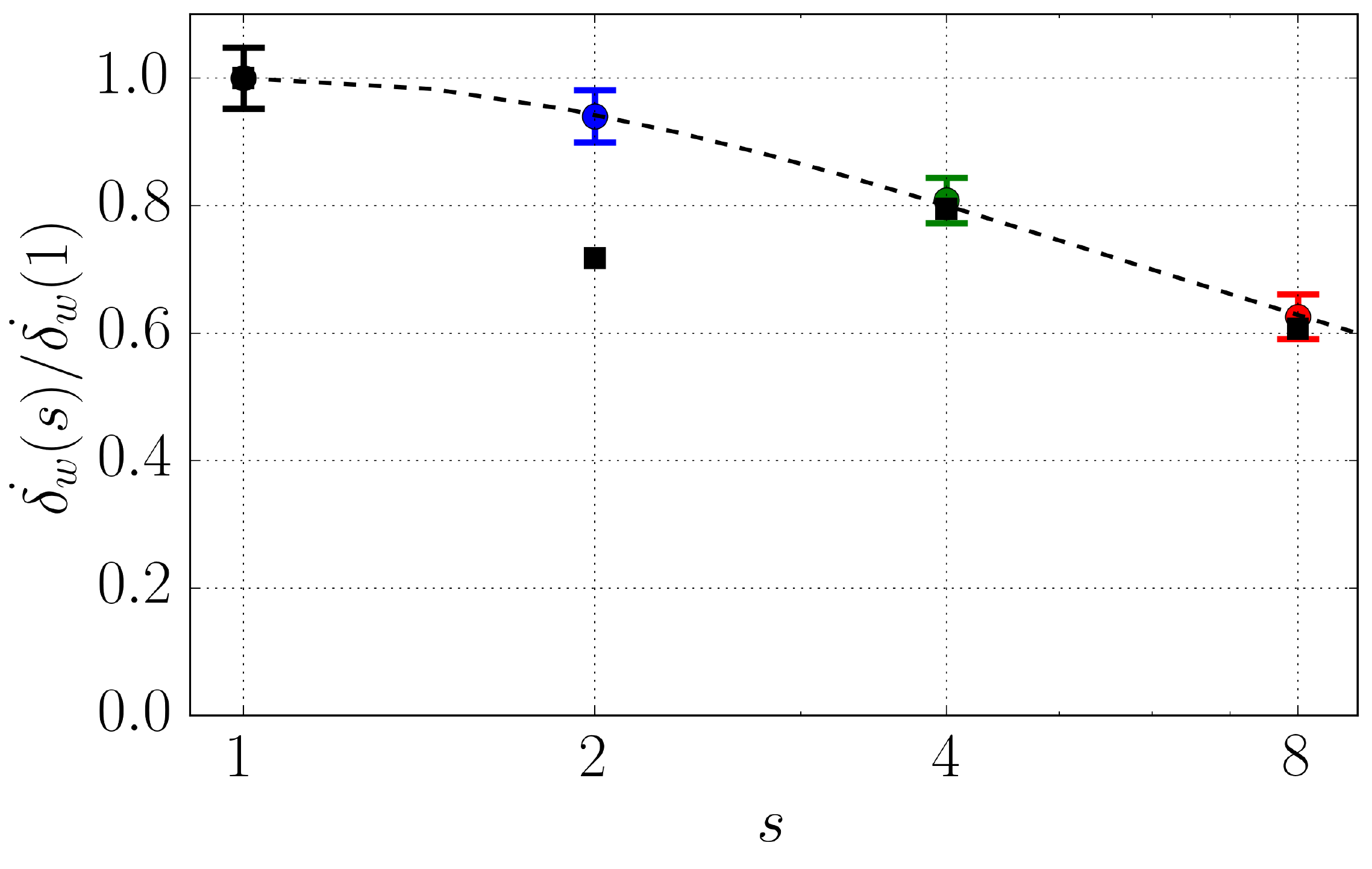}
\end{minipage}
\end{center}
\caption{Mixing-layer growth rate as a function of the density ratio. Growth rate based on ({\em a}) momentum 
thickness 
$\dot{\delta}_m$, and ({\em b}) vorticity thickness $\dot{\delta}_w$, normalised by the growth rate for 
$s=1.0$. In both panels the horizontal axis is in logarithmic scale. 
Colored dots with error bars stand for the present results, squares represent results for $M_c=0.7$ 
\citep{Pantano2002}. 
The dashed curve in ({\em b}) corresponds to equation (\ref{eq:ramshaw}), from  \citet{Ramshaw2000}. 
}
\label{fig:dotdmcomparison}
\end{figure}

As the density ratio increases, the values of $\dot\delta_m$ decrease. 
This 
reduction of the growth rate is quantified in figure \ref{fig:dotdmcomparison}({\em a}), in terms of the ratio of growth rates, $\dot{\delta}_m(s)/\dot{\delta}_m(1)$. 
For $s=8$, our results show that the growth rate of $\delta_m$ has been reduced by 60\% with respect to the growth rate of the case with $s=1$. 
A similar behaviour is observed for the subsonic cases of \cite{Pantano2002} at $M_c = 0.7$, also included in the figure. The ratio $\dot{\delta}_m(s)/\dot{\delta}_m(1)$ is very similar for the $M_c = 0$ and $M_c = 0.7$ cases for large density ratios, with significant differences for the smaller density ratio, $s=2$. Careful inspection shows that the density ratio $s=2$ is 
indeed somewhat anomalous in \citet{Pantano2002}, presenting a non-monotonic behaviour for some quantities 
(see for instance the growth rates and the profiles of Reynolds stress, as shown in table 6 and figure 18 respectively in their paper).

%

The results shown in figure \ref{fig:dotdmcomparison}({\em a}) for $\dot\delta_m$ are very similar to those obtained for $\dot\delta_w$, which are plotted in figure \ref{fig:dotdmcomparison}({\em b}). Again, our results are compared to the $M_c = 0.7$ cases of \cite{Pantano2002}, and the 
 
theoretical prediction by \cite{Ramshaw2000}. The latter is based on a model for the growth of the {\em visual} thickness of 
a variable density mixing layer at $M_c=0$, directly comparable to the present results. The model is obtained by extending  a linear stability analysis to the nonlinear regime through 
scaling hypothesis, leading after proper manipulation to 
\begin{equation}
\frac{\dot{\delta}_w(s)}{\dot{\delta}_w(1)}=\frac{2\sqrt{s}}{s+1}.
\label{eq:ramshaw}
\end{equation} 
Figure \ref{fig:dotdmcomparison}({\em b}) shows  a very good agreement between Ramshaw's model and our data. 
The agreement is also fairly good with the subsonic data of \cite{Pantano2002} at $M_c=0.7$,  
except for the case $s=2$ as it happened also for $\dot\delta_m$. 
It should be noted that, to the best of our knowledge, this is the first direct validation of the 
Ramshaw model with a variable density DNS at $M_c=0$. 

Overall, 
 the results presented in this subsection show that the growth rates of the $M_c=0$ cases are significantly higher than those reported by \citet{Pantano2002} for $M_c=0.7$, in agreement with previous works.
 However, the effect of $s$ on the reduction of the growth rate seems to be very similar at both Mach numbers, 
 except for maybe the low density ratio case, $s=2$. Also, the effect of $s$ seems to be stronger on $\delta_m$ than on $\delta_w$, with $\dot{\delta}_m(s=8)/\dot{\delta}_m(s=1) \approx 0.4$ and $\dot{\delta}_w(s=8)/\dot{\delta}_w(s=1)\approx0.6$.
As a consequence, the ratio between the two thicknesses, $D_w = \delta_w/\delta_m$, increases with $s$, as it can be observed in table \ref{tab:SimParams}. Note that since $\delta_w$ and $\delta_m$ grow linearly with time, $D_w \approx \dot\delta_w/\dot\delta_m$ for sufficiently long times.
For reference, \cite{Pantano2002} report a value of $D_w=5.0$ for a compressible mixing layer with $M_c=0.3$ and $s=1$, in good agreement with $D_w=4.83$ for our $s=1$ case.

\subsection{Mean density, velocity and temperature}
\label{sec:mean}

We now proceed to analyze the one point statistics 
 of the present DNS (mean values in this subsection, higher order moments in \S\ref{sec:rms}),  
averaging the data in the horizontal directions and in time, 
binning in $y/\delta_m(t)$. In all the vertical profiles presented in this section, a shadowing has been applied around plus/minus one 
standard deviation of the horizontally averaged data with respect to the mean, 
in order to show the uncertainty of the statistics.

%
\begin{figure}
\begin{center}
\begin{minipage}{0.49\linewidth}
\centerline{$(a)$}
\includegraphics[width=\linewidth]{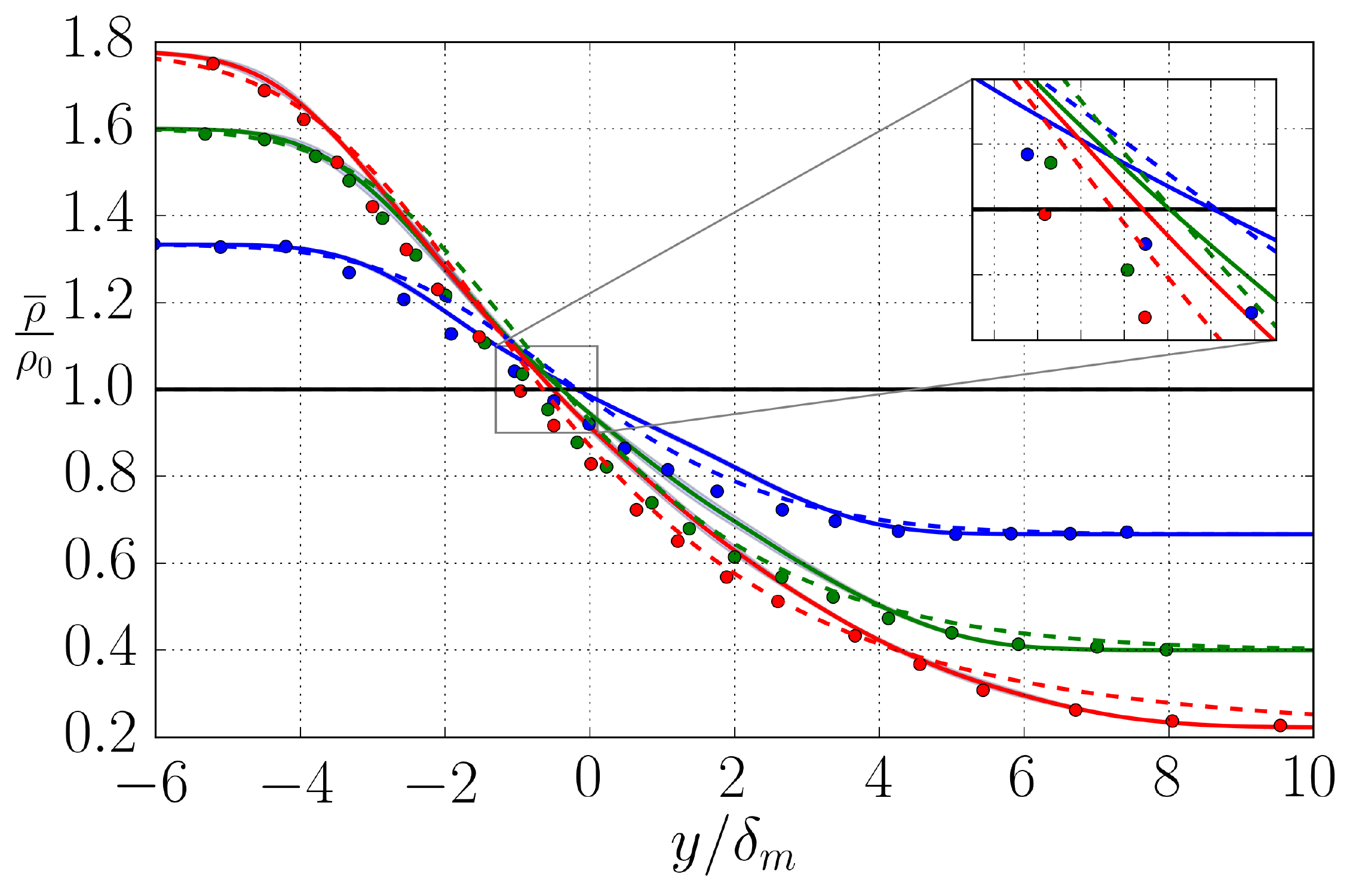}
\end{minipage}
\begin{minipage}{0.49\linewidth}
\centerline{$(b)$}
\includegraphics[width=\linewidth]{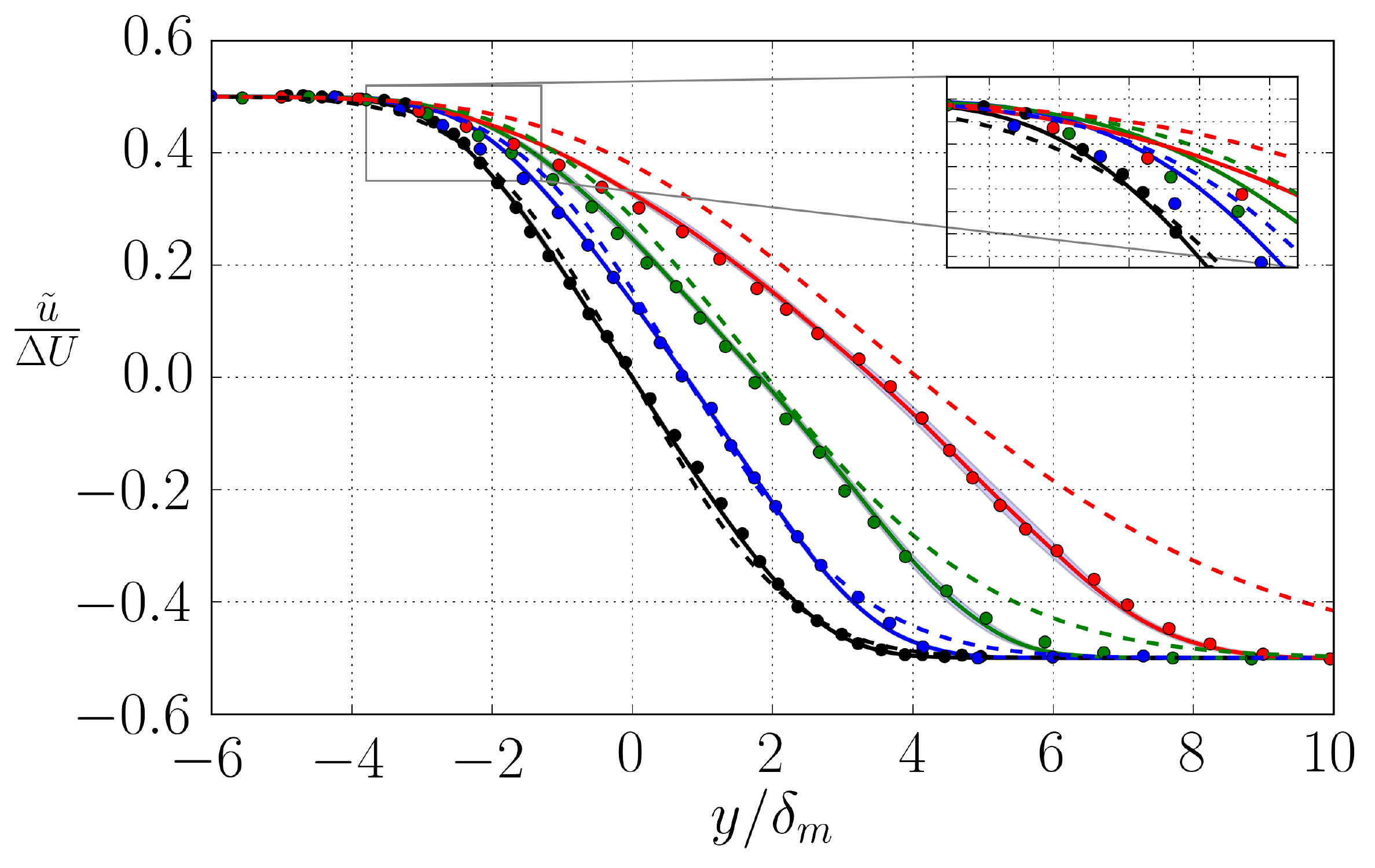}
\end{minipage}
\end{center}
\caption{$(a)$ Reynolds-average density profiles. $(b)$ Favre-averaged streamwise velocity profiles.
Different colors correspond to different density ratios: black, $s=1$; blue, $s=2$; green, $s=4$; and red, $s=8$. 
Solid lines are the present turbulent temporal mixing layers. Dashed lines are the laminar temporal mixing layers (see apendix \ref{sec:laminar}). 
Symbols: \citet{Rogers1994} for $s=1$, 
\citet{Pantano2002} for s=$2,4$ and 8.
\label{fig:profiles_mean}
}
\end{figure}
%

Figure \ref{fig:profiles_mean}({\em a}) shows  mean density profiles,  
comparing the present zero Mach results with the results of the subsonic mixing layer of \citet{Pantano2002} at $M_c=0.7$. The figure also includes for comparison the results from laminar temporal mixing layers, obtained as discussed in appendix \ref{sec:laminar}.
As the density ratio increases, the density mixing layer extends further into the low-density stream, with small variations in the position where $\overline{\rho}=\rho_0$. 
The profiles of the $M_c = 0.7$ and $M_c=0$ cases are qualitatively similar at any given density ratio, although there are some differences in the profiles in the central part of the mixing layer ($\abs{y} \lesssim 3\delta_m$). 
The agreement between the $M_c=0$ and $M_c=0.7$ cases is better for the Favre averaged velocity, shown in figure \ref{fig:profiles_mean}({\em b}). The only exception is maybe the region closer to the high-density free-stream, where the edge of the mixing layer seems sharper for the present simulations ($M_c=0$). The figure also includes the incompressible data of \cite{Rogers1994} for $s=1$, showing a very good agreement with our incompressible case.  

Besides some small changes in the shape of the profiles (which will be discussed later), the most apparent effect of the density ratio in $\overline{\rho}$ and $\tilde{u}$ is the shifting of the $\tilde{u}$ profile towards the low density side. 
Note that this effect is apparent in both turbulent cases ($M_c=0$ and $M_c=0.7$), as well as in the laminar self-similar profiles (dashed lines in figure \ref{fig:profiles_mean}).   
This shifting of the mean density and velocity profiles with the density ratio has already been reported in previous studies, both experimental and numerical,  
and it has been explained qualitatively in terms of the asymmetry in the momentum exchange of the large scales with the free-streams \citep{Brown1974} and their linear stability properties \citep{soteriou:1995}. 
Note that the mechanism proposed by \cite{Brown1974} acts in both,  turbulent diffusion (as originally proposed by the authors) and  mean velocity entrainment (i.e, $\langle \rho v \rangle$). 
While in turbulent mixing layers the turbulent diffusion is dominant over the mean velocity entrainment, the latter is important in laminar mixing layers. This could explain why figure \ref{fig:profiles_mean} also shows a clear shifting between $\tilde{u}$ and $\overline{\rho}$ for the laminar cases, as well as the results reported by  \cite{bretonnet:2007} in laminar mixing layers with density variations due to various effects:
different velocities, temperatures or species.

Besides the numerous qualitative observations of the shift between $\tilde u$ and $\overline{\rho}$ in the literature, few authors have tried to quantify it. 
In turbulent mixing layers, \citet{Pantano2002} proposed to quantify this shift 
using two semi-empirical relationships, $\overline{\rho}(\tilde{u})$ and $R_{12}(\tilde{u})$. They later used these relationships to estimate the reduction of the momentum thickness growth rate.  
 In laminar mixing layers, \citet{bretonnet:2007} proposed to characterize the drift as the distance between the inflection points of the velocity and density mean profiles. 

\begin{figure}
\begin{center}
\begin{minipage}{0.49\linewidth}
\centerline{$(a)$}
\includegraphics[width=\linewidth]{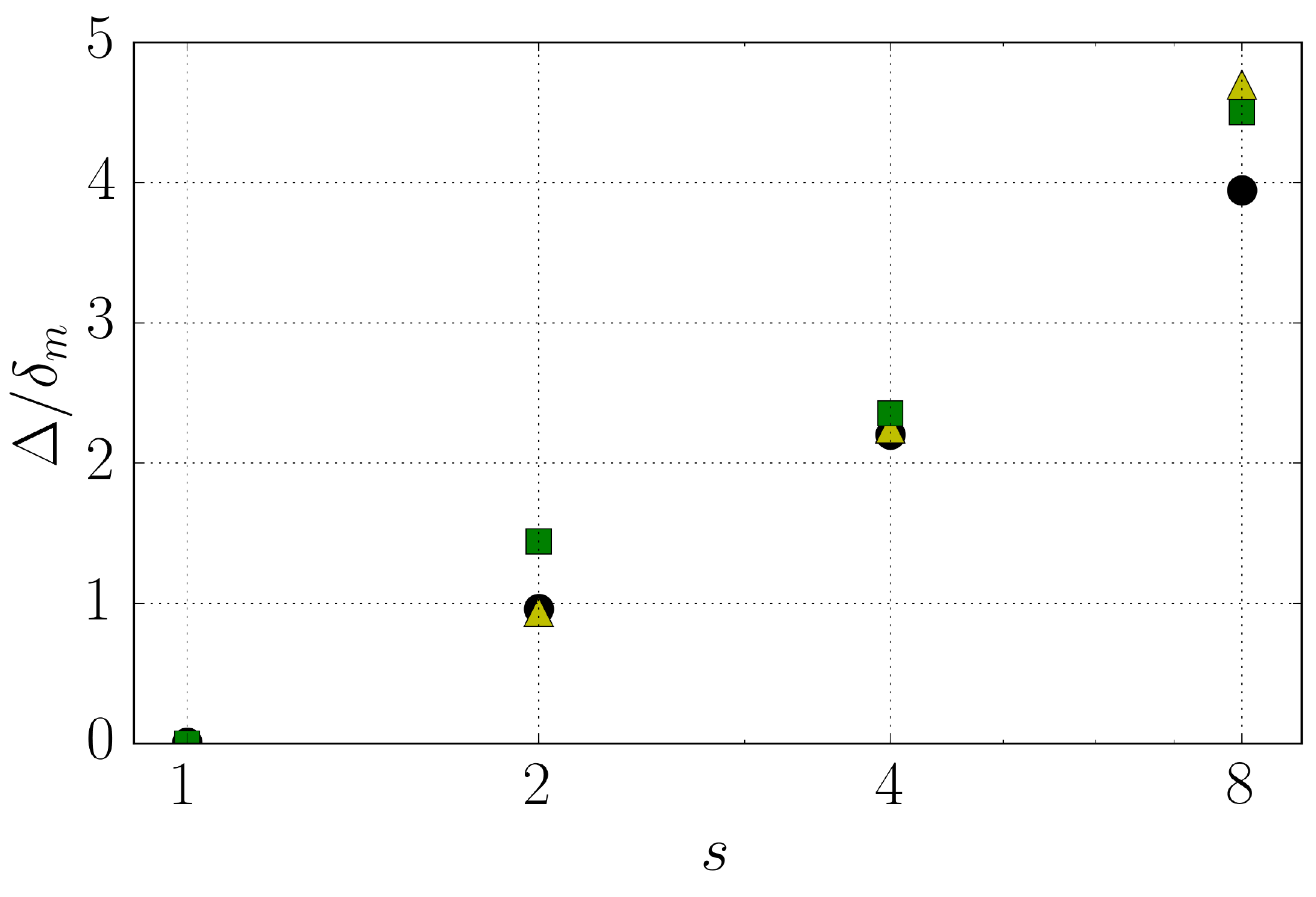}
\end{minipage}
\begin{minipage}{0.49\linewidth}
\centerline{$(b)$}
\includegraphics[width=\linewidth]{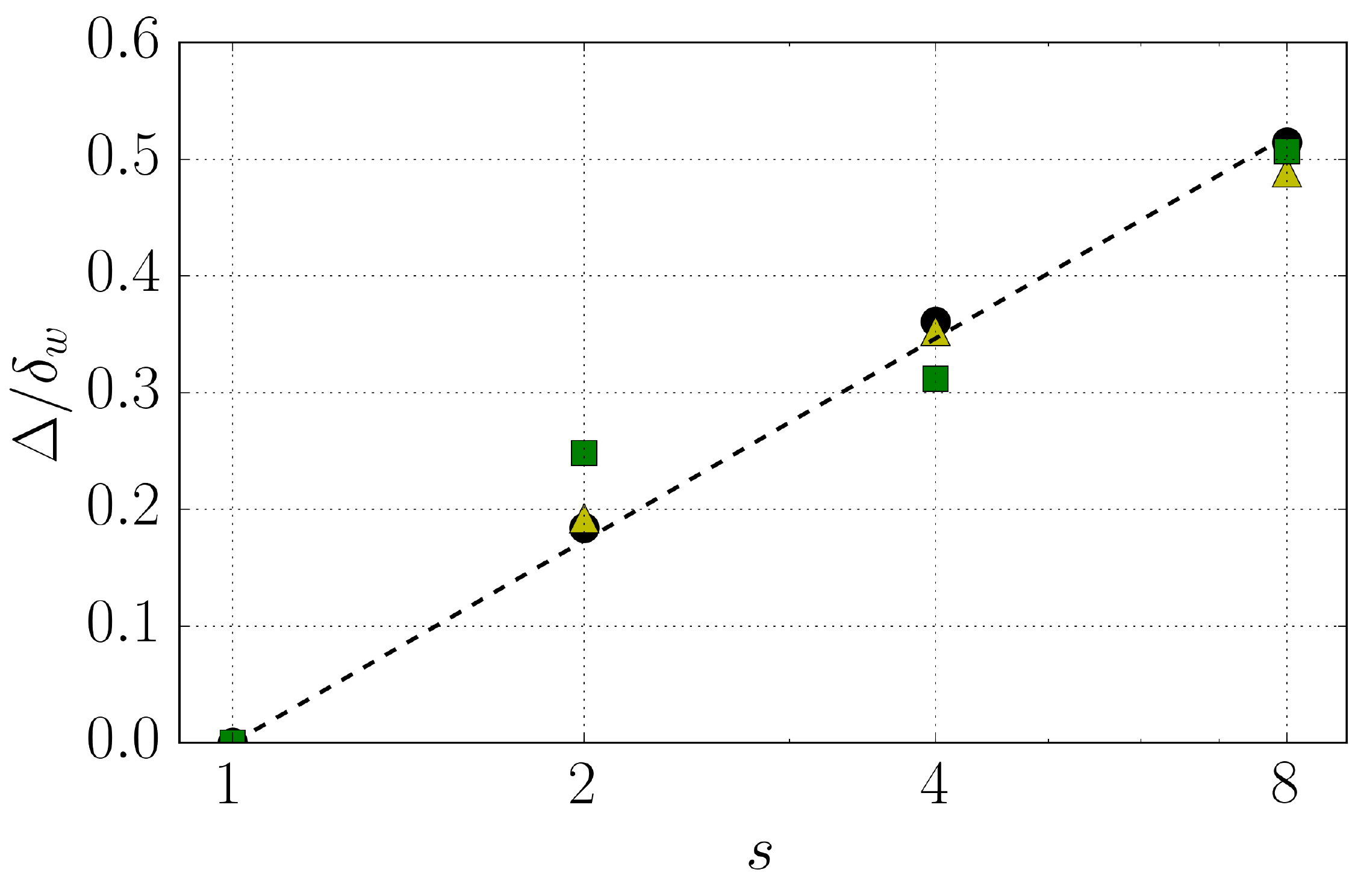}
\end{minipage}
\end{center}
\caption{Shifting of the mixing layer, normalized with $(a)$ the momentum thickness, $(b)$ the vorticity thickness. 
Circles for present DNS at $M_c = 0$. 
Squares for \citet{Pantano2002} at $M_c=0.7$. 
Triangles for laminar self-similar solutions.
The dashed line in ($b$) corresponds to $\Delta/\delta_w = 0.25 \log(s)$. 
\label{fig:shifting}
}
\end{figure}

Here we propose to quantify the shift using $\Delta$, which is defined as 
the distance between the $y$ locations where $\tilde{u}=0$ and $\overline{\rho} = \rho_0$, positive when $\tilde{u}$ is displaced towards $y$-positive (low-density side in our simulations). 
The main advantage of the present definition with respect to those used by \citet{Pantano2002} and \citet{bretonnet:2007} is that it can be easily computed from the mean profiles of velocity and density, without having to compute higher order derivatives. 
This distance is plotted in figure \ref{fig:shifting} as a function of the density ratio, for turbulent mixing layers with $M_c=0$ and $M_c=0.7$, and for the laminar self-similar solutions.  
The figure shows two possible scalings for $\Delta$, with $\delta_m$ (figure \ref{fig:shifting}{\em a}) and with $\delta_w$ (figure \ref{fig:shifting}{\em b}). The different datasets collapse better with the second scaling, especially for  $s=8$ cases, suggesting an empirical relation
\begin{equation} 
\Delta(s) = \delta_w(s) C_\Delta \log(s),  
\label{eq:empirical_shifting}
\end{equation}
with $C_\Delta = 0.25$. This empirical approximation yields correlation coefficients of $R^2=0.998$ for the present DNS results at $M_c=0$. Similar values of $C_\Delta$ are obtained for the other datasets in the figure. The results of \cite{Pantano2002} at $M_c=0.7$ yield  $C_\Delta=0.23$ and $R^2=0.956$, and the laminar self-similar solutions yield $C_\Delta =0.23$ and $R^2=0.994$. 
Finally, the good agreement between the laminar and turbulent data (and compressible and low-Mach number data) is consistent with the discussion in the previous paragraphs: the mixing layer is able to erode more easily the lighter free-stream, either by turbulent diffusion (in turbulent mixing layers) or by the mean entrainment (in laminar mixing layers). 
%

%
Although the present definition of shifting is not directly comparable to the one used by \cite{Pantano2002}, it is also possible to relate the present $\Delta$ to the ratio $\delta_m(s)/\delta_w(s)$. 
Lets assume that the mean density and velocity profiles are 
\begin{equation} 
\label{eq:prof_ideal}
\overline{\rho}  = \rho_0 + \frac{\rho_t-\rho_b}{2} F_\rho \left( \frac{y}{\delta_w} \right), 
\hspace{2mm}\mbox{and}\hspace{2mm} 
\tilde{u} = - \frac{\Delta U}{2} F_u \left(\frac{y}{\delta_w} - \frac{\Delta}{\delta_w} \right), 
\end{equation}
where $F_{u}(\xi)$ and $F_{\rho}(\xi)$ tend to $\pm1$ when $\xi \to \pm \infty$, and $\Delta$ is assumed to be a function of the density ratio, $s$. 
Note that this is equivalent to limiting the effect of $s$ to a shift between the profiles of $\overline{\rho}$ and $\tilde{u}$, with no explicit change in their shape.  
Introducing (\ref{eq:prof_ideal}) into 
(\ref{eq:momthickness}), it is possible to show that 
\begin{equation} 
\frac{\delta_m(s)}{\delta_w(s)} = \frac{\delta_m(1)}{\delta_w(1)} + \frac{\lambda(s)}{2} \int_{-\infty}^{\infty} \!\!\!F_\rho(\xi) \left[ 1 - \left(F_u\!\left(\xi - \frac{\Delta}{\delta_w}\right)\right)^2 \right] d\xi = 
\frac{\delta_m(1)}{\delta_w(1)} + \lambda(s) G\left(\frac{\Delta}{\delta_w} \right),
\label{eq:larga}
\end{equation}
where $\lambda(s) = (s-1)/(s+1)$. Note that by construction $G(0)=0$ and $G^\prime(0) < 0$. Hence, it is possible to simplify (\ref{eq:larga}) to 
\begin{equation} 
\frac{\delta_m(s)}{\delta_w(s)} = \frac{\delta_m(1)}{\delta_w(1)} - C \lambda(s) \frac{\Delta}{\delta_w}  + O\left( \frac{\Delta}{\delta_w} \right)^2.
\label{eq:short}
\end{equation}
Interestingly, piecewise linear expressions for $F_\rho$ and $F_u$ yield $C=1/3$ and a cubic leading order error in (\ref{eq:short}).

\begin{figure}
\begin{center}
\begin{minipage}{0.49\linewidth}
\centerline{$(a)$}
\includegraphics[width=\linewidth]{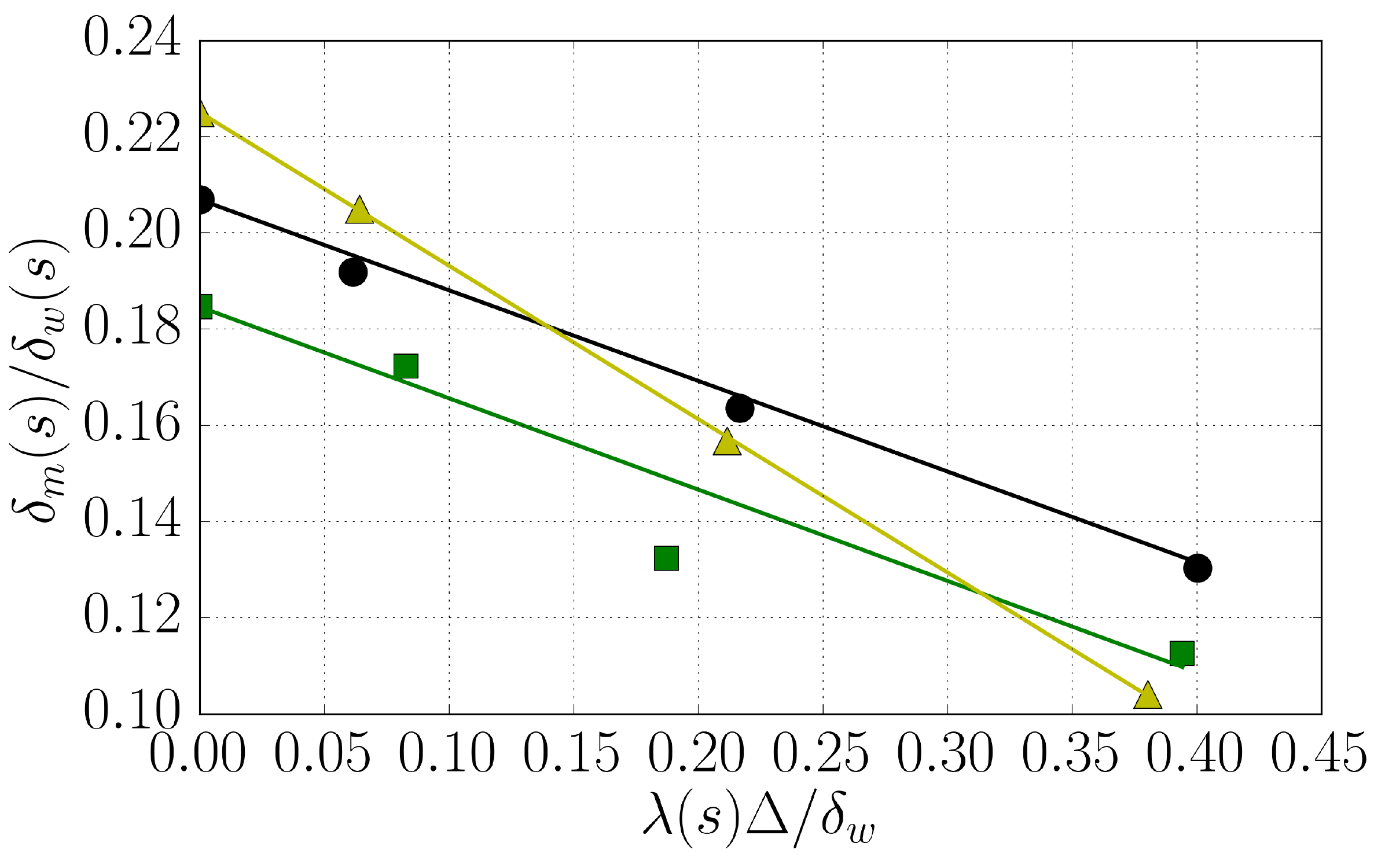}
\end{minipage}
\begin{minipage}{0.49\linewidth}
\centerline{$(b)$}
\includegraphics[width=\linewidth]{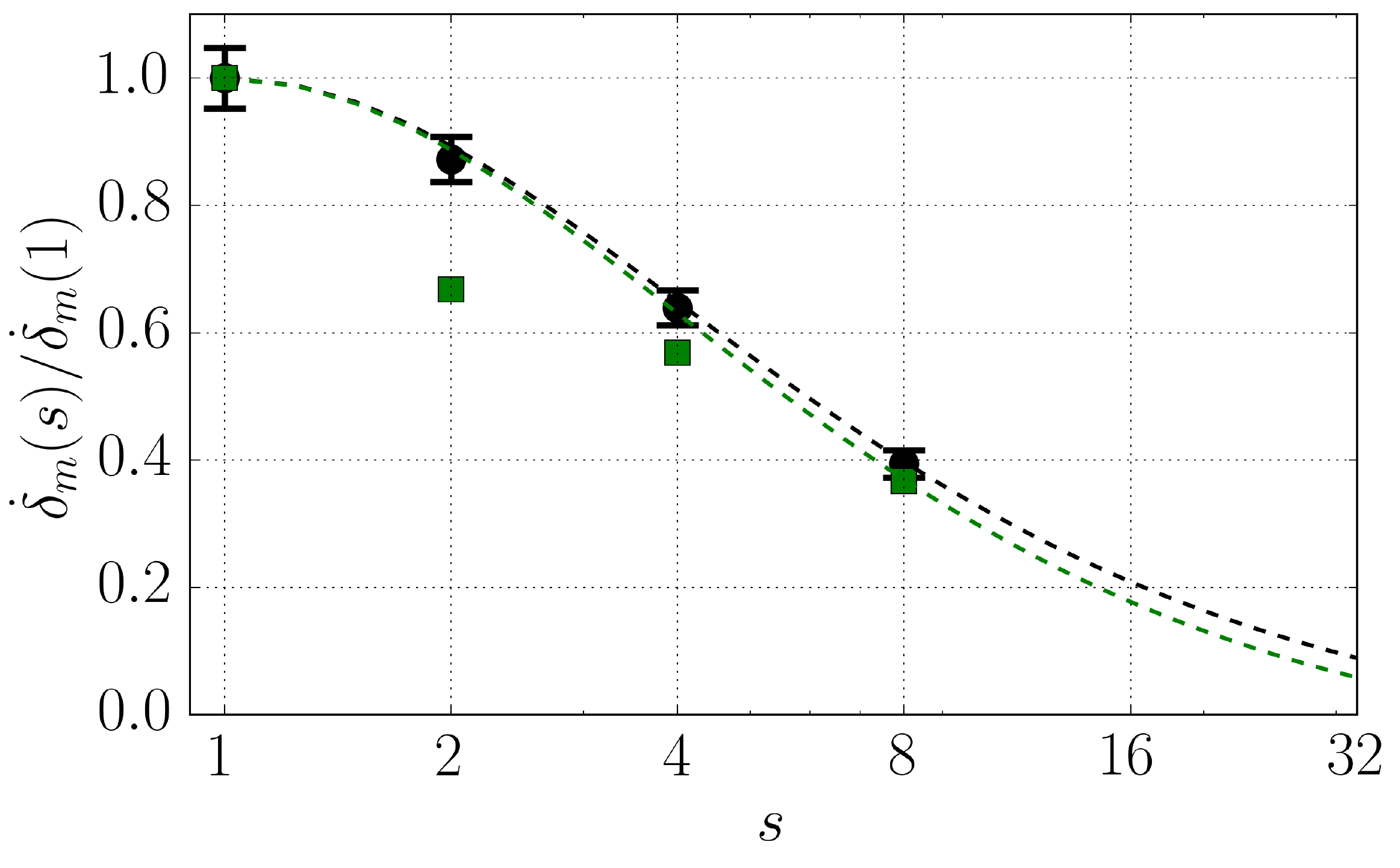}
\end{minipage}
\end{center}
\caption{Effects of $s$ and $\Delta$ on the reduction of the momentum thickness. 
$(a)$ $\delta_m/\delta_w$ versus $\lambda(s)\Delta/\delta_w$. $(b)$ $\dot\delta_m/\dot\delta_w$ versus $s$. 
In both panels, cicles are the present DNS at $M_c=0$, squares are \cite{Pantano2002} at $M_c=0.7$, 
and triangles are the self-similar solution for the laminar temporal mixing layer.
The solid lines in ($a$) correspond to equation (\ref{eq:short}) with: 
black, $C=0.188$; 
green, $C=0.190$; 
yellow, $C=0.32$. 
The dashed lines in ($b$) correspond to (\ref{eq:final_model}) with:
black, $C'=0.047$ and $\dot\delta_w(1)/\dot\delta_m(1) = 4.8$, 
green, $C'=0.047$ and $\dot\delta_w(1)/\dot\delta_m(1) = 5.4$. 
\label{fig:modelos}
}
\end{figure}

In order to estimate $C$ from the DNS data, figure \ref{fig:modelos}({\em a}) shows the ratio $1/D_w =  \delta_m/\delta_w$ as a function of $\lambda(s)\Delta/\delta_w$. 
The figure shows that $C=0.188$ for the present $M_c=0$ data, yielding a correlation coefficient between the data and the linear approximation equal to $R^2=0.998$. For the $M_c=0.7$ case, the ratio of the growth rates at $s=1$ is smaller, but the slope of the curve seems to be approximately the same ($C=0.190$, $R^2=0.920$), supporting the assumption that $F_\rho$, $F_u$ (and hence $C$) do not vary much with the density ratio. Note that for the laminar case, with notable differences in the shape of $\tilde{u}$ and $\overline{\rho}$ (and hence in $F_u$ and $F_\rho$), the value of the constant is $C=0.32$ and the linear approximation is exact ($R^2=1$). 

Finally, it is possible to combine (\ref{eq:ramshaw}), (\ref{eq:empirical_shifting}) and (\ref{eq:short}) to obtain a semi-empirical prediction of the reduction of the momentum thickness growth rate with the density ratio, 
\begin{equation} 
\frac{\dot\delta_m(s)}{\dot\delta_m(1)} \approx \frac{2\sqrt{s}}{s+1} \left( 1 - \frac{\dot\delta_w(1)}{\dot\delta_m(1)} C' \log(s) \right). 
\label{eq:final_model}
\end{equation} 
To obtain (\ref{eq:final_model}) we have also taken advantage of $1/D_w = \delta_m/\delta_w \approx \dot\delta_m/\dot\delta_w$, which is a reasonable approximation for sufficiently long times. 
The performance of this simple model for the reduction of the momentum thickness growth rate is evaluated in 
figure \ref{fig:modelos}({\em b}), where the dashed lines corresponds to equation (\ref{eq:final_model}) with $C' = C \cdot C_\Delta = 0.047$ and the appropriate value for $\dot\delta_w(1)/\dot\delta_m(1)$, black for $M_c=0$ and green for $M_c = 0.7$. The figure also includes the DNS data for both mach numbers. The agreement between the DNS data and the model is very good, except for the lower density ratios of the $M_c=0.7$ cases, which already showed differences when compared to the present $M_c=0$ cases in figure \ref{fig:dotdmcomparison}.

\begin{figure}
\begin{center}
\begin{minipage}{0.49\linewidth}
\centerline{$(a)$}
\includegraphics[width=\linewidth]{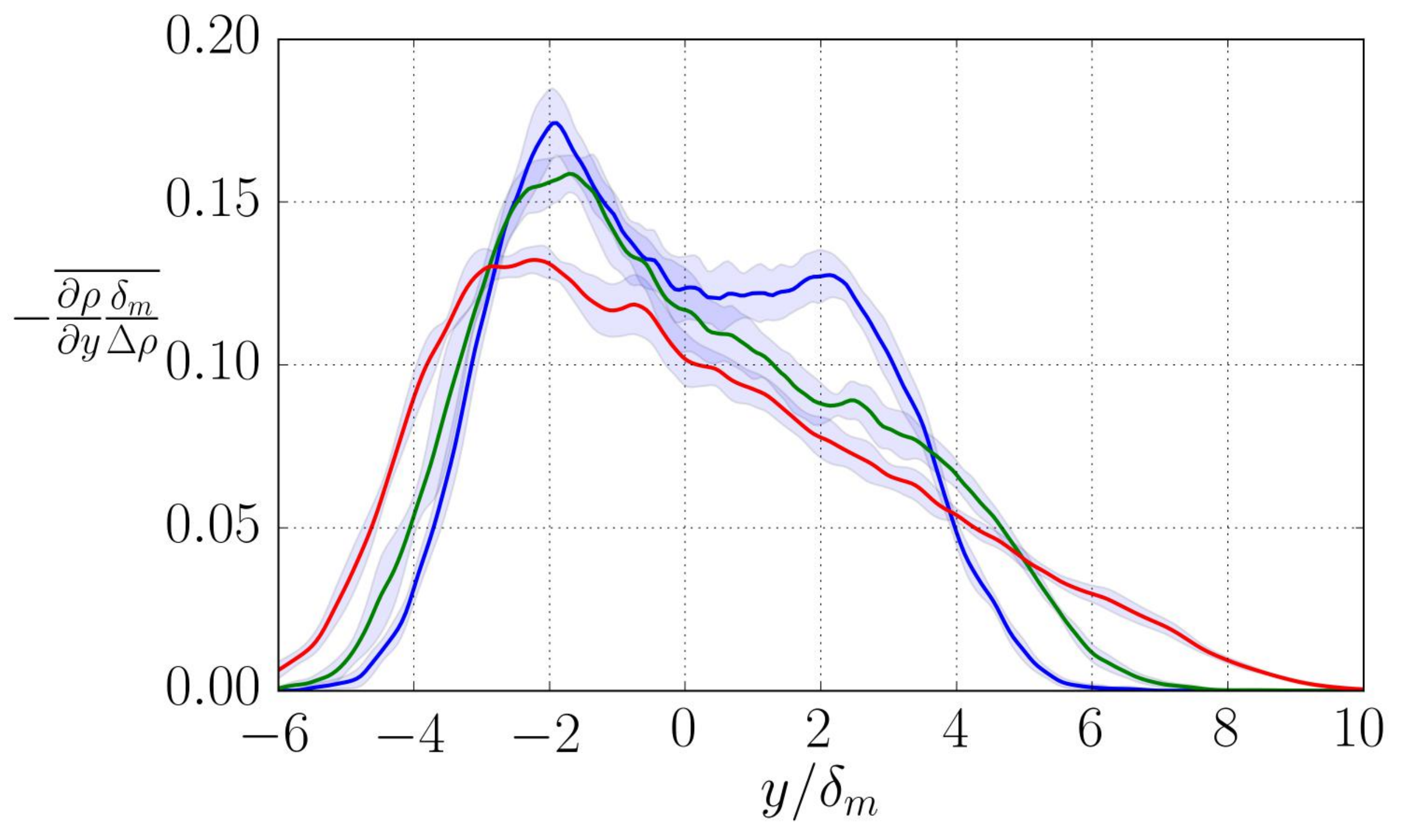}
\end{minipage}
\begin{minipage}{0.49\linewidth}
\centerline{$(b)$}
\includegraphics[width=\linewidth]{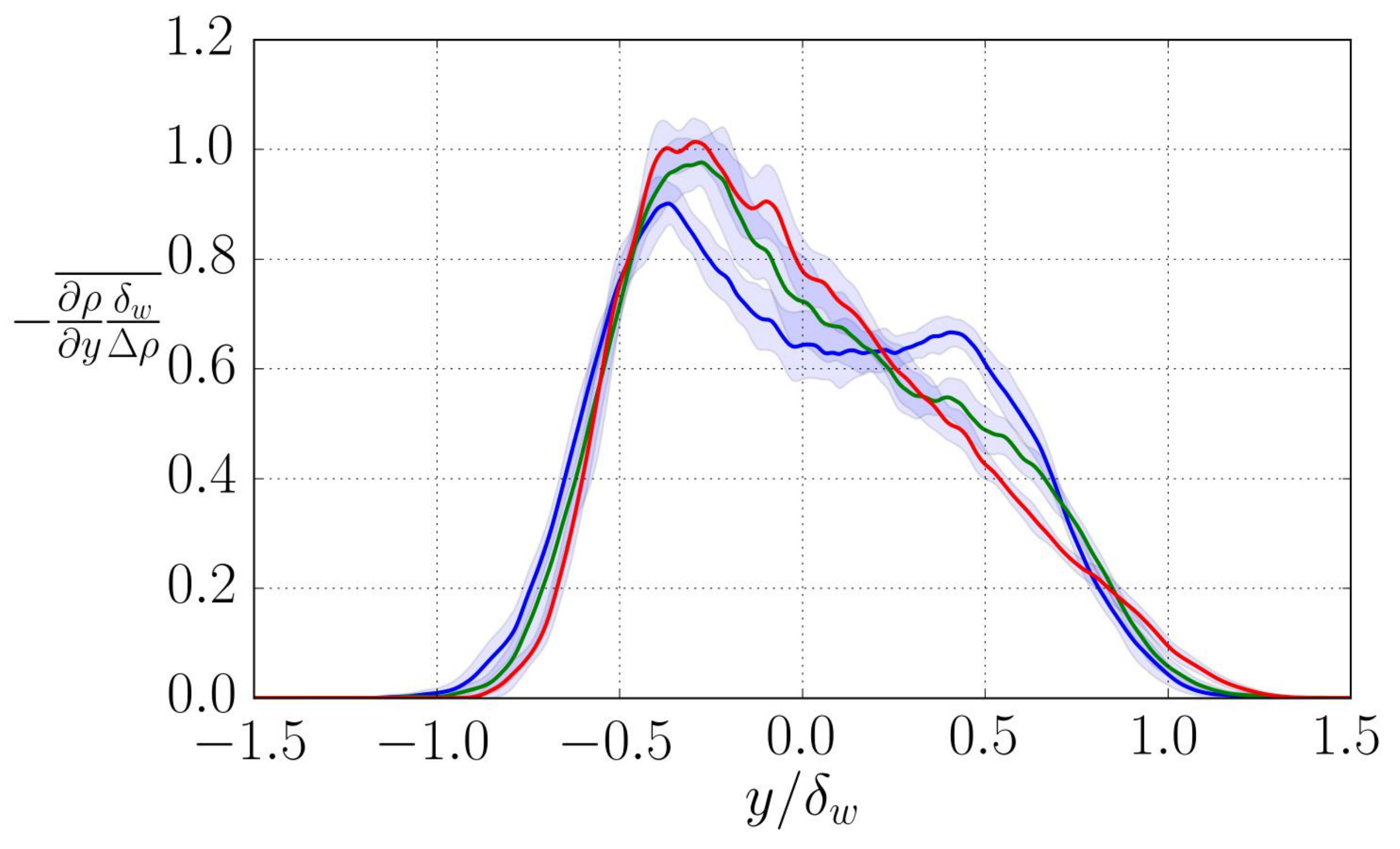}
\end{minipage}
\begin{minipage}{0.49\linewidth}
\centerline{$(c)$}
\includegraphics[width=\linewidth]{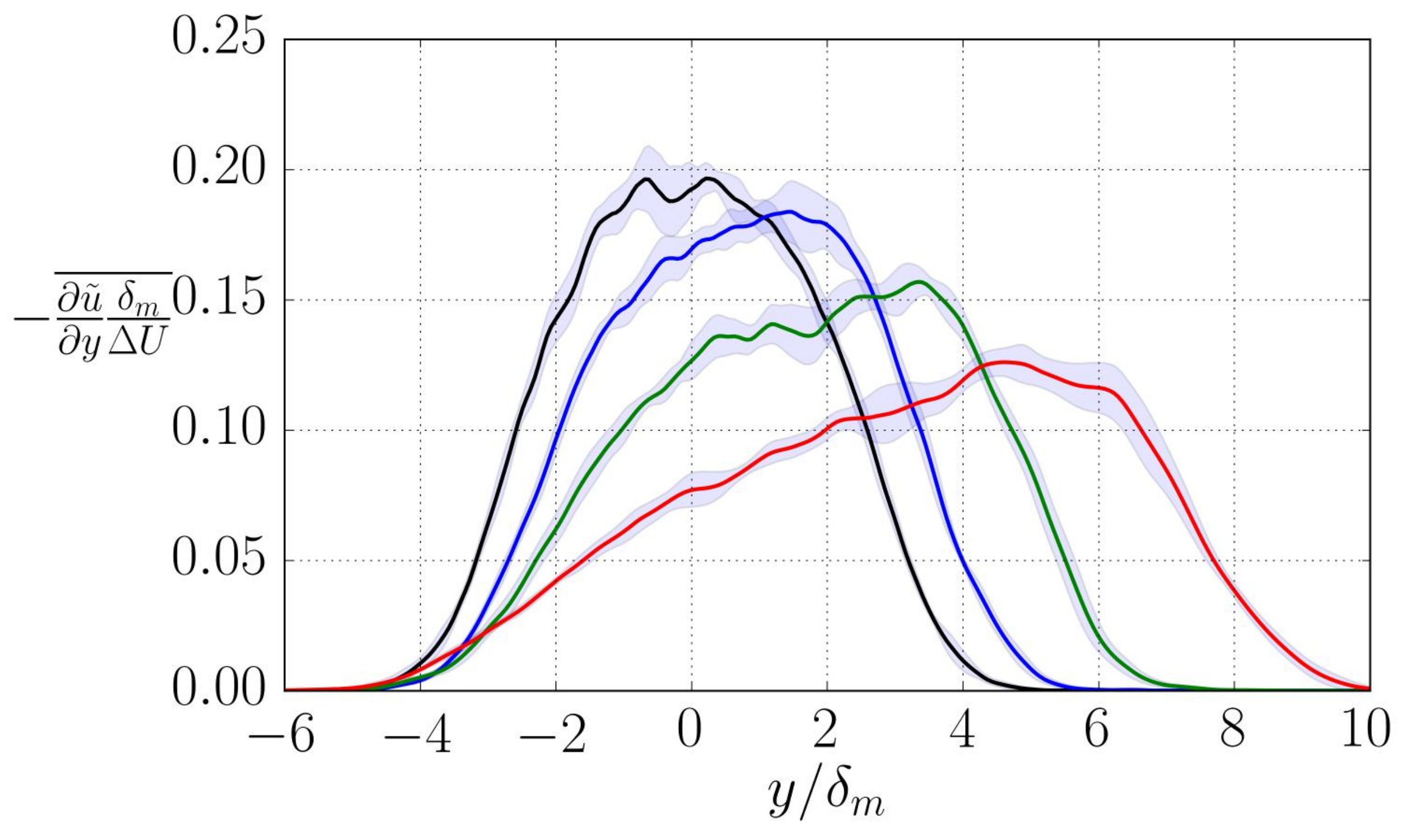}
\end{minipage}
\begin{minipage}{0.49\linewidth}
\centerline{$(d)$}
\includegraphics[width=\linewidth]{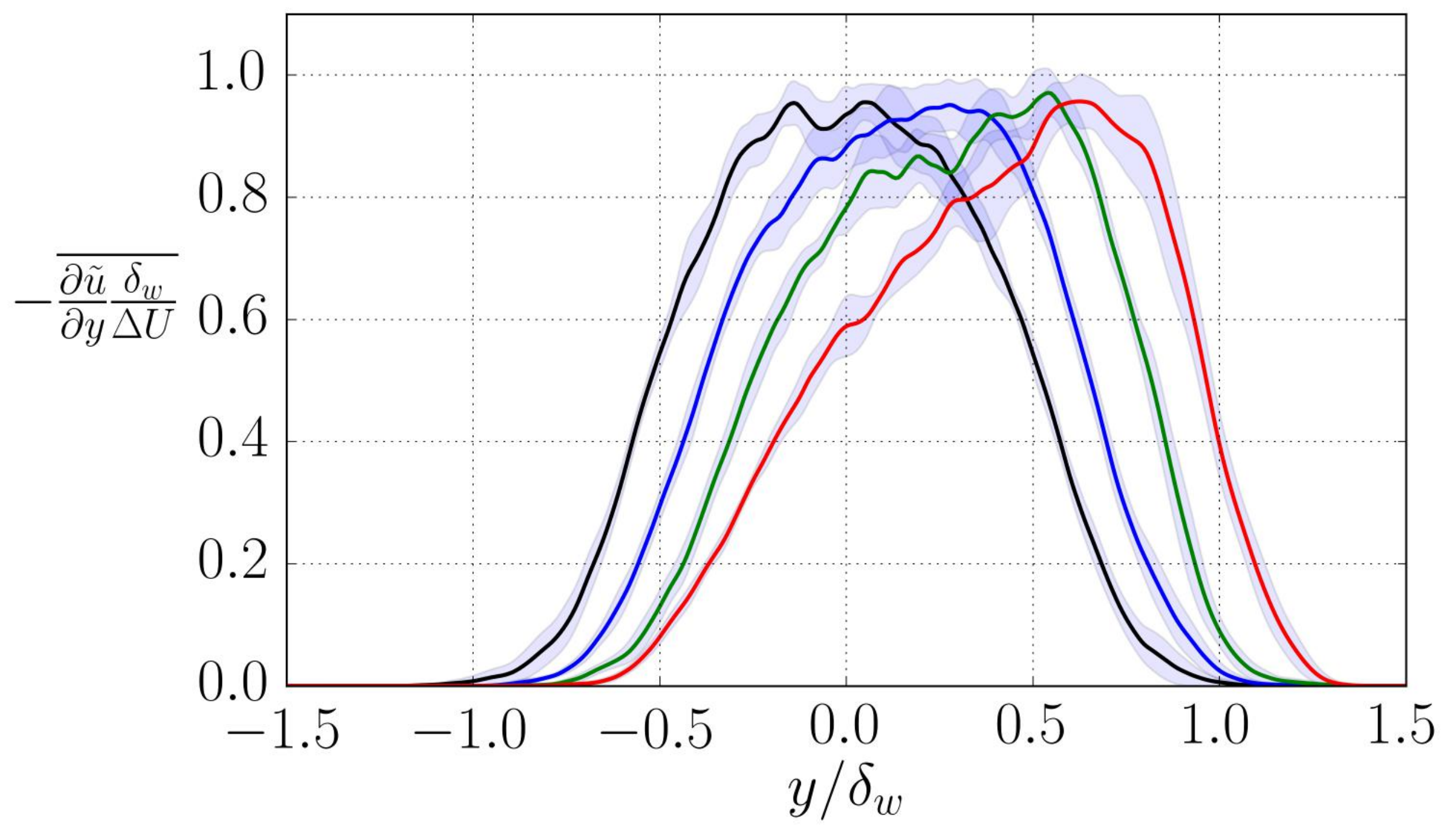}
\end{minipage}

\end{center}
\caption{$(a)$,$(b)$ Profiles of the vertical gradients of the Reynolds-averaged density. 
$(c)$,$(d)$ Profiles of the vertical gradients of the Favre-averaged streamwise velocity.
$(a)$,$(c)$ Normalized with the momentum thickness. 
$(b)$,$(d)$ Normalized with the vorticity thickness. 
Different colours correspond to different density ratios:
black, $s=1$; 
blue, $s=2$; 
green, $s=4$; 
red, $s=8$. 
\label{fig:profiles_gradients}
}
\end{figure}

In the previous discussion, the effect of $s$ on the shape of the profiles of $\overline{\rho}$ and $\tilde{u}$ has been neglected, resulting in a reasonable approximation for the reduction in the growth rate of the mixing layer with $s$. However, the density ratio has some effects in the shapes of $\overline{\rho}$ and $\tilde{u}$, which are responsible for changes in the structure of the turbulence in the mixing layer. These effects, which are difficult to evaluate in figure \ref{fig:profiles_mean}, are better observed in figure \ref{fig:profiles_gradients}, which shows the vertical gradients of the mean profiles with different normalizations. 
In particular, the gradients of the mean density normalized with $\Delta\rho=\rho_b-\rho_t$ and $\delta_w$ seem to collapse 
reasonably well 
(see figure \ref{fig:profiles_gradients}$b$), especially in  
the high density side (lower stream). More differences are visible near the low density side, where it is apparent that the gradients tend to become smoother with increasing $s$.
Indeed, for $s=8$, figure \ref{fig:profiles_gradients}$a$ and $b$ show that the gradient of $\overline{\rho}$ is roughly linear, so that $\overline{\rho}$ becomes roughly parabolic for $y \gtrsim -2\delta_m \approx -0.25\delta_w$.
Although outside of the scope of the present paper, it would be interesting to check whether the same linear region in $\partial \overline{\rho} /\partial y$ is obtained for higher density ratios. 
The shifting of the velocity profiles discussed above is clearly visible when looking at their corresponding
gradients, figures \ref{fig:profiles_gradients}($c$) and ($d$).  
For $\tilde{u}$ the change of shape of the profile results in the maximum gradients appearing nearer to the lower density side, with smoother gradients in the high density side. Indeed, opposite to what is observed for $\rho$, case $s=8$ seems to develop a nearly parabolic profile for $\tilde{u}$ towards the higher density side of the mixing layer ($y \lesssim 4\delta_m \approx y \lesssim 0.5\delta_w$).

\begin{figure}
\begin{center}
\begin{minipage}{0.49\linewidth}
\centerline{$(a)$}
\includegraphics[width=\linewidth]{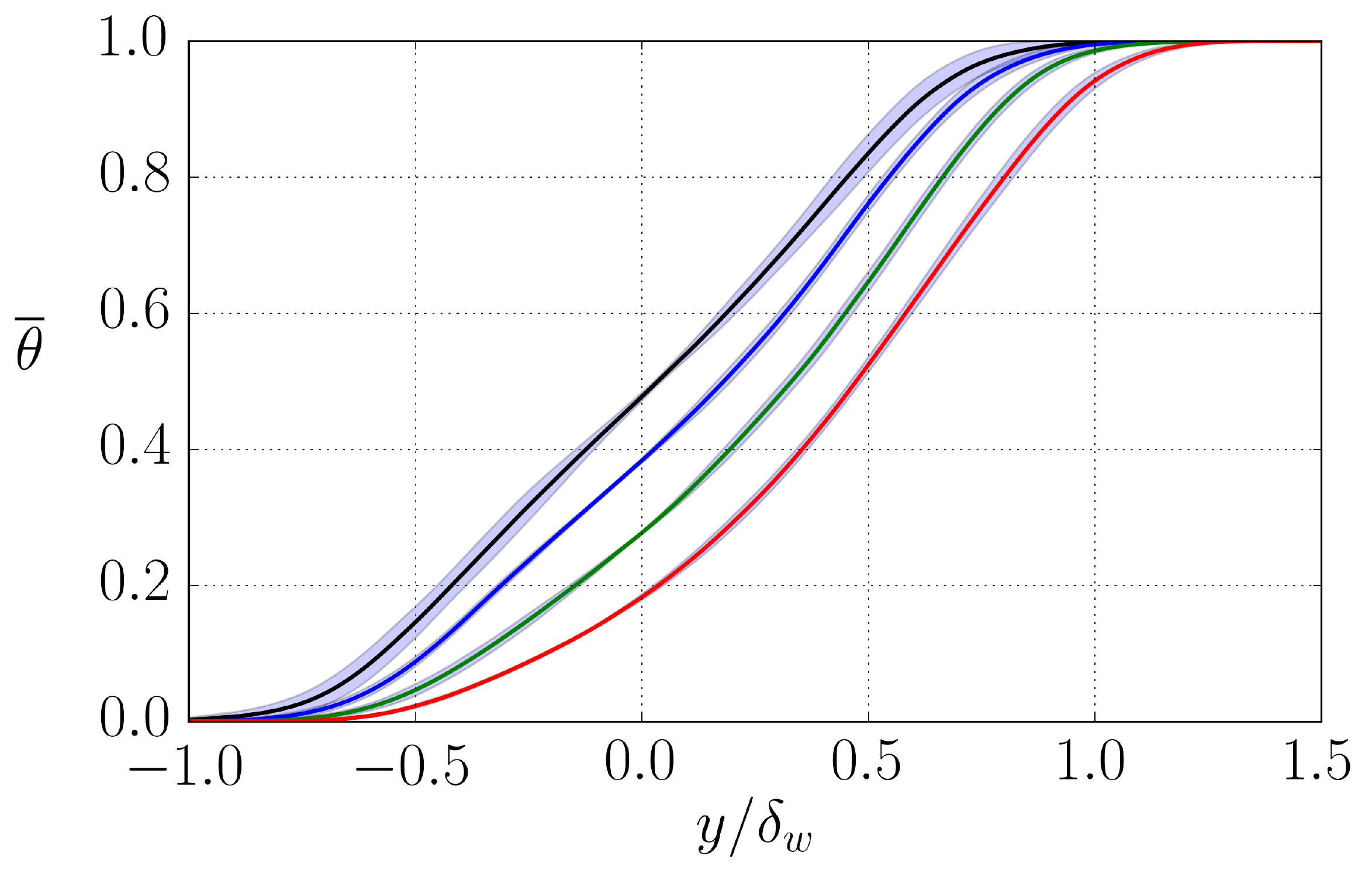}
\end{minipage}
\begin{minipage}{0.49\linewidth}
\centerline{$(b)$}
\includegraphics[width=\linewidth]{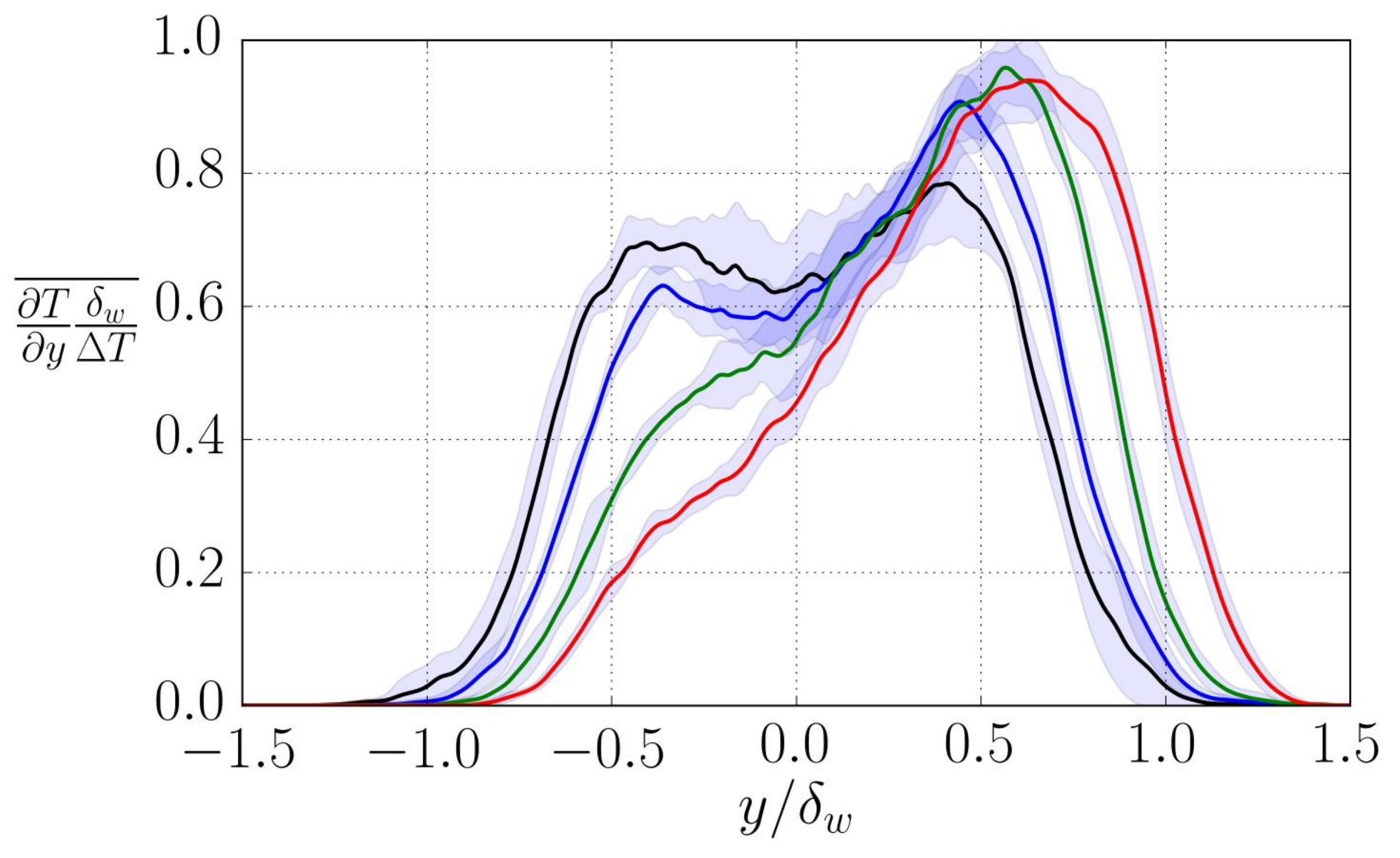}
\end{minipage}
\end{center}
\caption{$(a)$ Reynolds-averaged temperature profiles. 
$(b)$ Profiles of the vertical gradients of the Reynolds-averaged temperature.
Different colours correspond to different density ratios:
black, $s=1$; 
blue, $s=2$; 
green, $s=4$; 
red, $s=8$. \label{fig:profiles_T}}
\end{figure}

To finalize this subsection, we turn our attention to the mean temperature distribution, 
more especifically to the non-dimensional temperature jump $\overline{\theta} = (\overline{T}-T_b)/(T_t-T_b)$. 
It is interesting to study the temperature since it follows an advection-diffusion equation,
equation (\ref{eq:energy}). This allows the comparison of the variable density cases ($s=2$, 4 and 8) with the passive scalar simulated for the uniform density case ($s=1$). Note that although the temperature
is inversely proportional to the density (equation of state), the same is not true for the mean temperature and mean 
density.
Figure \ref{fig:profiles_T}({\em a}) shows the mean temperature profiles for all
cases and figure \ref{fig:profiles_T}({\em b}) the corresponding profiles of the 
vertical gradients of the mean temperature. 
The passive scalar shows a roughly symmetric distribution, with $\partial \overline{\theta} / \partial y$ peaking near the edges of the mixing layer ($\abs{y/\delta_w} \approx 0.5$). The small deviation with respect to a symmetric profile provides an impression about the convergence of the statistics. 

With increasing $s$, the mean temperature profiles shifts towards the upper stream (low density stream)
in a similar way as the Favre-averaged streamwise velocity. 
The profiles also become more asymmetric, which is 
more clearly visible in the mean temperature gradients 
shown in figure \ref{fig:profiles_T}({\em b}). 
As the density ratio increases, the gradients at the high density edge of the mixing layer are strongly damped, 
while the gradients at the low density edge are enhanced.

\subsection{Higher order statistics}
\label{sec:rms}
\begin{figure}
\begin{center}
\begin{minipage}{0.49\linewidth}
\centerline{$(a)$}
\includegraphics[width=\linewidth]{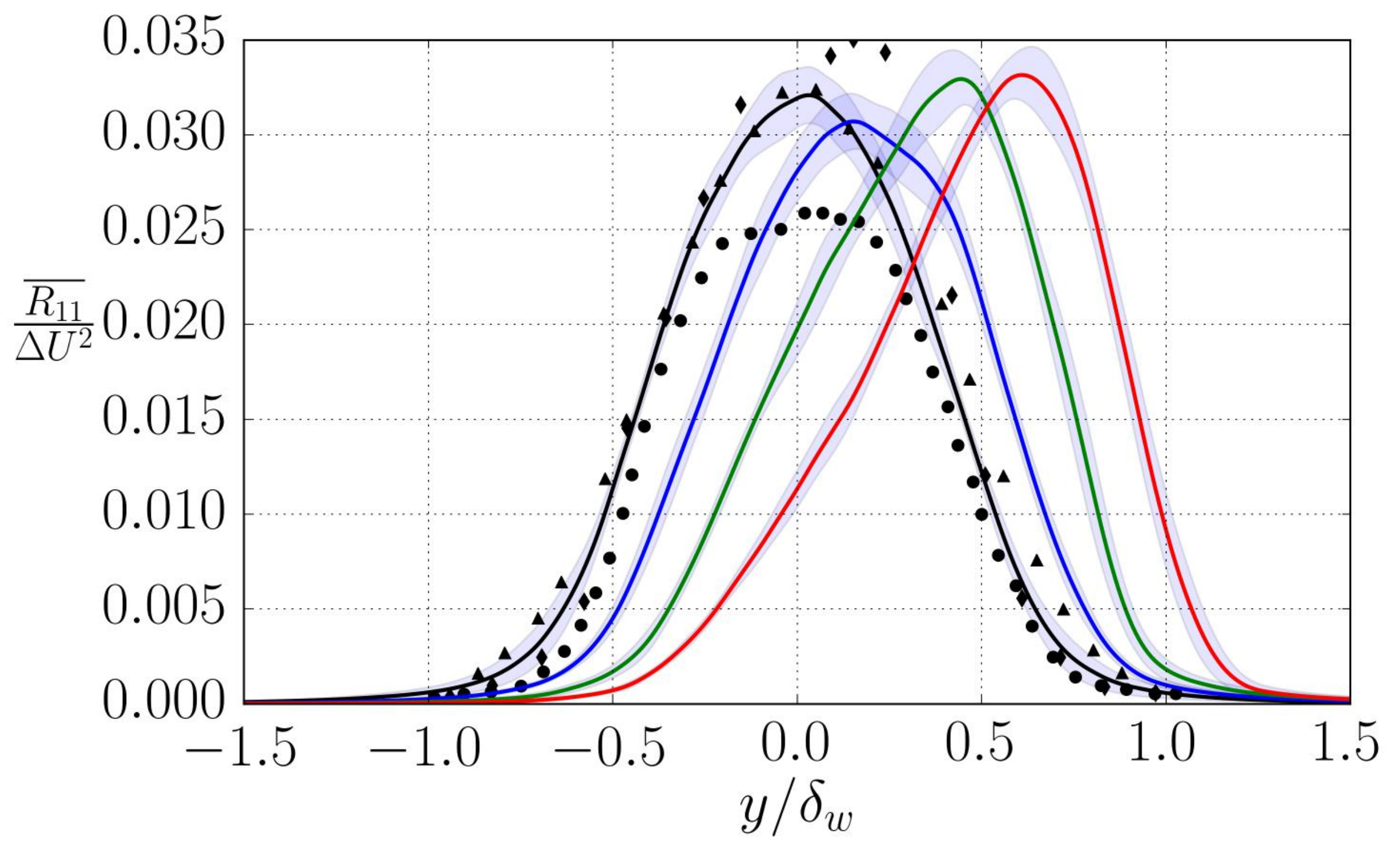}
\end{minipage}
\begin{minipage}{0.49\linewidth}
\centerline{$(b)$}
\includegraphics[width=\linewidth]{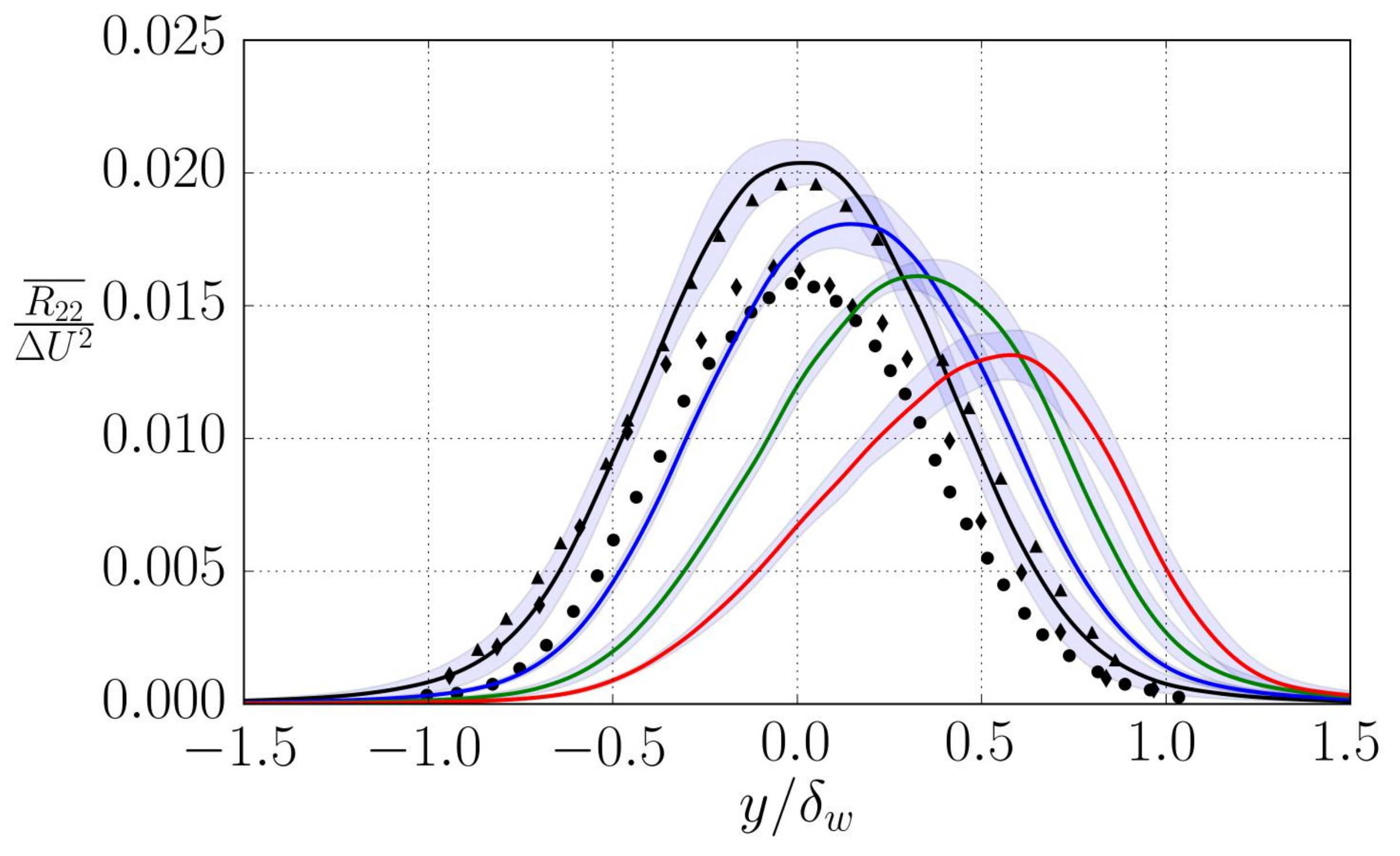}
\end{minipage}
\begin{minipage}{0.49\linewidth}
\centerline{$(c)$}
\includegraphics[width=\linewidth]{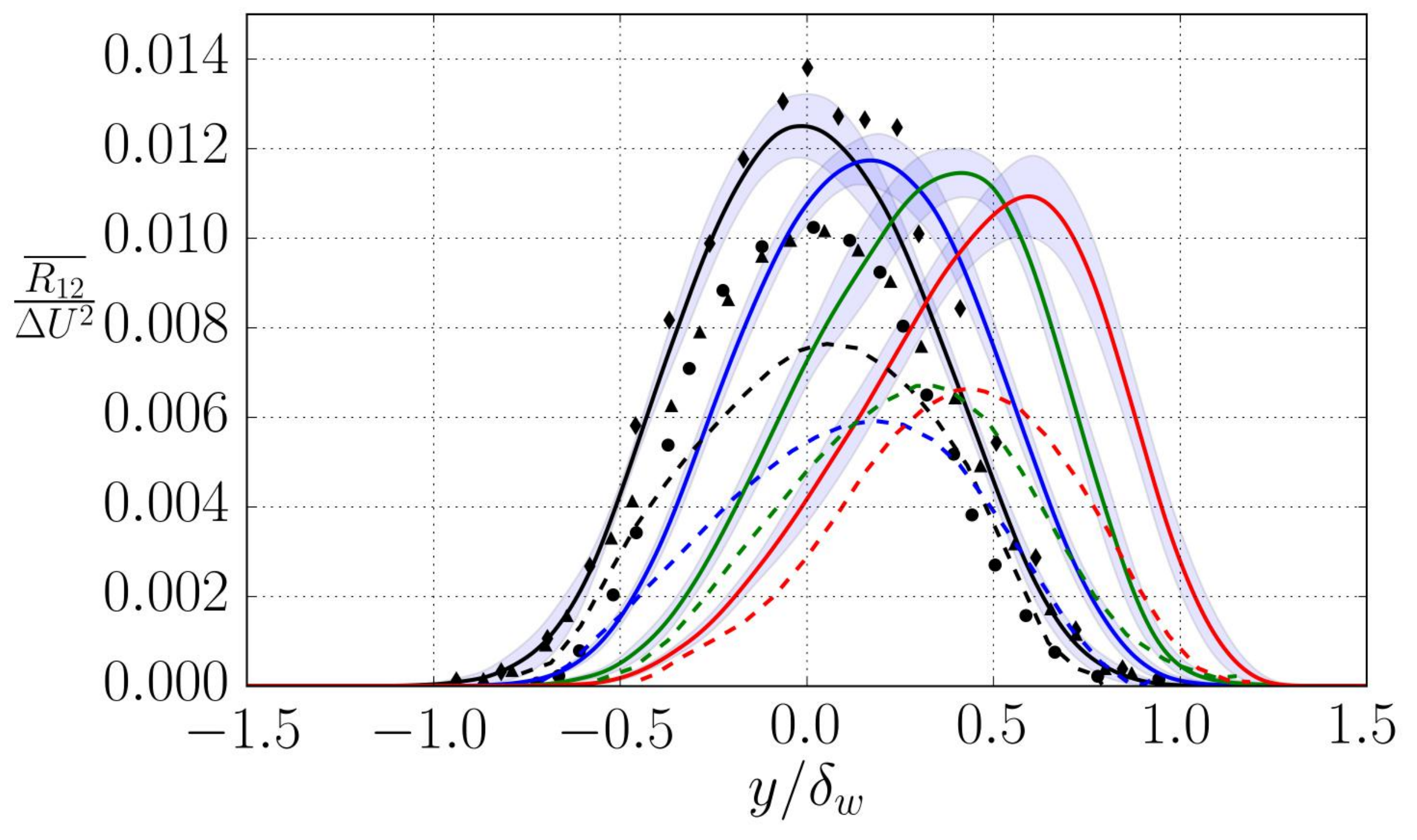}
\end{minipage}
\begin{minipage}{0.49\linewidth}
\centerline{$(d)$}
\includegraphics[width=\linewidth]{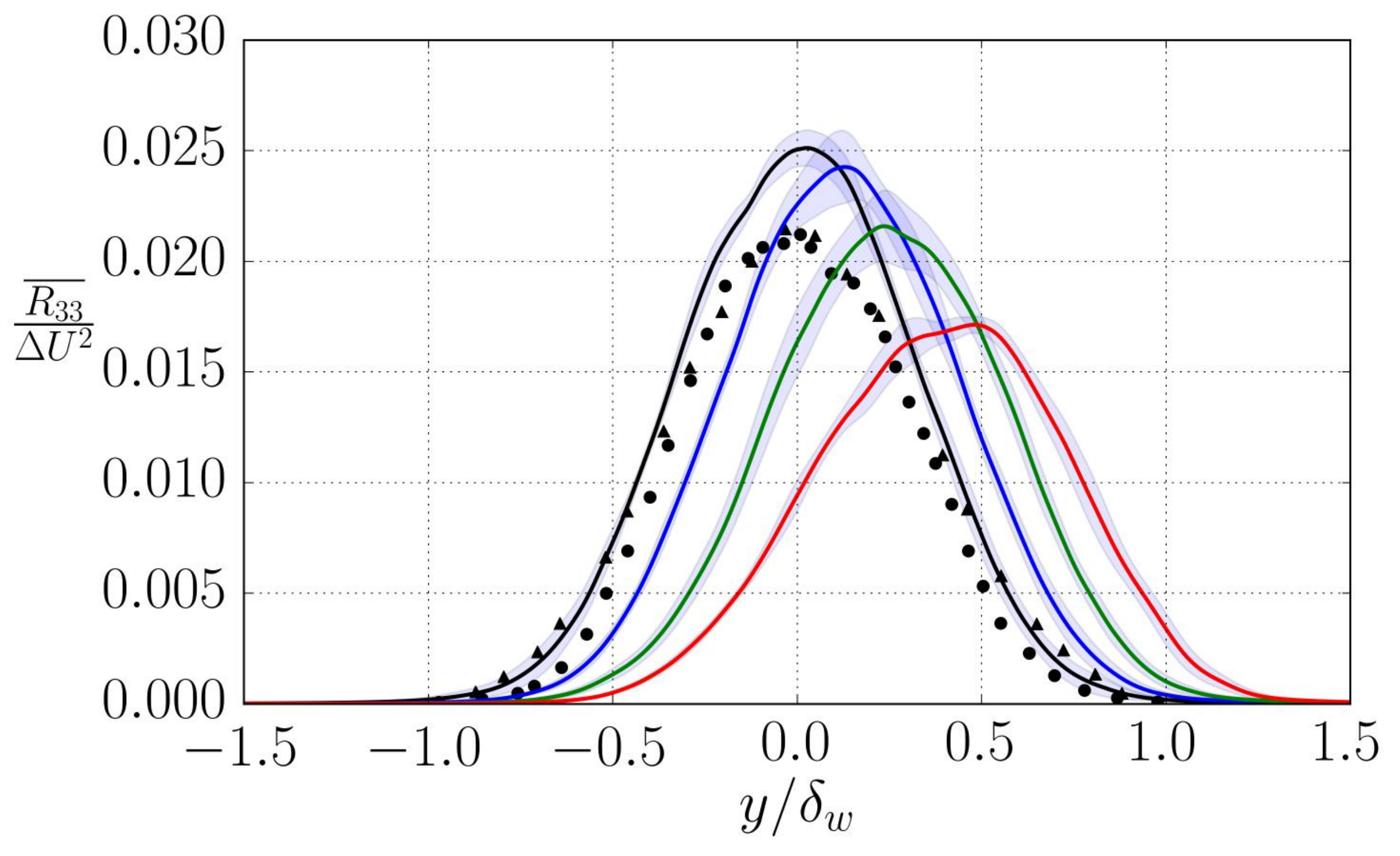}
\end{minipage}
\end{center}
\caption{Vertical profiles of  
({\em a}) $R_{11}/\Delta U^2$, ({\em b}) $R_{22}/\Delta U^2$,  
({\em c}) $R_{12}/\Delta U^2$, and  ({\em d}) $R_{33}/\Delta U^2$.
Different colours correspond to different density ratios: black, $s=1$; blue, $s=2$; green, $s=4$; and red, $s=8$. 
Solid lines are the present turbulent temporal mixing layers. 
Symbols are data from incompresible mixing layers: dots from simulations of \citep{Rogers1994}, triangles from experiments of \citep{Bell1990} and 
diamonds from experiments of \citep{Spencer:1971}.
Dashed  lines in ({\em c}) represent results from $M_c=0.7$ \citep{Pantano2002}.
}
\label{fig:Rij}
\end{figure}

\begin{figure}
\begin{center}
\begin{minipage}{0.49\linewidth}
\centerline{$(a)$}
\includegraphics[width=\linewidth]{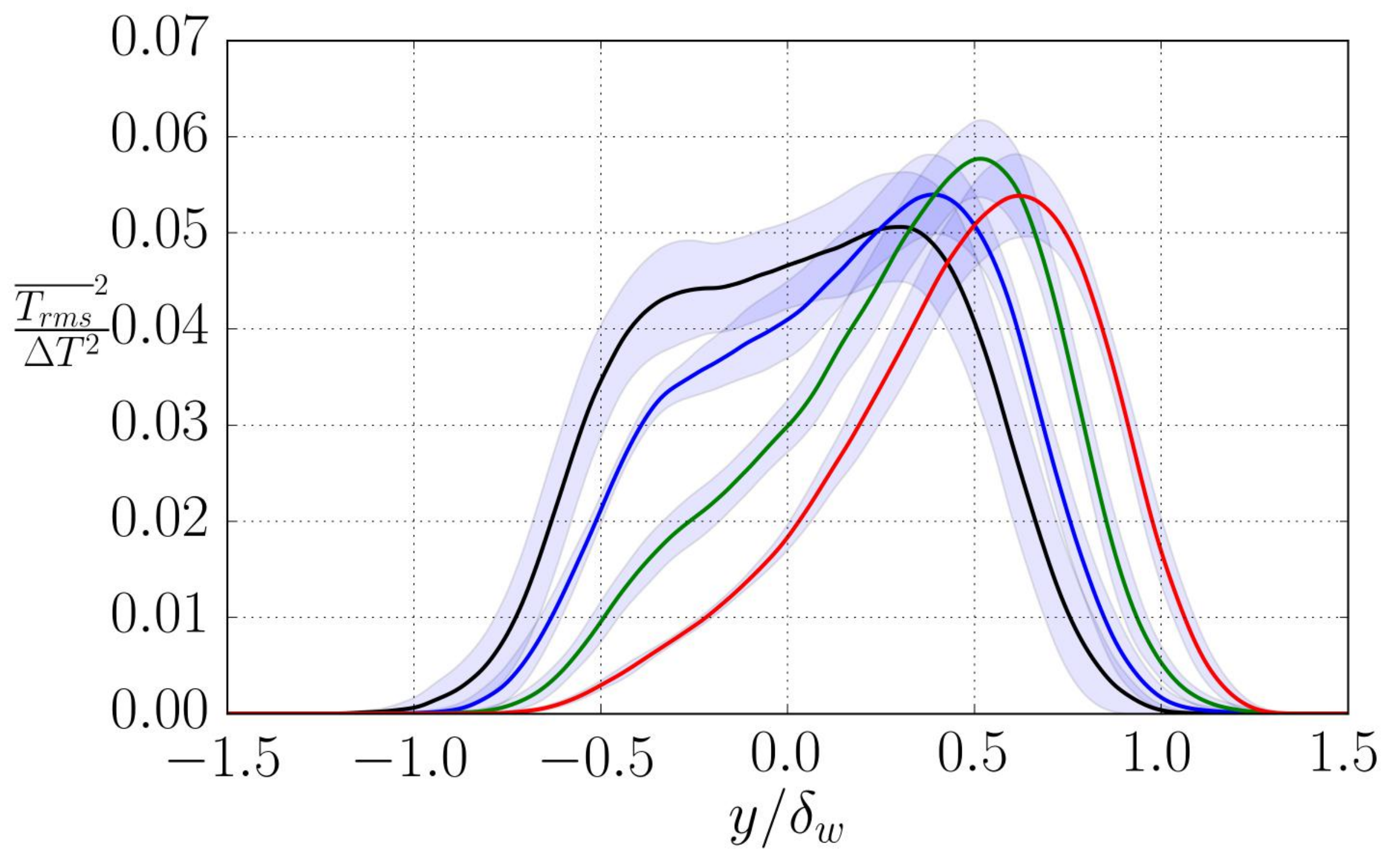}
\end{minipage}
\begin{minipage}{0.49\linewidth}
\centerline{$(b)$}
\includegraphics[width=\linewidth]{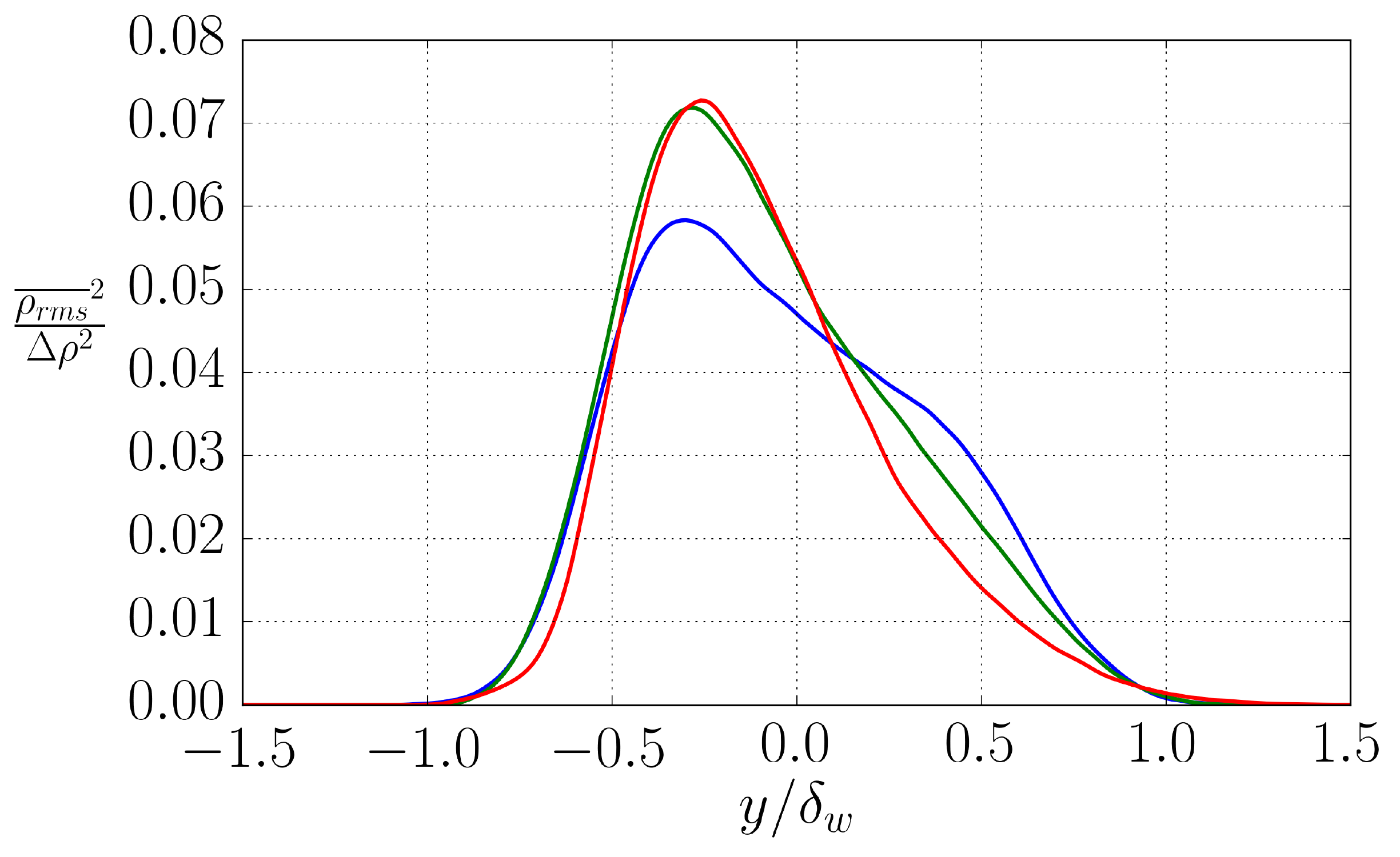}
\end{minipage}
\end{center}
\caption{
$(a)$ Profiles of $T_{rms}^2/\Delta T^2$. 
$(b)$ Profiles of $\rho_{rms}^2/\Delta \rho^2$. 
Different colours correspond to different density ratios:
black, $s=1$; 
blue, $s=2$; 
green, $s=4$; 
red, $s=8$. \label{fig:profiles_T2}
Both magnitudes calculated using Reynolds average.
}
\end{figure}

The shifts in the mean velocity and temperature, as well as the changes in their gradients, 
are also accompanied by changes in the root mean square of velocity and temperature fluctuations, which are analized in 
figures \ref{fig:Rij} and \ref{fig:profiles_T2}. 
In particular,
figure \ref{fig:Rij} displays the vertical profiles of the turbulent stress tensor, $R_{ij}$. 
The plots include the data for the incompressible mixing layer of \citet{Rogers1994}, and the experimental results of \citet{Bell1990} and \citet{Spencer:1971}. Both datasets  show profiles that are consistent with the shape of the present $s=1$ case, although there is considerable scatter between the three datasets. 
The scatter in $R_{12}$ (figure \ref{fig:Rij}$c$) is consistent with the scatter in the growth-rates of the mixing layers, since these two quantities are related through equation (\ref{eq:dotdeltam}). This could also explain the scatter in $R_{11}$,  $R_{22}$ and $R_{33}$ for the cases with $s=1$. 
As the density ratio increases, $R_{ij}$ tend to shift towards the low-density region, following the maximum gradient of
 $\tilde{u}$. Interestingly, while the peak values of $R_{22}, R_{12}$ and $R_{33}$
decrease with increasing $s$, the peak values of $R_{11}$ seem to remain roughly constant (at least within the uncertainty in the statistics, shown in the figure by the shaded areas around each curve). 
The high-speed data of \cite{Pantano2002} are also included in figure \ref{fig:Rij}$c$, and they also show a decrease of the peak values of $R_{12}$ with increasing $s$, although for the $M_c=0.7$ data the decrease is not monotonic as it is for the present $M_c=0$ results. Note also that, as expected, the $M_c=0.7$ profiles have lower maximum values, consistent with the lower growth rate of the subsonic mixing layers (as discussed in \S\ref{sec:growth} and in \citealp{Pantano2002}).   
Figure \ref{fig:profiles_T2} displays the profiles of the variance of the temperature, $T_{rms}^2$, 
and density, $\rho_{rms}^2$, normalized with the corresponding jumps across the mixing layer, $\Delta T = T_t - T_b$ and $\Delta \rho = \rho_b - \rho_t$, respectively. 

For $s=1$ the temperature corresponds to the passive scalar, which exhibits in figure \ref{fig:profiles_T2}({\em a}) the double-peak rms observed in high Reynolds numbers mixing layers by others (e.g., see \citealp{pickett2001passive}). 
When the density ratio is increased, the peak on the high density side gradually decreases, while the peak of $T_{rms}^2$ on the low-density side shifts with the mean temperature gradients (see figure \ref{fig:profiles_T}$b$).
This suggests that $T_{rms}^2$ is governed by the mean temperature gradient, in a similar way as $\overline{R_{ij}}$ are governed by $\partial \tilde{u}/\partial y$.
 Indeed, consistent with the mean temperature gradients, the peaks of $T_{rms}^2/\Delta T^2$ increase with $s$, except for maybe case $s=8$.
At the present moment, the reason for the non-monotonous behaviour of $s=8$ is unclear. It could be related to a decrease in the $Re_\lambda$ for this case. Another possible explanation could be the onset of interferences of the finite-size of the computational domain with the evolution of the mixing layer.

Not surprisingly, the behaviour of $\rho_{rms}^2$ shown in figure  \ref{fig:profiles_T2}({\em b}) suggest that $\rho_{rms}^2$ is governed by the the mean density gradients (figure \ref{fig:profiles_gradients}{\em b}), analogous to the behavior of temperature and velocity fluctuations. 
As $s$ increases, the rms around $y/\delta_w \approx -0.5$ (high density side) increases, while the fluctuations around $y/\delta_w \approx 0.5$ (low-density side) decrease. The behaviour is opposite to $\overline{R_{ij}}$ (which are more intense near the low density side), which is consistent with the arguments of \cite{Brown1974} for the shift and the asymmetry of the growth of the variable density mixing layers.

\begin{figure}
\begin{center}
\begin{minipage}{0.49\linewidth}
\centerline{$(a)$}
\includegraphics[width=\linewidth]{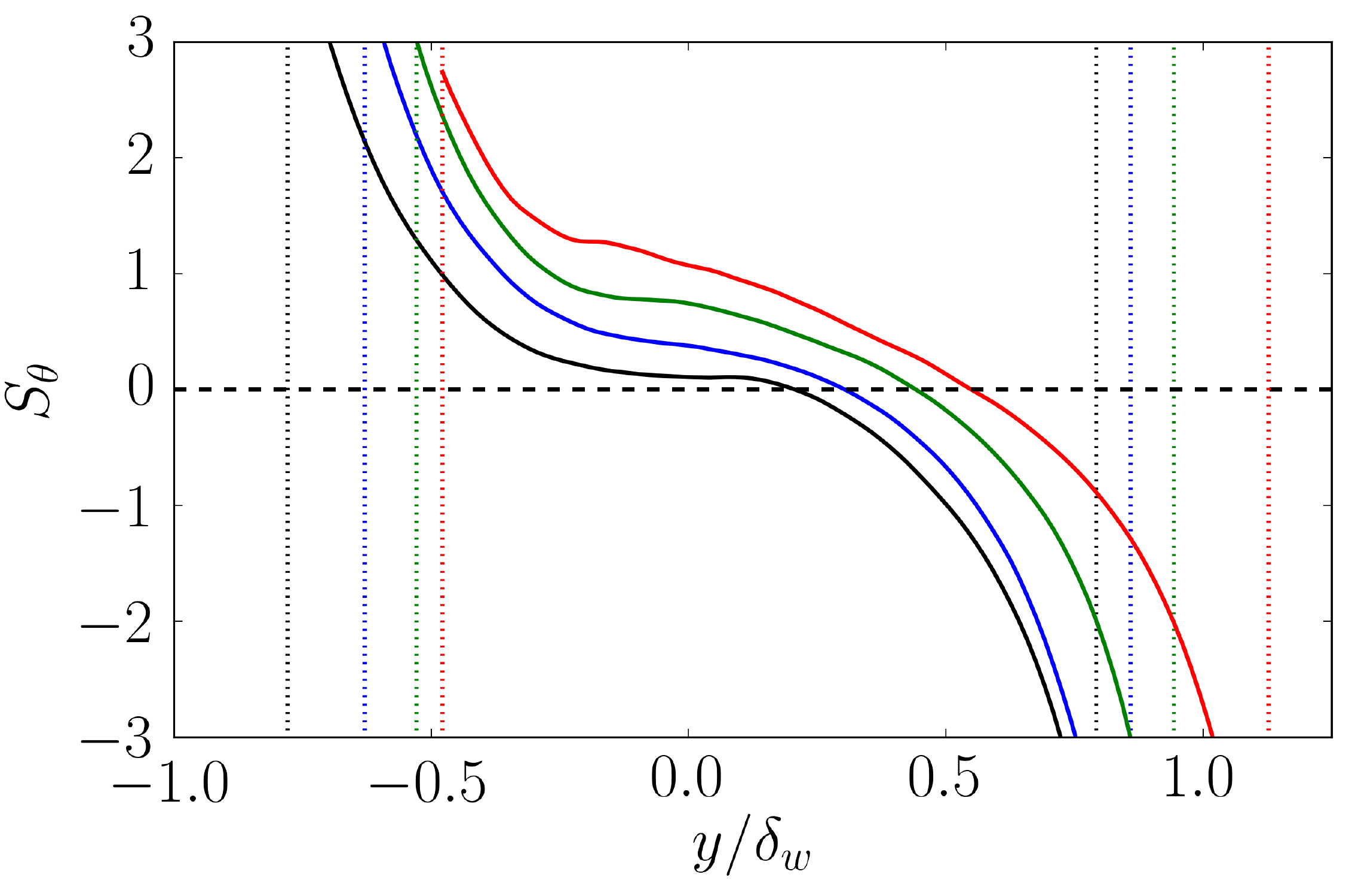}
\end{minipage}
\begin{minipage}{0.49\linewidth}
\centerline{$(b)$}
\includegraphics[width=\linewidth]{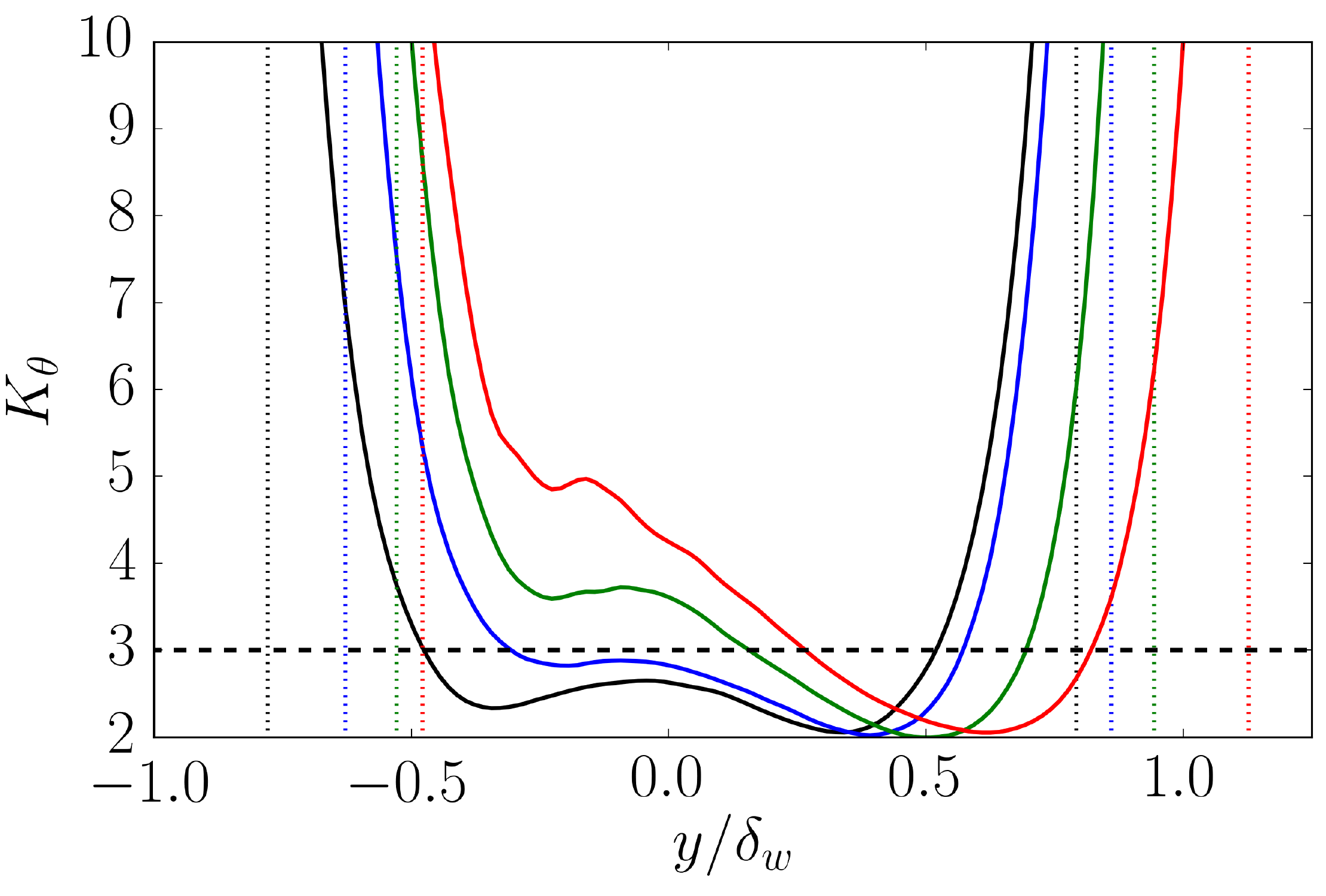}
\end{minipage}
\begin{minipage}{0.49\linewidth}
\centerline{$(c)$}
\includegraphics[width=\linewidth]{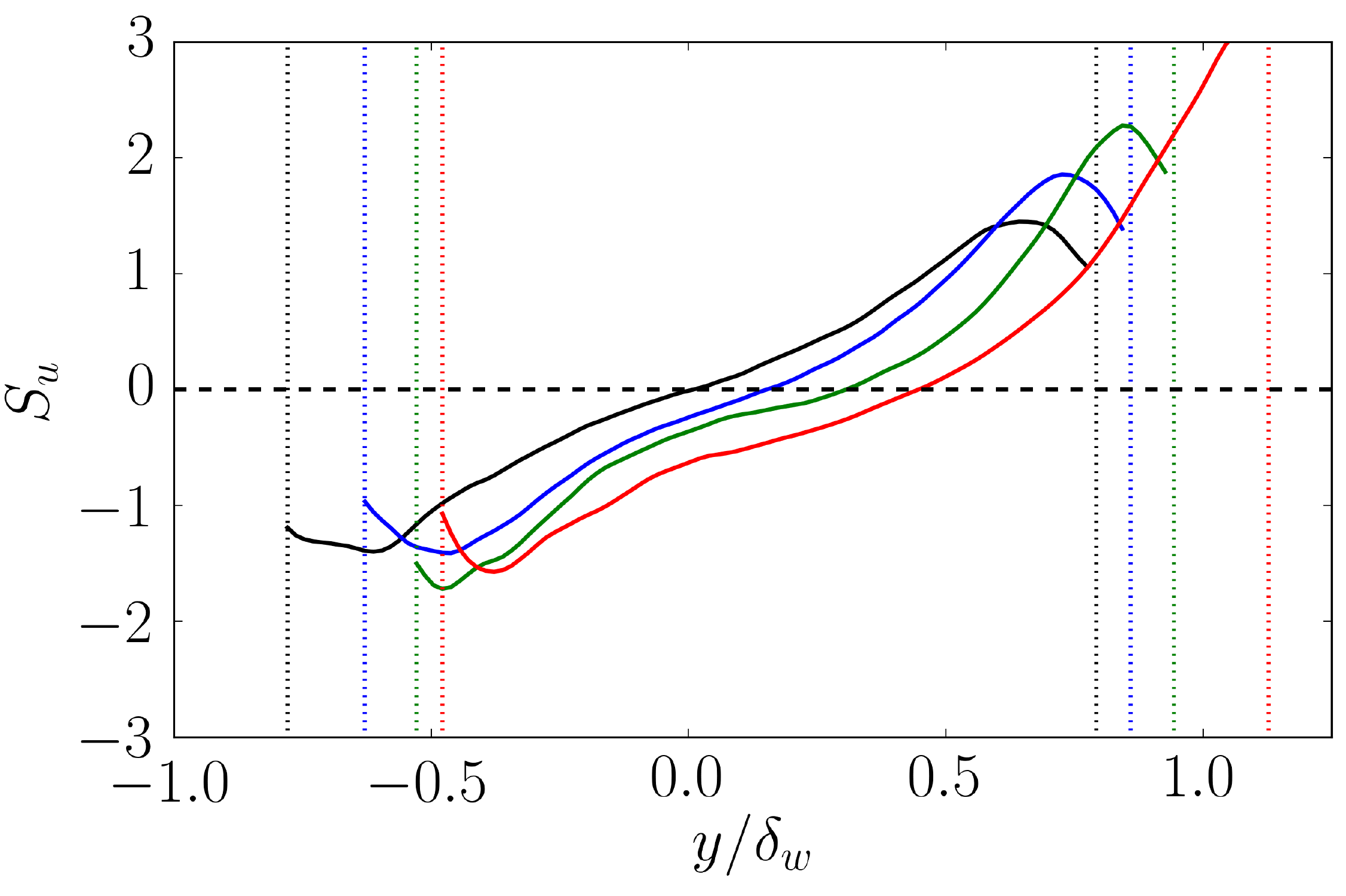}
\end{minipage}
\begin{minipage}{0.49\linewidth}
\centerline{$(d)$}
\includegraphics[width=\linewidth]{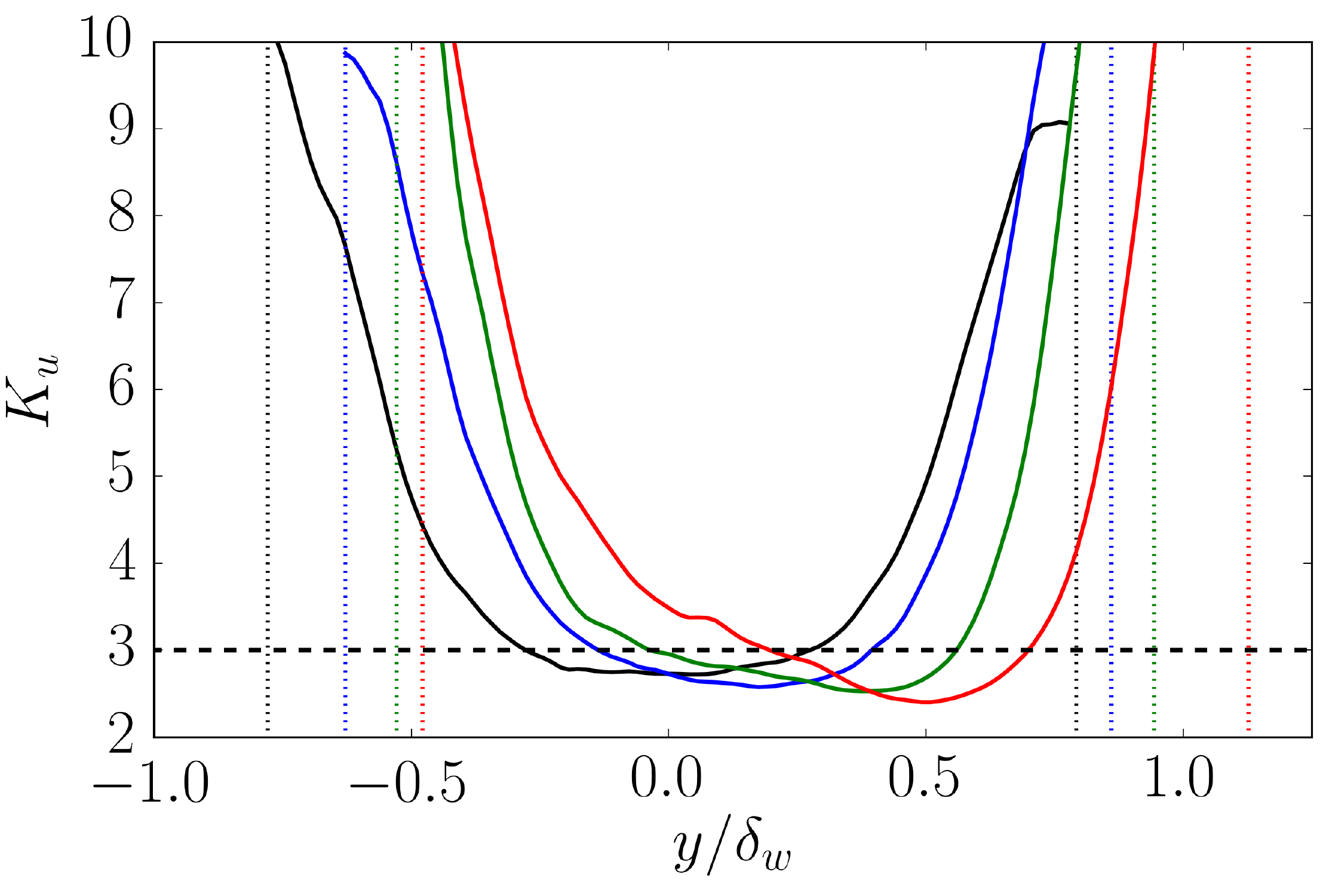}
\end{minipage}
\begin{minipage}{0.49\linewidth}
\centerline{$(e)$}
\includegraphics[width=\linewidth]{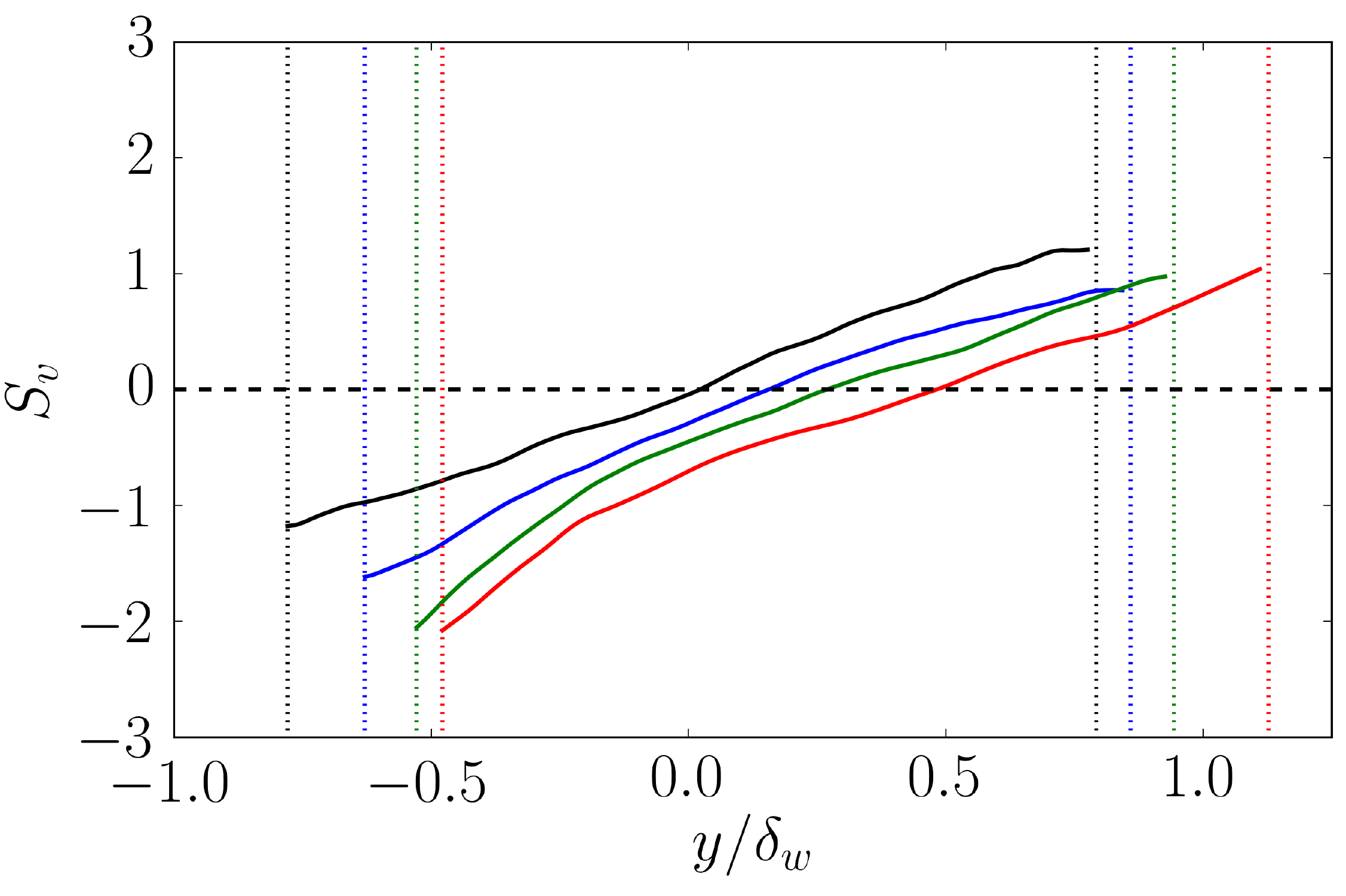}
\end{minipage}
\begin{minipage}{0.49\linewidth}
\centerline{$(f)$}
\includegraphics[width=\linewidth]{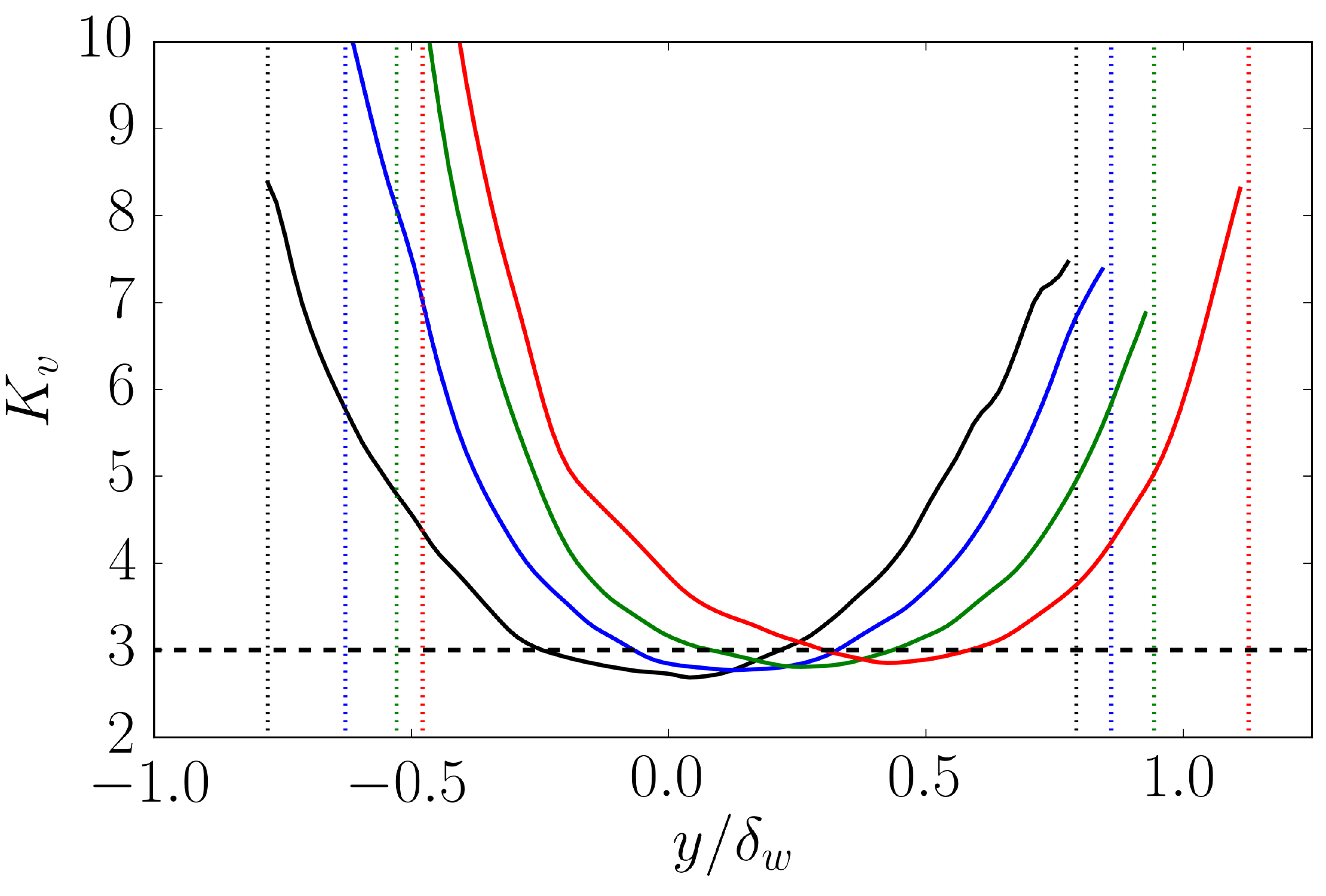}
\end{minipage}
\end{center}
\caption{ 
({\em a}) Skewness distribution and ({\em b}) Kurtosis distribution of temperature $\theta$; 
({\em c}) Skewness distribution and ({\em d}) Kurtosis distribution of streamwise velocity $u$;
({\em e}) Skewness distribution and ({\em f}) Kurtosis distribution of vertical velocity $v$. 
Different colours correspond to different density ratios: black, $s=1$; blue, $s=2$; green, $s=4$; and red, $s=8$. 
}
\label{fig:stats}
\end{figure}

Finally, figure \ref{fig:stats} shows the profiles of the 
skewness, $S$, and kurtosis, $K$,  of the temperature and the velocity field.
Since these profiles are more noisy than the second order moments beyond the
edge of the mixing layer, figure \ref{fig:stats} only shows them
in the region limited by 98\% of the free stream velocity, 
indicated with vertical dotted lines.
For reference,  
the horizontal dashed lines represent the expected value for a Gaussian distribution, i.e. $S=0$ and $K=3$.
Due to the symmetry of the configuration for the passive scalar case, $s=1$, we expect an antisymmetric distribution
for the skewness and a symmetric distribution for the kurtosis. 
Deviations from this symmetry in figure \ref{fig:stats} are small and provide an impression of the convergence of the statistics. 
Note also that the almost linear profile of  $\overline{\theta}$ in the center of the mixing layer results in $S_\theta \approx 0$ for the case with 
$s=1$ (recall the broad maximum of the vertical gradient of $\overline{\theta}$ in figure \ref{fig:profiles_T}). 

\cite{Carlier:2015} measured the skewness and kurtosis in a spatially-developing mixing layer. 
Their neutral case is comparable to the present passive scalar case. 
They distinguish between two zones. 
First, a mixed region in the central part,  
characterized by a moderate slope of the temperature skewness profile 
and an almost constant value of all kurtosis profiles. The value of $K$ in this region 
is somewhat smaller than the Gaussian value.
Secondly, the entrained region in the outer part
that presents higher slopes of the temperature skewness profile than the mixed, region and also steep
gradients of all kurtosis profiles. 
All these features are clearly observed in the present profiles for the passive scalar case.

Overall, increasing $s$ results in a shift of the profiles of $S$ and $K$ to the low density side, for both temperature and velocity. This is especially clear in $S_u$, $S_v$ and $K_v$, which show small variations on the shape of the profiles (see figures \ref{fig:stats}$c,e$ and $f$). For the skewness of the temperature (see figure \ref{fig:stats}$a$) we can observe the same shift, and a gradual increase of $S_\theta$ on the high density half of the central region of the mixing layer. This is probably a consequence of the narrowing of the maximum of $\partial \overline{\theta}/\partial y$ with $s$, and its displacement towards the high temperature (low density) side: 
a sharper edge on the high temperature side makes it more likely for a pocket of high temperature fluid to be entrained into the mixing layer, biasing $S_\theta$ towards positive values. 
It is also interesting to observe that, on top of the shifting, $K_\theta$ and $K_u$ show some changes in their shape with $s$. In particular, both kurtosis become larger in the high density half of the mixing layer ($y\lesssim 0$).  
This can be interpreted as an increase in the intermittency of $u$ and $T$, and it suggests that mixing becomes more difficult near the high density region as $s$ increases, in agreement with the qualitative arguments of \cite{Brown1974} regarding the reduced velocity fluctuations near the denser stream.  As a result, the size of the well mixed region (i.e., with values of $K$ below the Gaussian threshold) is reduced.

\subsection{Turbulence structure} 
\label{sec:spectra}

\begin{figure}
\begin{center}
\begin{tikzpicture}
    \node[anchor=south west,inner sep=0] at (0,0) {\includegraphics[width=1.0\textwidth]{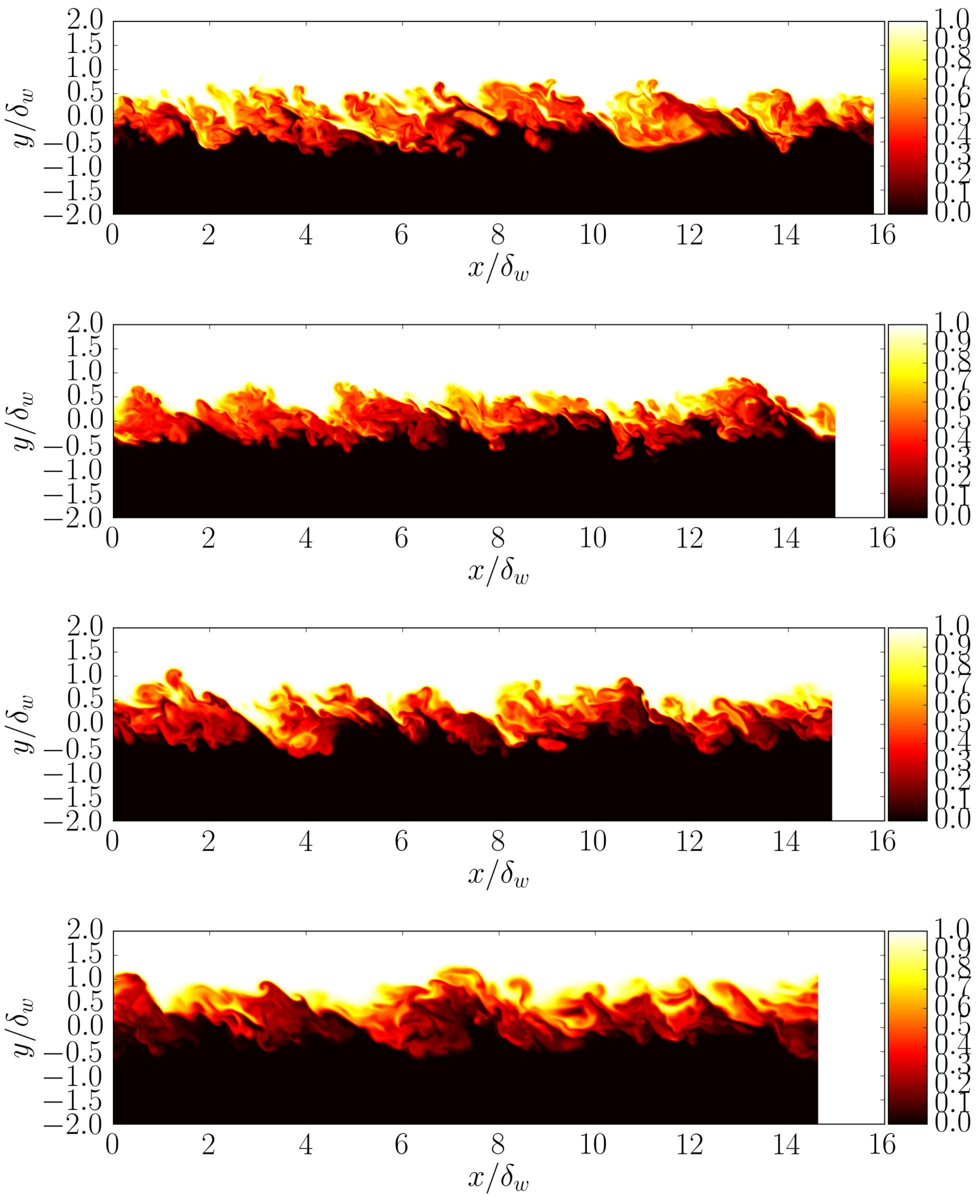}};
    \draw[-] (0.5,20.5) node[left] {(a)};
    \draw[-] (0.5,15) node[left] {(b)};
    \draw[-] (0.5,10) node[left] {(c)};
    \draw[-] (0.5,4.5) node[left] {(d)};
\end{tikzpicture}

\caption{Visualization of $\theta$ on an $xy$-plane, at the beginning of  the self-similar evolution. The corresponding density ratios and times are 
({\em a}) $s=1$, $t\Delta U/\delta_m^0 = 400$; 
({\em b}) $s=2$, $t\Delta U/\delta_m^0 = 418$; 
({\em c}) $s=4$, $t\Delta U/\delta_m^0 = 455$; 
and ({\em d}) $s=8$, $t\Delta U/\delta_m^0 = 570$.}
\label{fig:T_xyplanes}
\end{center}
\end{figure}
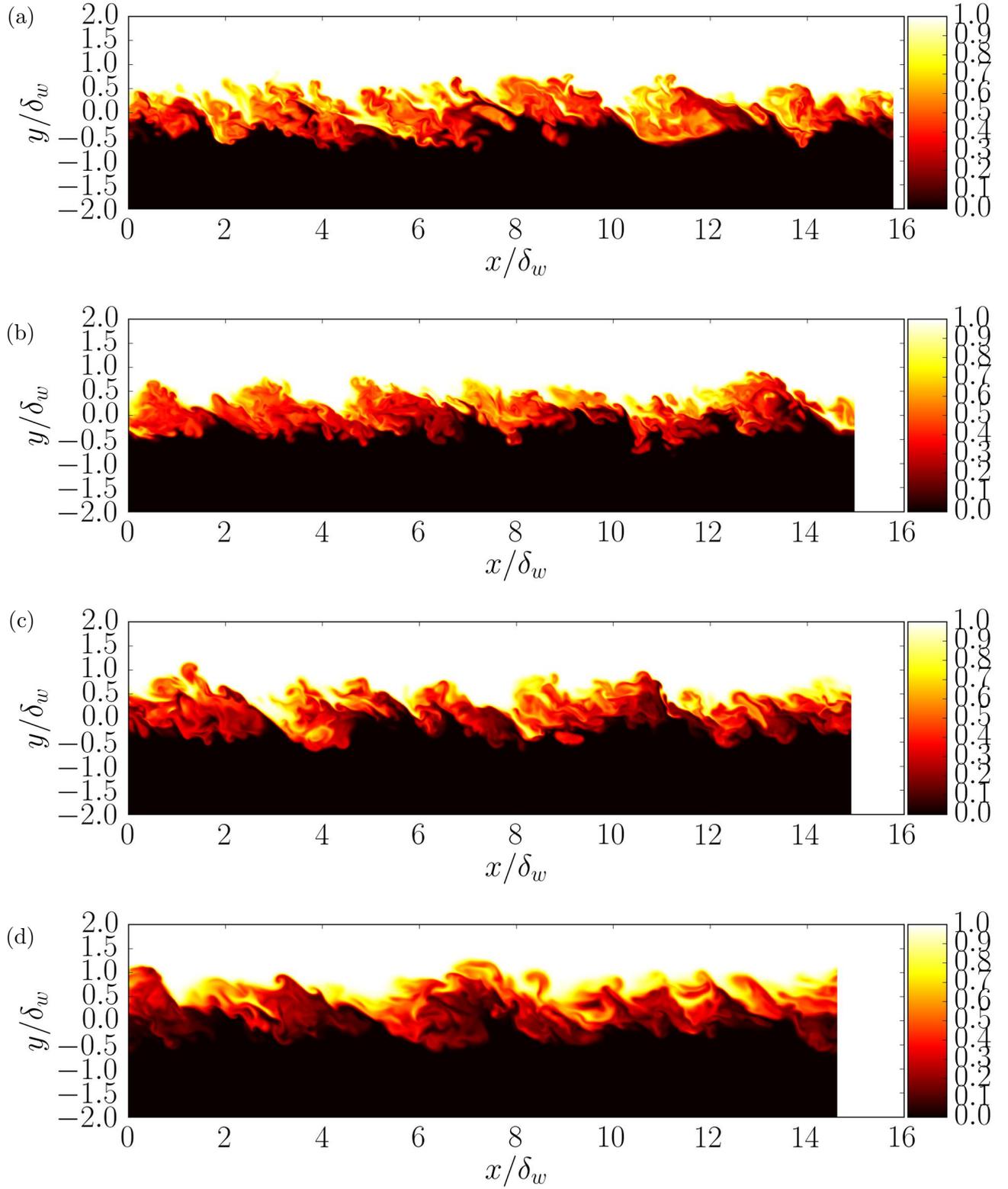

\begin{figure}
\begin{center}
\begin{tikzpicture}
\node[anchor=south west,inner sep=0] at (0,0) {\includegraphics[width=0.9\linewidth]{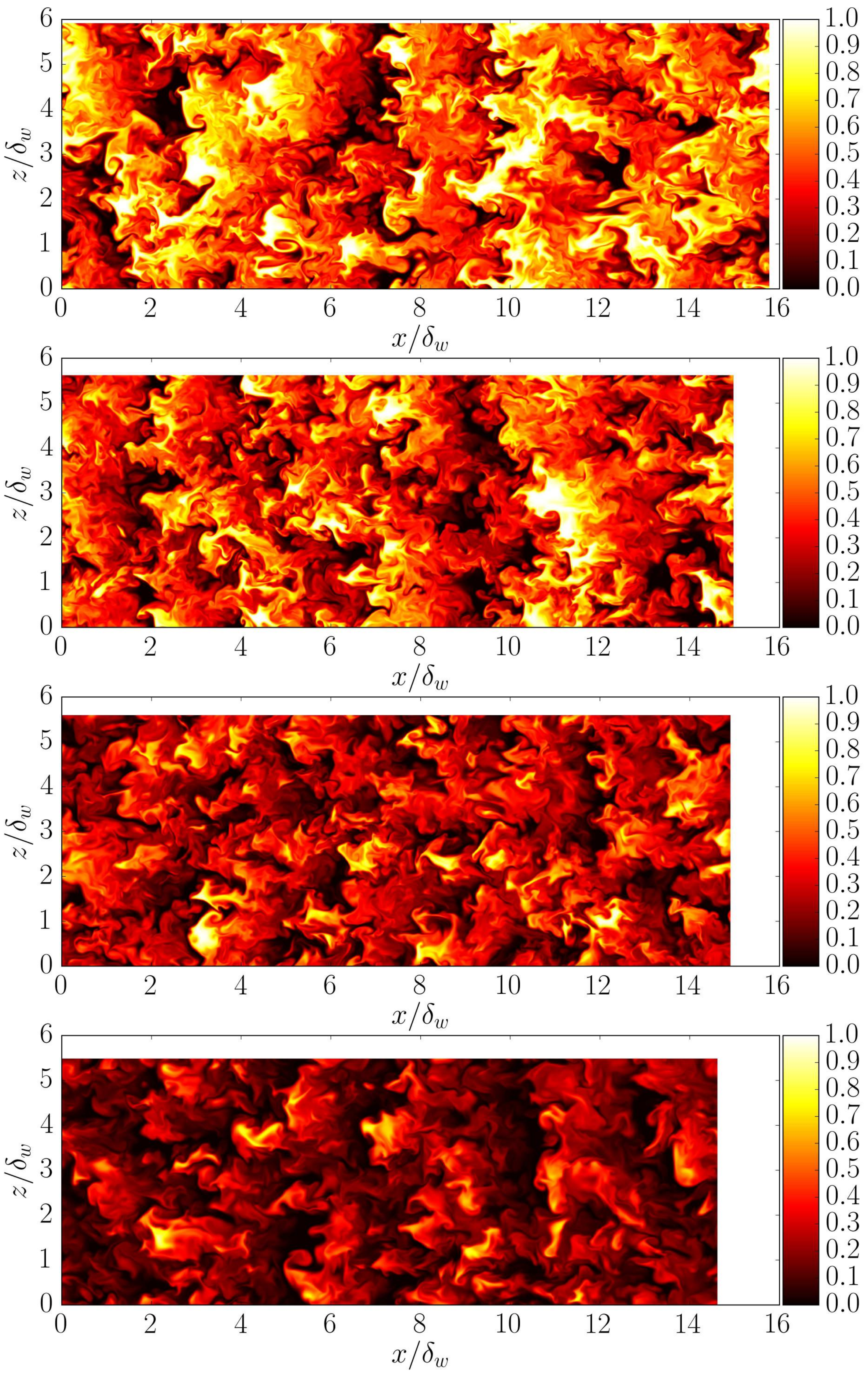}};
    \draw[-] (0,24) node[left] {(a)};
    \draw[-] (0,18) node[left] {(b)};
    \draw[-] (0,12) node[left] {(c)};
    \draw[-] (0,6) node[left] {(d)};
\end{tikzpicture}
\caption{Visualization of $\theta$ on an $xz$-plane at $y=0$, at the beginning of  the self-similar evolution. The corresponding density ratios are
({\em a}) $s=1$,
({\em b}) $s=2$,
({\em c}) $s=4$,
({\em d}) $s=8$.
Times as in figure \ref{fig:T_xyplanes}.}
\label{fig:T_xz_y0}
\end{center}
\end{figure}

\begin{figure}
\begin{center}
\begin{tikzpicture}
    \node[anchor=south west,inner sep=0] at (0,0) {\includegraphics[width=1\textwidth]{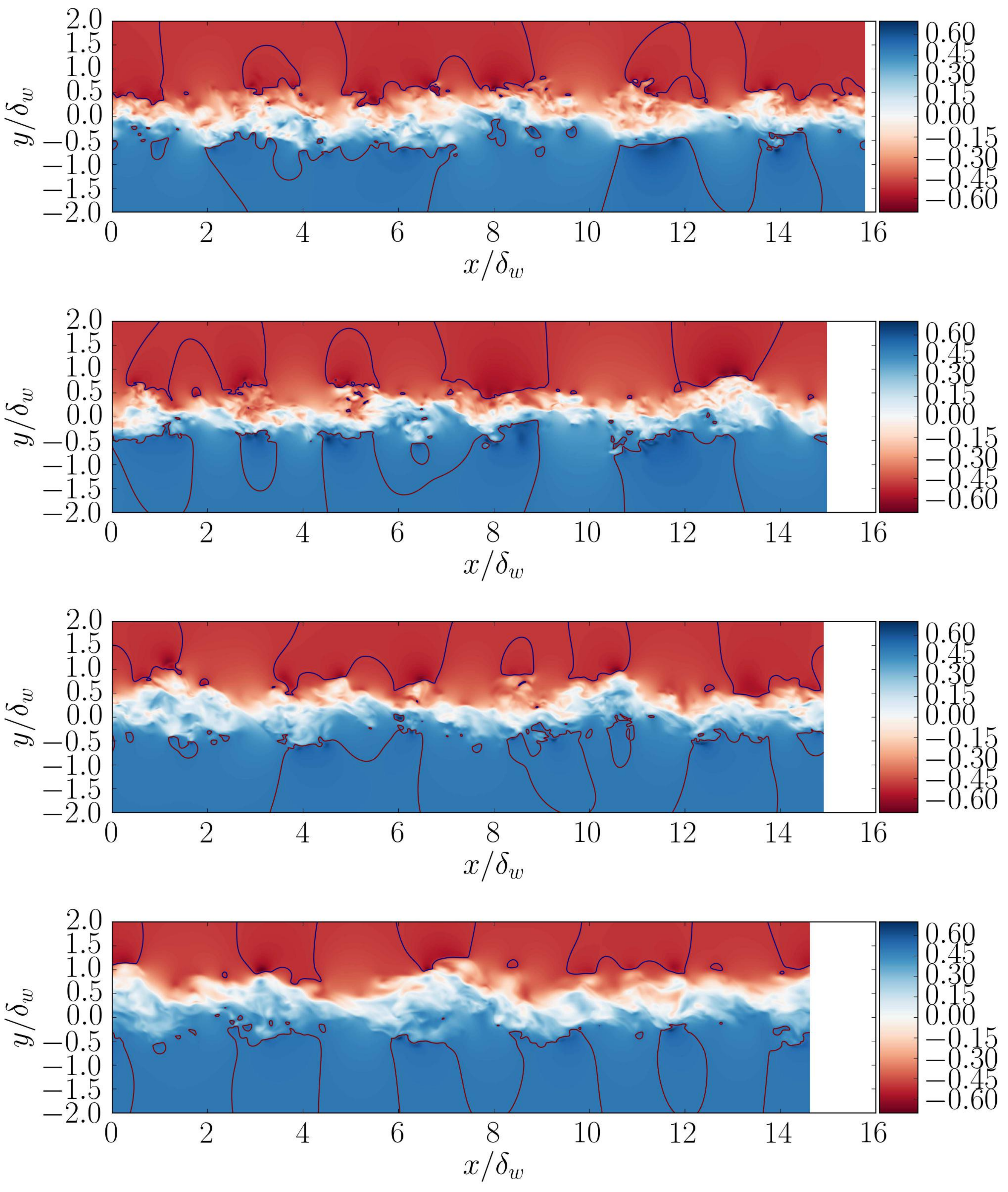}};
    \draw[-] (0.5,19.5) node[left] {(a)};
    \draw[-] (0.5,14.5) node[left] {(b)};
    \draw[-] (0.5,9.5) node[left] {(c)};
    \draw[-] (0.5,4.5) node[left] {(d)};
\end{tikzpicture}

\caption{Visualization of streamwise velocity on an $xy$-plane, at the beginning of  the self-similar evolution. The corresponding density ratios are
({\em a}) $s=1$, 
({\em b}) $s=2$, 
({\em c}) $s=4$, 
({\em d}) $s=8$.
Times as in figure \ref{fig:T_xyplanes}.
The black lines show contours of $u=\pm\Delta U/2$.}
\label{fig:u_xyplanes}
\end{center}
\end{figure}
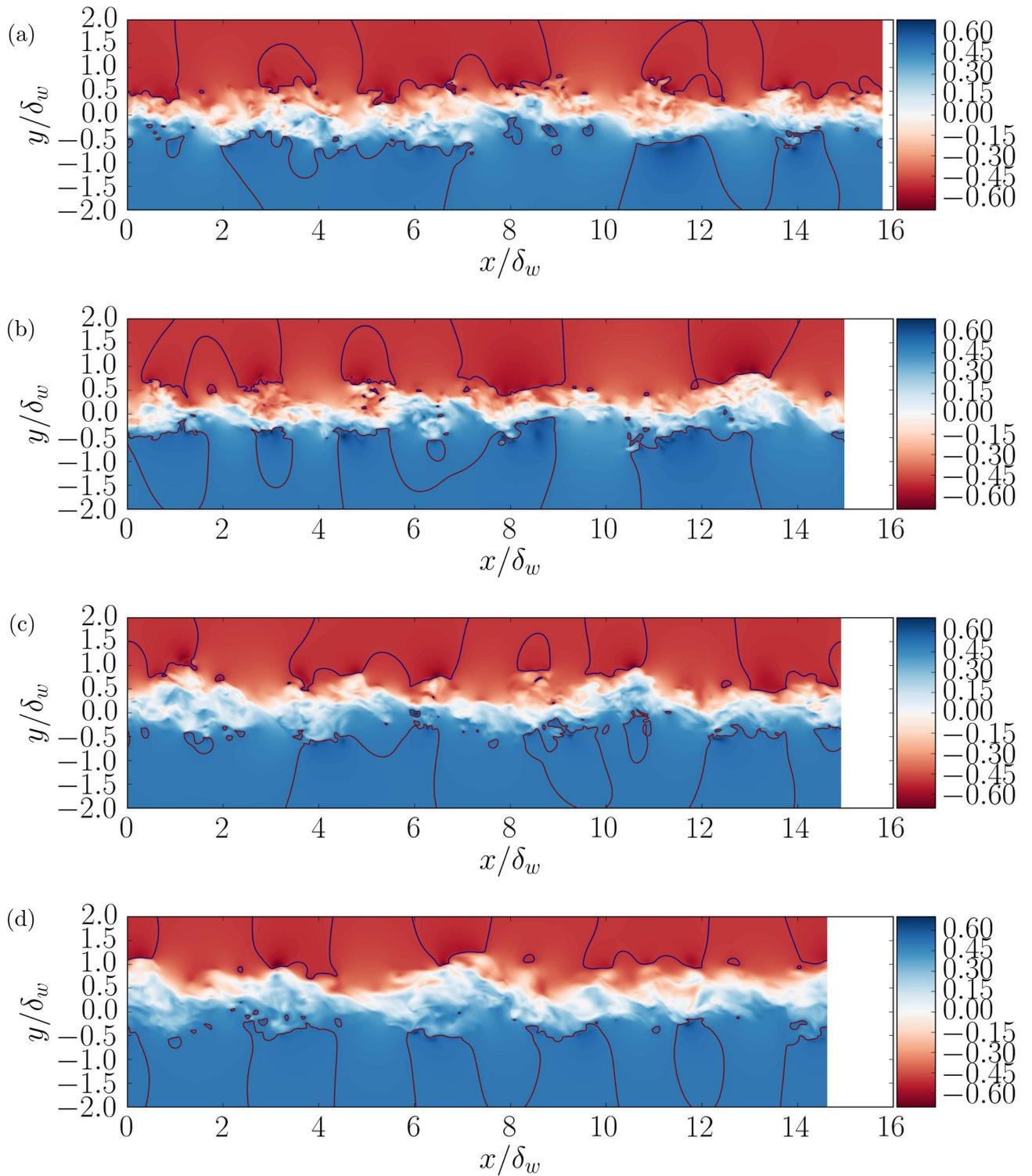

\begin{figure}
\begin{center}
\begin{tikzpicture}
\node[anchor=south west,inner sep=0] at (0,0) {\includegraphics[width=0.9\linewidth]{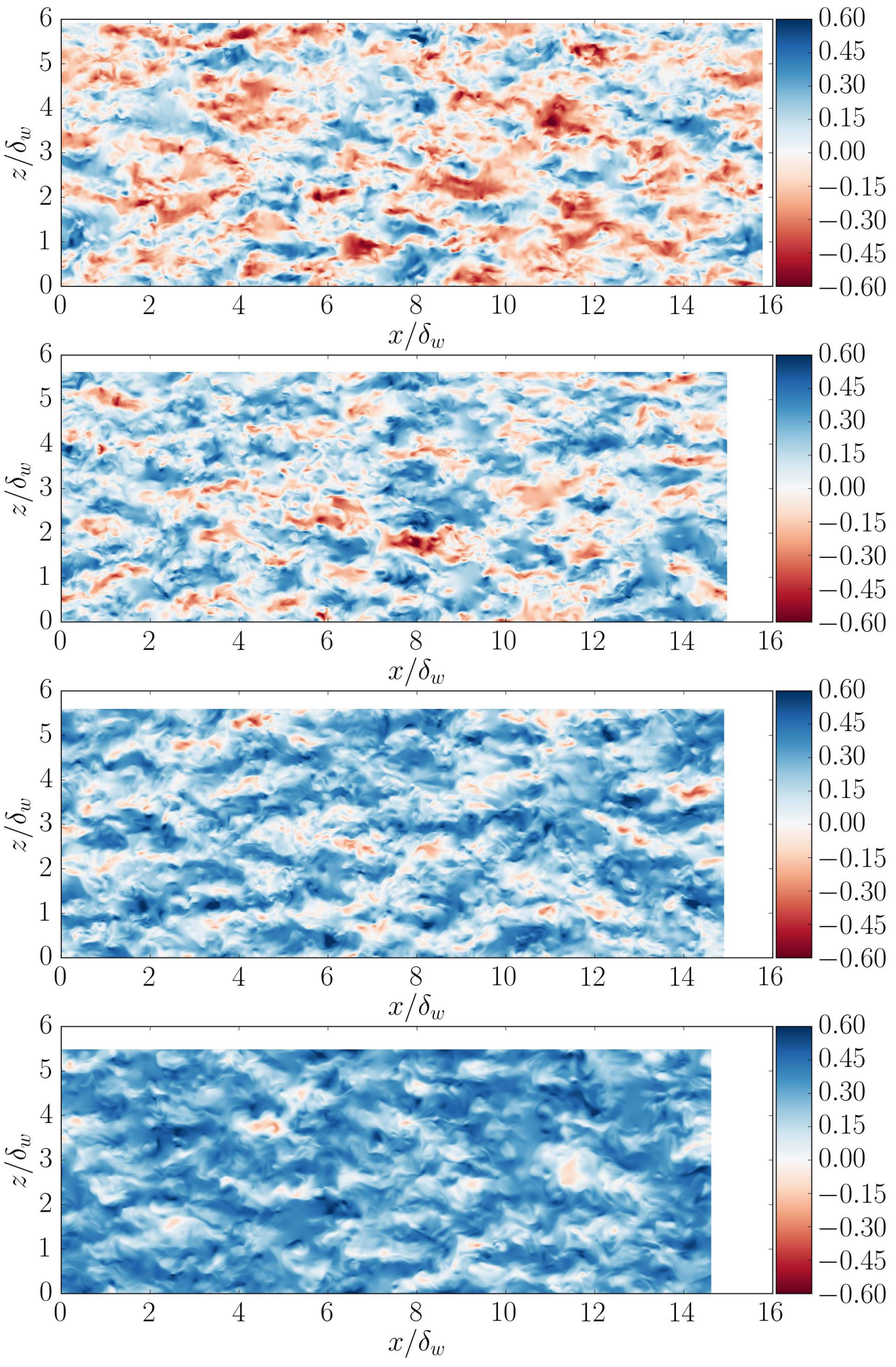}};
    \draw[-] (0,22.5) node[left] {(a)};
    \draw[-] (0,17.) node[left] {(b)};
    \draw[-] (0,11.5) node[left] {(c)};
    \draw[-] (0,5) node[left] {(d)};
\end{tikzpicture}
\caption{Visualization of streamwise velocity on an $xz$-plane at $y=0$, at the beginning of  the self-similar evolution. The corresponding density ratios are
({\em a}) $s=1$,
({\em b}) $s=2$,
({\em c}) $s=4$,
({\em d}) $s=8$.
Times as in figure \ref{fig:T_xyplanes}.}
\label{fig:u_xz_y0}
\end{center}
\end{figure}

We provide now visualizations to obtain an impression of the changes in the turbulent structures of the mixing layer induced by the density ratio. 
Instantaneous fields of the temperature and velocity field are shown 
 using vertical planes (figures \ref{fig:T_xyplanes} and \ref{fig:u_xyplanes} for $\theta$ and $u$, respectively) and horizontal planes 
(figure \ref{fig:T_xz_y0} and \ref{fig:u_xz_y0} for $\theta$ and $u$ at the plane $y=0$, respectively).
Note that, all visualizations correspond to the same time snapshot and same plane locations. 
For case $s=1$, in which the temperature is a passive scalar, the 
visualization in figure \ref{fig:T_xyplanes}$(a)$ 
shows the typical features of a turbulent mixing layer, with patches of mixed fluid in the central region
alternating with patches of unmixed fluid that are entrained from both streams.
The presence of quasi-2D rollers is visible in both temperature (figure \ref{fig:T_xyplanes}$a$) and velocity (figure \ref{fig:u_xyplanes}$a$) visualizations, but maybe more clearly so in the midplane visualization of the temperature shown in figure \ref{fig:T_xz_y0}$(a)$. 

On the other hand,  the presence of the quasi-2D rollers in the $u$ velocity (figure \ref{fig:u_xyplanes}$a$) is masked by the formation of more elongated structures, 
similar to the streaky structures observed in other free and wall-bounded turbulent shear flows \citep{lee1990structure, flores2010hierarchy, sekimoto2016direct}.

Increasing the density ratio produces small changes in the flow visualizations. The quasi-2D rollers are also observed for $s=2$, 4 and 8 in both temperature (figure \ref{fig:T_xyplanes}$b,c$ and $d$) and velocity (figure \ref{fig:u_xyplanes}$b,c$ and $d$). 
Also, in agreement with the results discussed in section \ref{sec:mean},
the mixing layer shifts upwards (towards
the low density side) with increasing $s$, as it can be observed in figure \ref{fig:T_xyplanes} and \ref{fig:u_xyplanes}. 
In addition, the temperature field becomes somewhat smoother at the small scales. 
This fact is reflected in the lower value of $Re_\lambda$ obtained in the cases
with large $s$, as shown in table \ref{tab:SimParams}. 

The shift of the mixing layer is also apparent in the visualization of the $y=0$ plane shown in 
figures  \ref{fig:T_xz_y0} and \ref{fig:u_xz_y0}. With increasing $s$ the temperature field at this height 
is increasingly dominated by patches of fluid entrained from the lower stream, 
while the mean value of the $u$ field drifts to positive values. 
The footprint of the quasi-2D rollers is also clear in the temperature field (figure \ref{fig:T_xz_y0}) for all density ratios, 
while this footprint becomes less apparent in the $u$ velocity as $s$ increases (figure \ref{fig:u_xz_y0}). 
 
Finally, it is interesting to observe in figure \ref{fig:u_xyplanes} that the turbulence within the mixing layer produce irrotational perturbations into the free-stream, with characteristic sizes of the order of $\delta_w$. 
This potential perturbations are relatively weak, and are highlighted in figure \ref{fig:u_xyplanes} by contours of $u=\pm\Delta U$ (in black).

In order to quantify the changes in the structure of the turbulent motions in the mixing layer due to the density ratio, we proceed to analyse the one dimensional spectra of velocity and temperature fluctuations: 
$E_{ii}(k_x,y)$ and $E_{ii}(k_z,y)$ for $i=u, v$ and $T$ (no summation). 
These spectra are computed during runtime, as functions of $k_x\delta_m^0$, $k_z\delta_m^0$, $y/\delta_m^0$ and $t$. Then, during post-processing, these spectra are interpolated into wavenumbers and vertical distances normalized with $\delta_w(t)$, and averaged (ensemble and in time) for the self-similar evolution of the mixing layer. 
The smallest wavenumbers considered in the interpolation are $k_x^0\delta_w \approx 0.4-0.5$ and $k_z^0\delta_w \approx 1.1-1.3$, depending on the density ratio. 
 
\begin{figure}
\begin{center}
\begin{minipage}{0.49\linewidth}
\centerline{$(a)$}
\includegraphics[width=\linewidth]{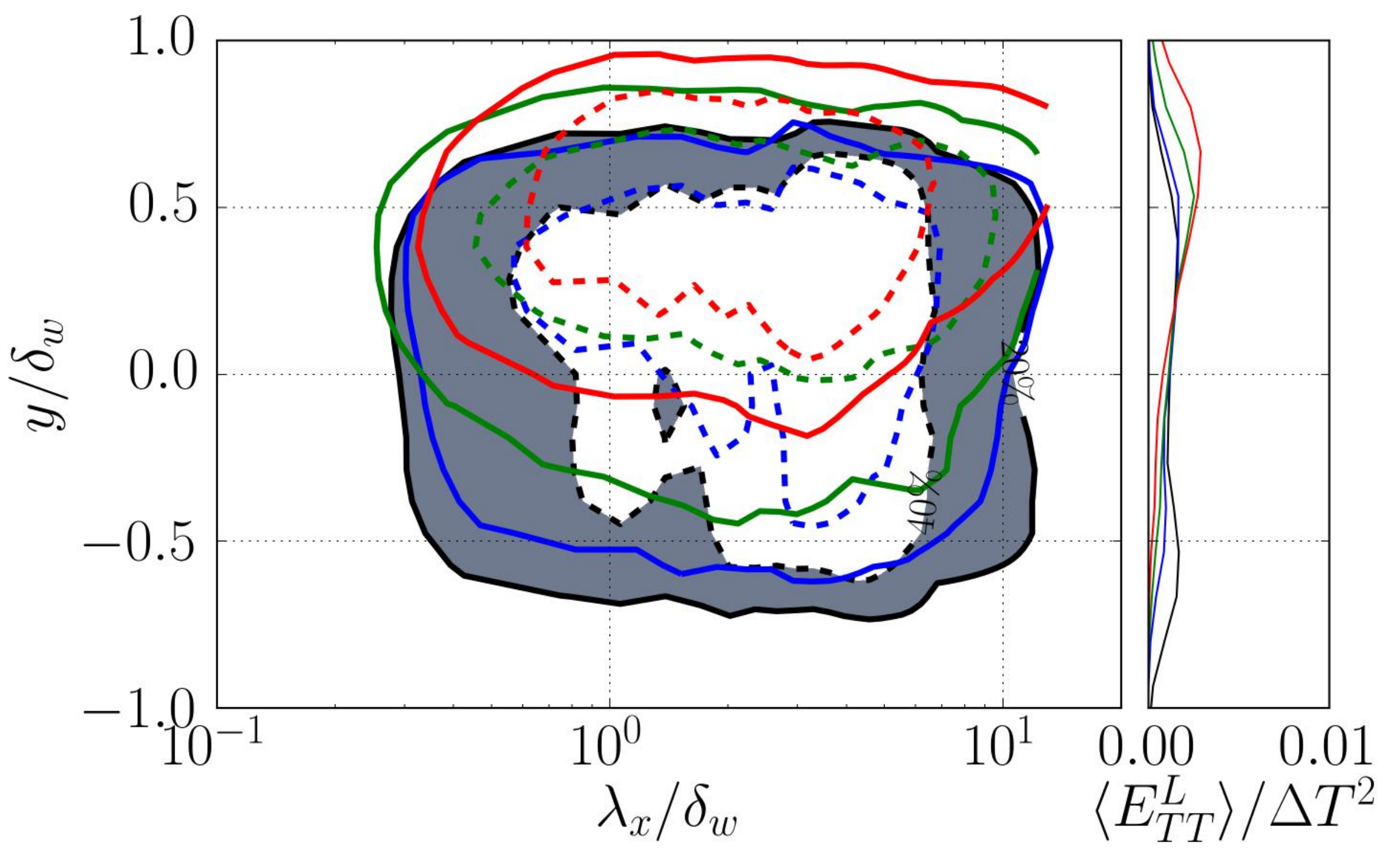}
\end{minipage}
\begin{minipage}{0.49\linewidth}
\centerline{$(b)$}
\includegraphics[width=\linewidth]{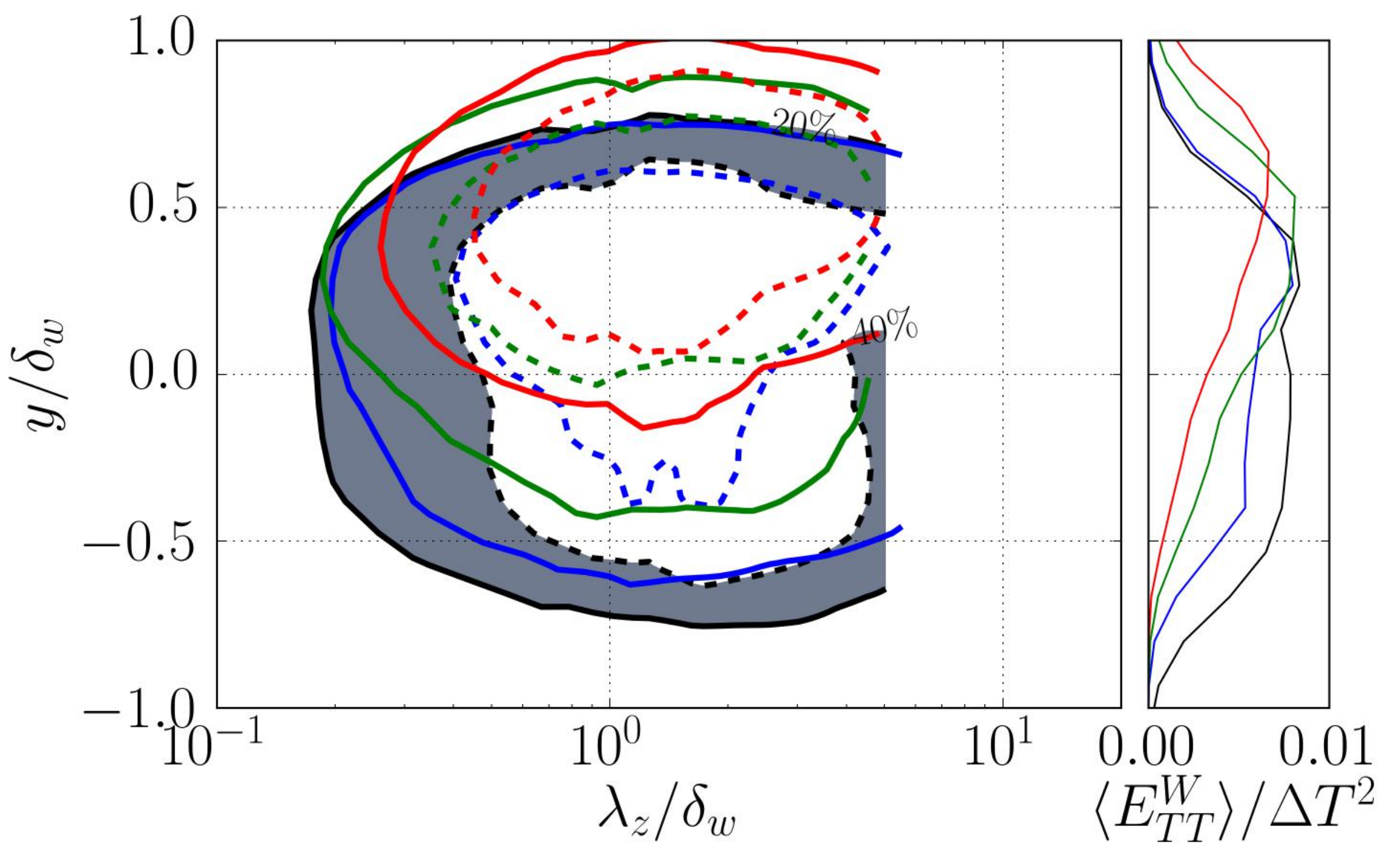}
\end{minipage}
\begin{minipage}{0.49\linewidth}
\centerline{$(c)$}
\includegraphics[width=\linewidth]{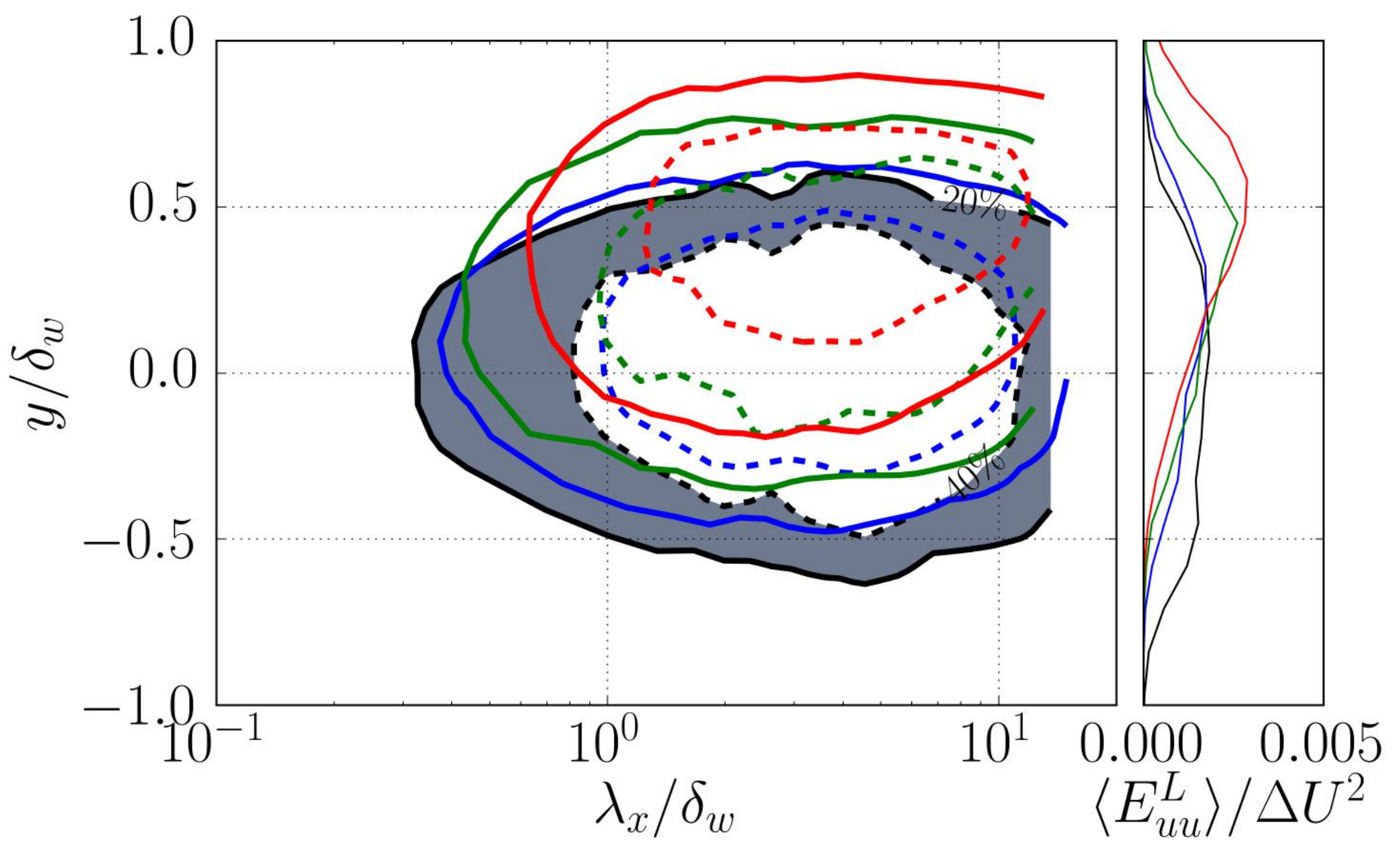}
\end{minipage}
\begin{minipage}{0.49\linewidth}
\centerline{$(d)$}
\includegraphics[width=\linewidth]{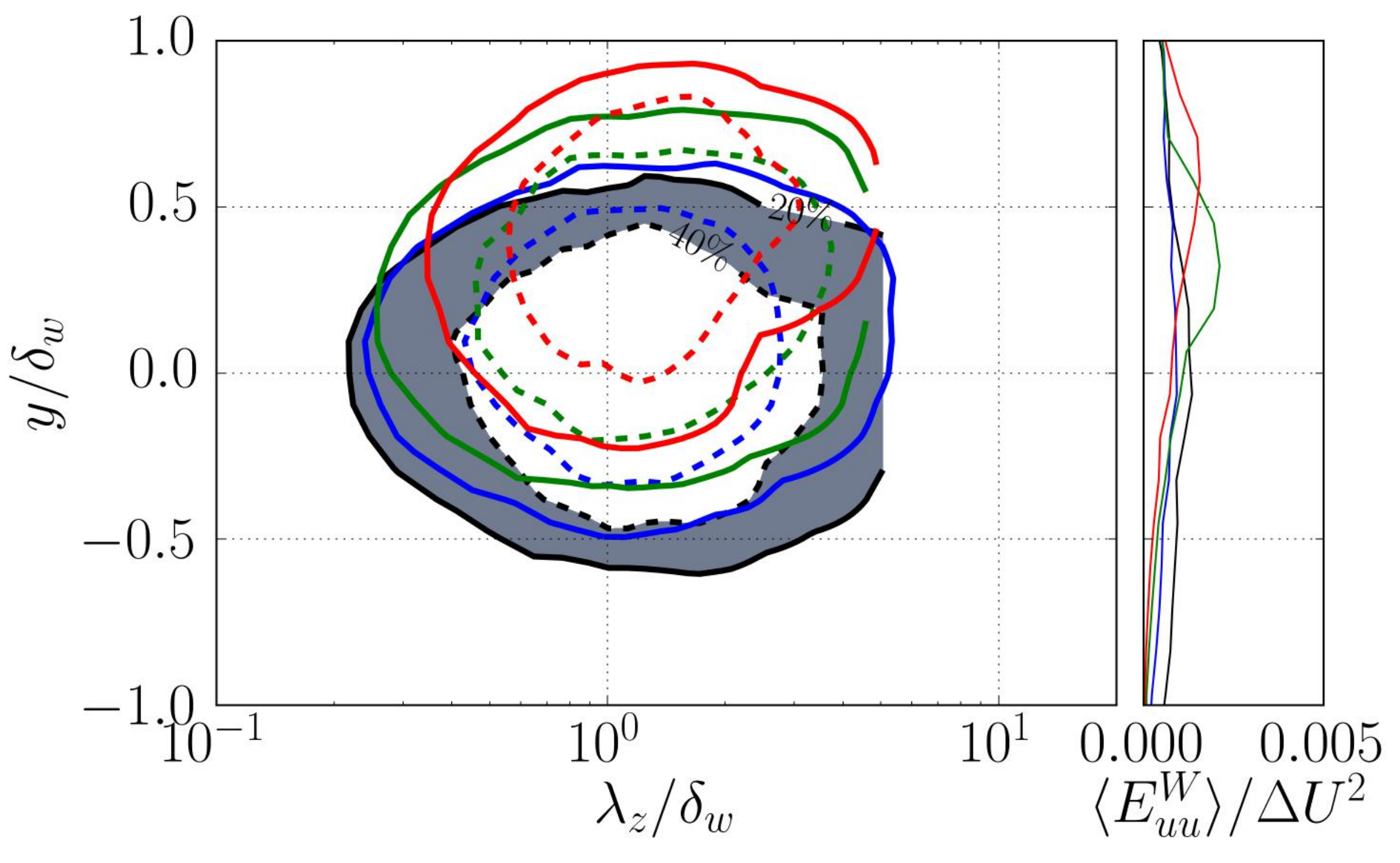}
\end{minipage}
\begin{minipage}{0.49\linewidth}
\centerline{$(e)$}
\includegraphics[width=\linewidth]{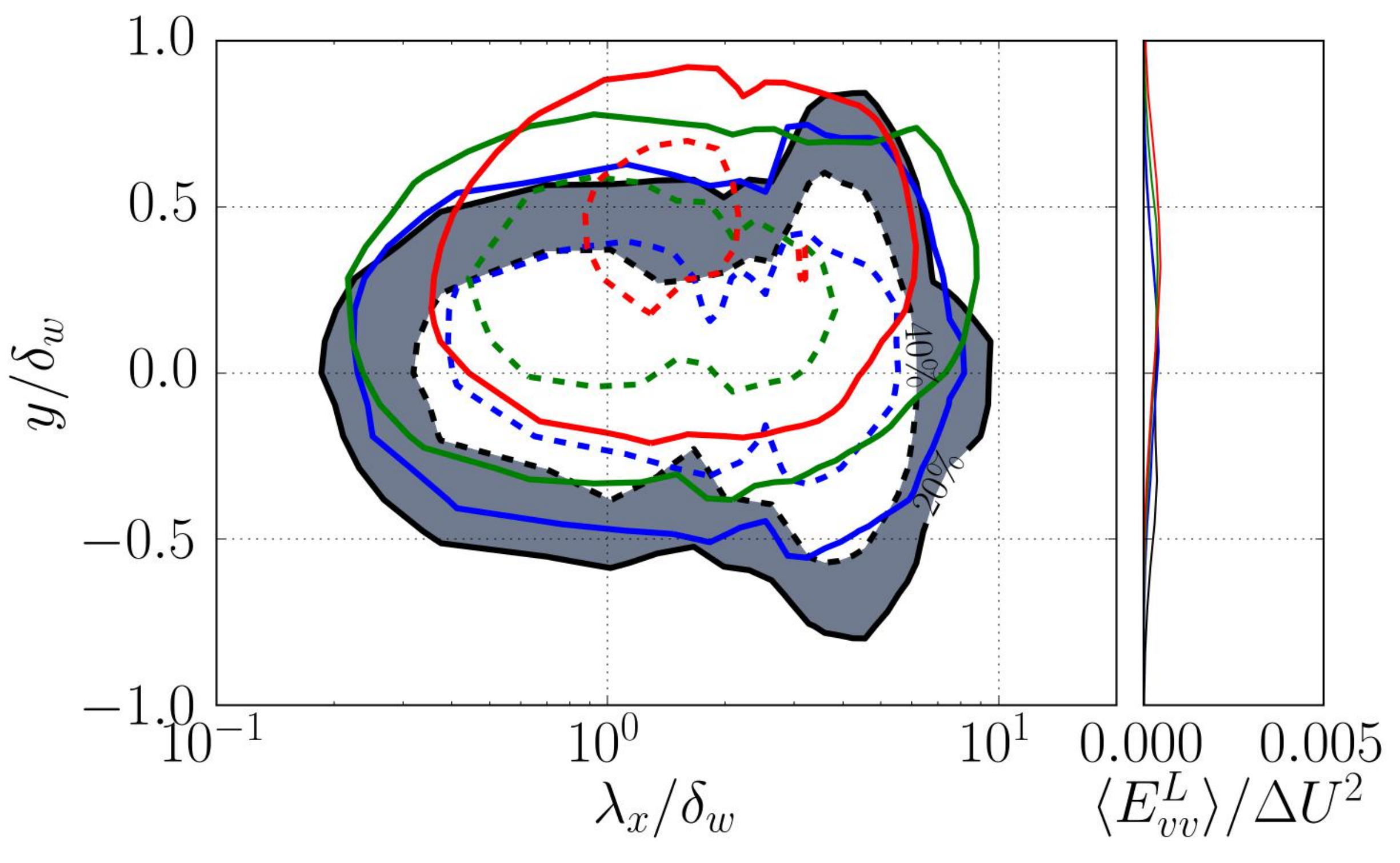}
\end{minipage}
\begin{minipage}{0.49\linewidth}
\centerline{$(f)$}
\includegraphics[width=\linewidth]{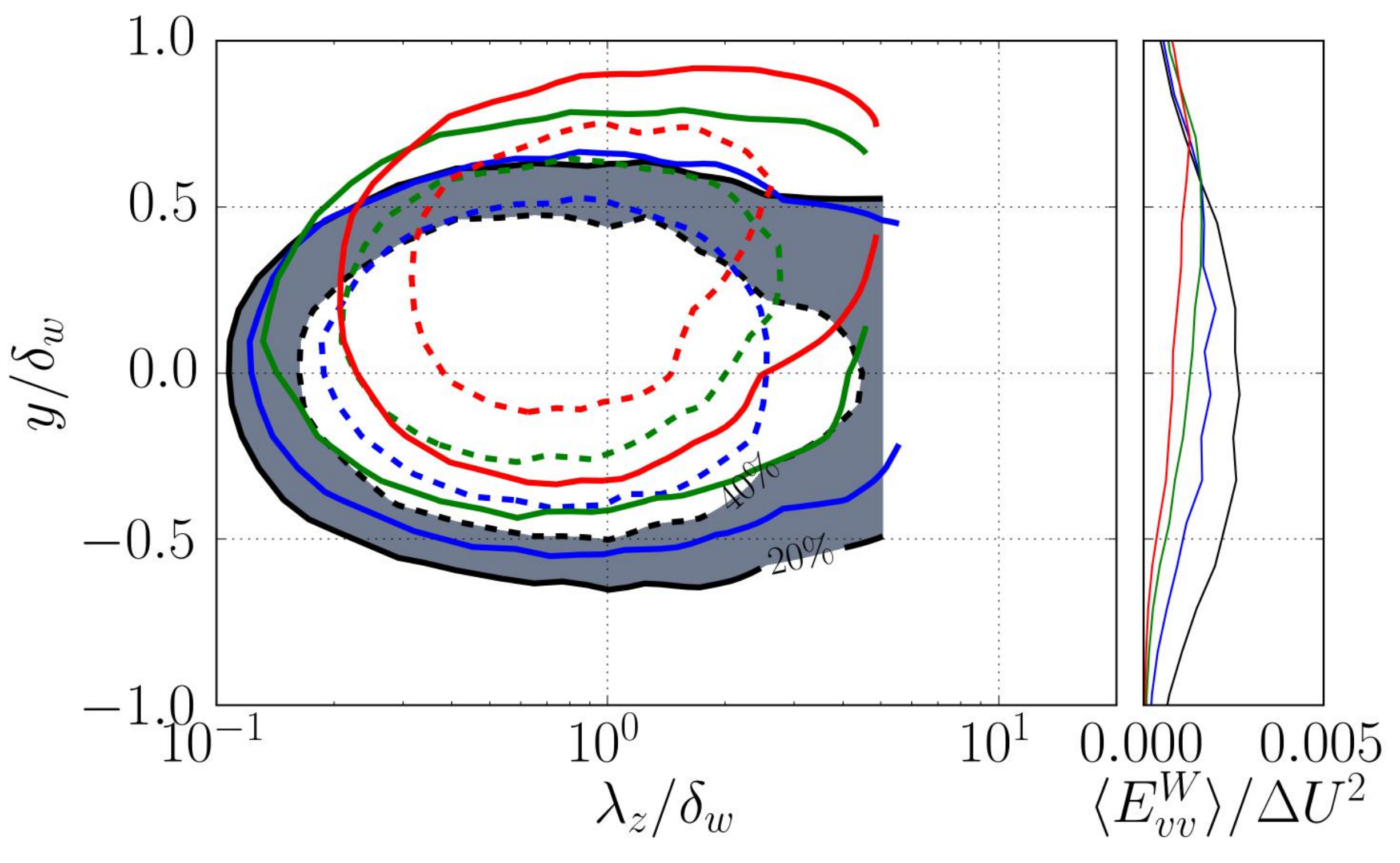}
\end{minipage}
\end{center}
\caption{
Vertical distribution of the premultiplied spectral energy distribution of velocity and temperature.  
({\em a}) $k_x E_{TT}(\lambda_x,y)$. 
({\em b}) $k_z E_{TT}(\lambda_z,y)$. 
({\em c}) $k_x E_{uu}(\lambda_x,y)$. 
({\em d}) $k_z E_{uu}(\lambda_z,y)$. 
({\em e}) $k_x E_{vv}(\lambda_x,y)$. 
({\em f}) $k_z E_{vv}(\lambda_z,y)$. 
The inset to the right of each panel shows the energy in wavenumbers not included in the corresponding panel (see text for discussion). 
The contours plotted correspond to 20\% (solid) and 40\%(dashed) of the maxima of all the spectra shown in each panel. 
Different colours correspond to different density ratios: black with shading, $s=1$; blue, $s=2$; green, $s=4$ and red, $s=8$. 
}
\label{fig:spectra}
\end{figure}

Figure \ref{fig:spectra} shows the premultiplied spectra ($k_x E_{ii}$ and $k_z E_{ii}$), as a function of the vertical position in the mixing layer and the streamwise or spanwise wavelength, $\lambda_x = 2\pi/k_x$ and $\lambda_z=2\pi/k_z$. The spectra is premultiplied by the wavenumber so that, when plotted in log-scale for the wavelength, the area under the surface corresponds to the actual energy content of a given range of wavelengths.  
The contours plotted in the figure correspond to 20\% and 40\% of the maxima among all cases, so that they represent equal levels of energy density for all cases. 
The small inset to the right of each panel shows the energy in wavenumbers smaller than $k_x^0$ and $k_z^0$, 
\begin{equation} 
E_{ii}^{L}(y)=\sum_{k_x=0}^{k_x^0}  E_{ii}(k_x,y) 
\hspace{4mm}\mbox{and}\hspace{4mm}
E_{ii}^{W}(y)= \sum_{k_z=0}^{k_z^0} E_{ii}(k_z,y).  
\end{equation}

From a physical point of view, these two quantities roughly corresponds to the energy in structures that are infinitely long ($E_{ii}^L$) or wide ($E_{ii}^W$). 

For the incompressible case, figure \ref{fig:spectra} shows that the spectra of $u$ tend to be longer than wide, while the spectra of $v$ and $T$ tend to be wider than long, 
consistently with the visualizations shown in figures \ref{fig:T_xz_y0} and \ref{fig:u_xz_y0}. 
Indeed, both $T$ and $v$ show considerably more energy on structures that are wide ($\lambda_z > 2\pi/k_z^0 \approx 5\delta_w$) than in structures that are long ($\lambda_x > 2\pi/k_x^0 \approx 12\delta_w$), which is shown by $E_{vv}^W > E_{vv}^L$ and $E_{TT}^W > E_{TT}^L$. 
It is also apparent in figure \ref{fig:spectra} that the spectra of $v$ is shifted towards smaller scales with respect to the spectra of $u$ and $T$, both in $\lambda_x$ and $\lambda_z$. In terms of the vertical extension of the spectra, \ref{fig:spectra}($a$) and ($b$) show that the temperature spreads over $\abs{y} \lesssim 0.8\delta_w$, while $u$ and $v$ are limited to a narrower region ($\abs{y} \lesssim 0.5\delta_w$), in agreement with the results shown in figure \ref{fig:Rij}. Interestingly, figure \ref{fig:spectra}($e$) shows that $E_{vv}$ has a larger spread in the vertical direction, at about $\lambda_x \approx 4\delta_w$. Careful inspection of figures \ref{fig:spectra}($e$) and ($f$) shows that those peaks correspond to infinitely wide structures ($k_z=0$): 
note that $E_{vv}(\lambda_z,y)$ at $y=0.7\delta_w$ has little energy (i.e., below the 20\% contour), while $E_{vv}^W$ at the same height is still important (i.e., around 50\% of the maximum of $E_{vv}^W$).
 
Although not shown here, instantaneous visualizations of $v$ show that these wavelengths ($\lambda_x \approx 4\delta_w$, $\lambda_z \to \infty$) roughly correspond to  potential perturbations of $v$ into the free-stream,
 analogous to the potential perturbations of $u$ highlighted in figure \ref{fig:u_xyplanes}.

As the density ratio increases, figures \ref{fig:spectra}($a$) and ($b$) show that the spectra of the temperature gradually shifts towards the low density side (see the contours of 20\% in the figures). The shift occurs first on the high density edge of the spectrum ($y<0$), and a bit later in low density side ($y>0$). Note that for $y/\delta_w \gtrsim 0.25$, there are little differences between the spectra of the $s=1$ and $s=2$ cases, consistent with the agreement of $T_{rms}^2$ in figure \ref{fig:Rij}($d$) in these same vertical locations. In terms of the effect of $s$ in the streamwise and spanwise wavelengths, figure \ref{fig:spectra}($a$) shows that the longest scales in the high density side are gradually inhibited ($\lambda_x/\delta_w\approx 5-10, y\approx 0$). The same effect, although weaker, is also present in the spanwise wavelengths (figure \ref{fig:spectra}$b$). 
In terms of the small scales, figures \ref{fig:spectra}($a$) and ($b$) suggest that the effect of $s$ is stronger on $\lambda_z$ than on $\lambda_x$. This could be related to the fact that the small scales in $x$ are not only due to turbulent fluctuations (i.e., vortices), but to the formation of sharp gradients $\partial T/\partial x$, due to the roll-up of the shear layer (see blue lines in figures \ref{fig:T_xyplanes} and \ref{fig:T_xz_y0}). 

The behaviour of the spectra of $u$ and $v$ in figures \ref{fig:spectra}($c-f$) is qualitatively similar to that discussed for $T$, with all spectra shifting towards the low-density side, with a gradual reduction of the energy in small scales (both $\lambda_x$ and $\lambda_z$). There is also a clear reduction of the energy of large scales near the high-density edge of the mixing layer, more apparent for wide ($\lambda_z/\delta_w \gtrsim 3-5$) structures than for long structures ($\lambda_x/\delta_w \gtrsim 5-10$). The $u$ and $v$ spectra of cases $s=1$ and $s=2$ also agree reasonably well near the low-density edge of the mixing layer ($y\gtrsim 0.25\delta_w$), except for the $v$ spectrum at about $\lambda_x \approx 4\delta_w$, suggesting that even a small change on the density ratio has an important effect on the potential perturbations of the mixing layer into the free-stream.

\section{Conclusions} \label{sec:conclusions}

In this paper we have presented results from direct numerical simulations of temporal, turbulent mixing layers with variable density. The simulations are performed in the low-Mach number limit, so that temperature and density fluctuations develop while the thermodynamic pressure remains constant. Four different density ratios are considered, $s=1$, 2, 4 and 8, which are run in large computational boxes until they reach an approximate self-similar evolution. To give an impression of the turbulence in these mixing layers, during the self-similar evolution the Reynolds numbers based on the Taylor micro-scale vary between $Re_\lambda =$ 140-150 for the case $s=1$, and $Re_\lambda =$ 85-95 for the case with the highest density ratio, $s=8$. 

The results of the simulations show that, in agreement with turbulent mixing layers with higher velocities (and  convective Mach number, $M_c =0.7$), the growth rate of the momentum thickness decreases with the density ratio. 
Note that at a given density ratio, the momentum thickness of the low-Mach number mixing layer will grow faster than the subsonic one. 
However, the ratio between the growth rate for large density ratios and the growth rate of the $s=1$ case seems to be independent 
of the flow speed in the range considered. For example, for $s=8$ a 60\% growth reduction with respect to $s=1$ is obtained for both the present
low Mach number case and the $M_c=0.7$ case.

In terms of the visual thickness of the mixing layer, the effect of the density ratio 
in the growth reduction with respect to the $s=1$ case 
is smaller, and our results agree with previous theoretical models for $M_c=0$ and with the data of high-speed mixing layers. 
However, the growth rate reduction for low density ratio ($s=2$) is not the same in the $M_c=0$ and in the $M_c=0.7$ 
cases from \citet{Pantano2002}. 
This discrepancy could mean that compressibility effects are more dominant for low
 density ratios, but additional analysis of the subsonic cases is required to confirm this conjecture.

The Favre averaged profiles of velocity show that with increasing density ratio, the gradients shift towards the low density side. The behaviour is analogous to that observed in high-speed mixing layers. Indeed, the velocity and density profiles of  our low-Mach number cases agree qualitatively well with the high-speed cases when the vertical distance is normalised with $\delta_m$. There are some small differences in the mean velocities near the low density stream, and the density profiles of the high-speed cases seem to be displaced with respect to the low-Mach number profiles.

We have quantified the shifting of the Favre-averaged velocity profiles
 with respect to the density profiles as the distance ($\Delta$) between the locations where velocity and density are equal to the mid value between the free-streams.
Using our data, we have also obtained an empirical relationship between $\Delta$ and $s$, which we have used to obtain a semi-empirical model for the reduction of momentum thickness growth rate with the density ratio, see equation (\ref{eq:final_model}). This model uses the 
theoretical prediction of the reduction of the vorticity thickness growth rate due to \citet{Ramshaw2000}. 
From a physical point of view, the model assumes that the only effect of the density ratio is a shift in the velocity profile, with no change on the shape of the density and velocity profiles. 
Our data for $M_c=0$ and the data of \citet{Pantano2002} for $M_c=0.7$ are in good agreement with the model prediction, except for maybe the $M_c=0.7$ case at low density ratios ($s\approx 2$).
It would be interesting to check the validity of the model prediction for higher density ratios.

The fluctuation profiles of the low-Mach number cases show that, as expected, the fluctuations follow the gradients. 

While velocity and temperature shift towards the low-density region, density fluctuations and gradients seem to concentrate near the high-density edge of the mixing layer, consistently with the quantitative arguments of \cite{Brown1974} for the asymmetric growth of variable density mixing layers.
 
The analysis of the skewness and the kurtosis of the fluctuations shows that increasing the density ratio, the well mixed region that appears in the central region of the case $s=1$ becomes narrower, since mixing becomes more difficult
near the high density side as the density ratio is increased. 

Finally, the flow structures have been analyzed using flow visualizations and premultiplied spectra. The spectra shows that with increasing
density ratio there is a shift of the turbulent structures towards the low density side, while the longest scales in the high density side are gradually inhibited. 
A gradual reduction of the energy in small scales with increasing density ratio is also observed. This effect is consistent with the reduction
of $Re_\lambda$ with increasing density ratio mentioned above. 
\\

\section*{Acknowledgments}

This work was supported by the Spanish MCINN through project CSD2010-00011.
 The computational resources provided by RES (Red Espa\~nola de Supercomputaci\'on) and by  
XSEDE (Extreme Science and Engineering Discovery Environment) are gratefully acknowledged.

\appendix
\section{Numerical method}
\label{sec:appendix}

In this section, we describe the equations and algorithms implemented in the in-house code employed in this work
for solving temporal mixing layers under the low-Mach number approximation.
The governing equations for a variable-density flow with constant fluid properties
under the low-Mach number approximation can be written in the following dimensionless form,
\begin{align}
 \ppt{\rho}+\ppi{(\rho u_i)} &  = 0, \label{eq:continuity2} \\
 \ppt{(\rho u_i)}+\ppj{(\rho u_i u_j)} &  =-\ppi{p^{(1)}}+\frac{1}{Re}\ppj{\tau_{ij}},
\label{eq:momentum2} \\
 \ppt{T}+ u_i \ppi{T} &  =  \frac{T}{Pe}  \nabla^2{T},
\label{eq:energy2}\\
 \rho T & = 1,
\label{eq:EOS12}
\end{align}
where, all variables are non-dimensionalized by the initial momentum thickness $\delta_m^0$, 
the characteristic velocity $\Delta U$ and the physical magnitudes at the reference temperature, $T_0$, 
and pressure, $p^{(0)}=\rho_0 R T_0$, 
namely $\rho_0$, $\mu $, $C_{p}$ and $\kappa$. Therefore, the dimensionless numbers appearing here are defined as,
\begin{equation}
Re = \frac{\rho_0 \Delta U \delta_m^0}{\mu},
\end{equation}
\begin{equation}
Pe = \frac{\rho_0 C_{p} \Delta U \delta_m^0}{\kappa} = Pr Re. 
\end{equation}
Note that $p^{(1)}$ in eq. (\ref{eq:momentum2}) is the mechanical pressure, different from the
thermodynamic pressure, $p^{(0)}$, as discussed in the introduction.
In order to eliminate the mechanical pressure $p^{(1)}$ from the equations, 
first a Helmholtz decomposition is applied  to the momentum vector  
\begin{equation}
\rho \vec{u} = \vec{m} + \nabla \psi, \label{eq:decomposition}
\end{equation}
with $\vec{m}$ being a divergence-free component, so that
\begin{equation}
\ppx{m_x} + \ppy{m_y} + \ppz{m_z} =0, \label{eq:div_m_vector} 
\end{equation}
and $\nabla \psi$ is a curl-free component. 
Similar to the formulation developed by \cite{Kim1987} for incompressible flow, 
we define 
\begin{align}
\phi & = \nabla^2 m_y, \label{eq:def_phi} \\  
\Omega_y & = \left. \nabla \times \vec{m} \right|_y=\ppz{m_x} - \ppx{m_z}. \label{eq:def_Omey}
\end{align} 
The evolution equations for these two variables are obtained by proper manipulation of 
eqs. (\ref{eq:continuity2}-\ref{eq:momentum2}). This leads to a system of four evolution equations
for the variables $\phi$, $\Omega_y$, $T$ and $\rho$ together with the equation of state (\ref{eq:EOS12}),
 \begin{align}
 \frac{\partial{\phi}}{\partial t} &= F(\rho,u_j) = \pppx{N_y}+\pppz{N_y}-\ppy{} \left( {\ppx{N_x}}+{\ppz{N_z}} \right)-\frac{1}{Re} \left. \nabla^2 ({\nabla \times \vec{\omega}})\right|_y,
\label{eq:phieq}\\
 \ppt{\Omega_y}                    &=  M(\rho,u_j) = \ppz{N_x}-\ppx{N_z}+\frac{1}{Re}\left( {\nabla^2 \omega_y} \right),
\label{eq:omegay2}\\
 \ppt{T}                           &=  E(\rho,u_j) = -\vec{u} \cdot \nabla T+\frac{T}{Pe}\nabla^2{T},
\label{eq:energy3}\\
\ppt{\rho}                          &=  C(\rho,u_j) = -\nabla \cdot (\rho \vec{u}) =-\nabla^2 \psi, 
 \label{eq:cont3} 
\end{align}
where $N_i =-\partial(\rho u_i u_j)/\partial x_j$ and $\vec{\omega}$ is the vorticity. 

The manipulations to obtain eqs. (\ref{eq:phieq}-\ref{eq:omegay2}) involve taking spatial derivatives of the momentum equations.
In this process, information concerning the horizontally averaged momentum vector is lost, 
requiring additional equations to keep this effect. 
Averaging eq. (\ref{eq:momentum2}) over the homogeneous directions
$x$ and $z$, we obtain equations for $\langle \rho u\rangle$ and $\langle \rho w\rangle$,
\begin{align}
\ppt{\langle \rho u \rangle} & =-\ppy{ \langle \rho u v \rangle}+\cfrac{1}{Re} \ppy{ \langle \tau_{xy} \rangle}, 
         \label{eq:mode0u}\\
\ppt{\langle \rho w \rangle} & =-\ppy{\langle \rho v w \rangle}+\cfrac{1}{Re} \ppy{\langle \tau_{zy} \rangle}. 
         \label{eq:mode0w}
\end{align}
Averaging eq. (\ref{eq:continuity2}) over the homogeneous directions $x$ and $z$ and integrating in $y$
we obtain an equation for $\langle \rho v \rangle$,
\begin{equation}
\int_{-\infty}^{y} \ppt{\langle \rho \rangle} dy =- \int_{-\infty}^{y}  \ppy{\langle \rho v \rangle} dy = 
\langle \rho v \rangle_b - \langle \rho v \rangle (y).
\label{eq:rhov}
\end{equation}
Note that $\langle \rho u \rangle, \langle \rho v\rangle$ and $\langle \rho w\rangle$ correspond to the $k_x=0$ and $k_z=0$ modes of the Fourier expansions in $x$ and $z$. These variables are in principle functions of $y$ and $t$. 

The algorithm to solve eqs. (\ref{eq:phieq}-\ref{eq:cont3}) is split into two parts. First, we employ 
an explicit, low-storage, 3-stage Runge-Kutta scheme for eqs. (\ref{eq:phieq}-\ref{eq:energy3}), 
that for the $i$-th stage reads
\begin{eqnarray}
\phi^i  &=& \phi^{i-1} + \gamma_i \Delta t  F(\rho,u_j)^{i-1}	+ \epsilon_i \Delta t  F(\rho,u_j)^{i-2},		 		\nonumber \\
\Omega_y^i  &=& \Omega_y^{i-1} + \gamma_i \Delta t  M(\rho,u_j)^{i-1}	+ \epsilon_i \Delta t  M(\rho,u_j)^{i-2},	\nonumber \\
T^i  &=& T^{i-1} + \gamma_i \Delta t  E(\rho,u_j) ^{i-1}	+ \epsilon_i \Delta t E(\rho,u_j) ^{i-2},
\label{eq:explicit}
\end{eqnarray}
where $\gamma_i=(8/15,5/12,3/4)$  and $\epsilon_i=(0,-17/60,-5/12)$ 
are the coefficients of the explicit scheme \citep{Spalart1991}. For eq. (\ref{eq:cont3}) 
we employ an implicit, low-storage, 3-stage Runge-Kutta scheme, that for the $i$-th stage reads

\begin{equation}
\rho^{i} = \rho^{i-1} - \Delta t\left( \alpha_i \nabla^2 \psi^{i-1} + \beta_i \nabla^2 \psi^{i} \right),
\label{eq:implicit}
\end{equation}
where $\alpha_i=(5/66,17/15,1/22)$ and $\beta_i=(151/330,-1,19/66)$ 
are the coefficients of the implicit scheme, 
optimized to enhance the stability of the code in a similar way as \citet{jang:2007}.  
Note that this equation is a Poisson problem for $\psi^i$ if $\rho^i$ is known. 
However, from the point of view of mass conservation, it is beneficial to express $\rho^i-\rho^{i-1}$ in terms of the temperature, and use the fact that $\nabla^2 \psi = -\partial\rho/\partial t = T^{-2}\partial T/\partial t = T^{-2} E(\rho,u_j)$ , yielding 
\begin{equation} 
\nabla^2 \psi^i = \frac{1}{\beta_i\Delta t} \frac{T^{i}-T^{i-1}}{T^i T^{i-1}}   - \frac{\alpha_i}{\beta_i} \left( \frac{E(\rho,u_j)}{T^2} \right)^{i-1}.
\label{eq:implicit2}
\end{equation}
With this formulation, we are assuring that the energy equation acts as a constraint for the continuity equation (as suggested by \citealp{nicoud:2000}), keeping both equations synchronised at every time step. 

From eq. (\ref{eq:explicit}) we obtain $\phi^i$, $\Omega_y^i$ and $T^i$. Using eq. (\ref{eq:EOS12}) we obtain
$\rho^i$, and solving the Poisson problem (\ref{eq:implicit}) we obtain $\psi^i$.
In order to compute the right hand side of eqs. (\ref{eq:phieq}-\ref{eq:energy3}) the velocity and the vorticity
are needed. The velocity is constructed as follows. First, knowing $\phi$  we solve the Poisson problem 
eq. (\ref{eq:def_phi}) to obtain $m_y$. Knowing $\Omega_y$, we can solve eqs. (\ref{eq:div_m_vector})
and (\ref{eq:def_Omey}) to obtain $m_x$ and $m_z$. Finally, knowing $\psi$ and $\rho$, from the 
definition, eq. (\ref{eq:decomposition}), we obtain the velocity field and by differentation the vorticity field.

\subsection{Boundary conditions: entrainment}
\label{sec:bcs}

As discussed in the main text, the velocity and density fluctuations should tend to zero as $y\rightarrow \pm \infty$.
We impose
\begin{equation}
\begin{array}{llll}
\rho=\rho_b,\, &u=\Delta U/2,\,  &w=0 & \mbox{ at } y\rightarrow -\infty,  \\
\rho=\rho_t,\, &u=-\Delta U/2,\, &w=0 & \mbox{ at } y\rightarrow \infty. 
\end{array}
\end{equation}
Due to the entrainment there is a non-zero value of $\langle \rho v\rangle$ at $y\rightarrow \pm \infty$.
Integrating eq. (\ref{eq:rhov}) from $-\infty$ to $\infty$ we obtain the total mass outflow, $\Phi$, as
\begin{equation}
\Phi=\int_{-\infty}^{\infty} \ppt{\langle \rho \rangle} dy = \langle \rho v \rangle_b-\langle \rho v \rangle_t.
\label{eq:Phi}
\end{equation}
It is possible to express the total mass outflow as a function of the vertical entrainment ratio, 
$E_v=-\langle v\rangle_b/\langle v\rangle_t$, as
\begin{equation}
\Phi =  \langle \rho  v \rangle_b \left( 1 + \frac{1}{E_v s} \right). \label{eq:Phi_Ev}
\end{equation}
\citet{Dimotakis1986} suggests that, for a variable-density temporal mixing layer, the entrainment ratio
should be equal to the square root of the density ratio, an argument attributed to \cite{Brown1974b}.
Using this result and computing during runtime the value of $\Phi$ we obtain 
$\langle \rho  v \rangle_b$ from  eq.(\ref{eq:Phi_Ev}) and $\langle \rho  v \rangle_t = \langle \rho  v \rangle_b - \Phi$.
Note that during the self-similar evolution, since $\rho$ should scale with $\rho_b-\rho_t$ and the thickness
of the layer grows linearly with time, the value of $\Phi$ should remain constant. Therefore, during the
self-similar evolution the values of $\langle \rho  v \rangle_t$ and $\langle \rho  v \rangle_b$ should be
constant as well.

\section{Variable density laminar mixing layer: self-similar solution}
\label{sec:laminar}

In this appendix we present the procedure followed to obtain a self-similar solution for a laminar temporal mixing layer. 
The configuration is the same discussed in the body of the paper for the turbulent mixing layer: two opposing streams with a velocity difference $\Delta U$ and a density ratio $s$. The differences with respect to equations (\ref{eq:continuity}-\ref{eq:EOS1}) is that the spanwise velocity is $w=0$, and that the rest of the fluid variables are only functions of the vertical coordinate, $y$, and time, $t$. Then, the equations governing the problem are
\begin{eqnarray}
\ppt{\rho} + \ppy{\rho v} & = 0 & ,  \label{eq:NSlam_cont} \\
\ppt{u} + v \ppy{u} & = T \frac{\mu}{\rho_0 T_0} \pppy{u} & ,  \\
\ppt{T} + v \ppy{T}   & =  T \frac{ k}{\rho_0 C_p T_0} \pppy{T}  \label{eq:NSlam_ener}&, 
\end{eqnarray}
plus the equation of state $\rho T = \rho_0 T_0$.  
In these equations $\mu$ is the dynamic viscosity, $\kappa$ is the thermal conductivity and $C_p$ is the specific heat at constant pressure. Note that the vertical component of the momentum equation is not included, since it introduces an additional unknown, the mechanical pressure $p^{(1)}(y,t)$. 
The boundary conditions are the same as for the turbulent mixing layer, with velocity and density (temperature) going to the free-stream values when $y \to \pm\infty$. 

In order to solve the system of coupled partial differential equations given by (\ref{eq:NSlam_cont}-\ref{eq:NSlam_ener})  we define the density-weighted vertical coordinate, 
\begin{equation} 
\xi = \frac{1}{\rho_0} \int_{-\infty}^y \rho dy.
\end{equation}
We also define a characteristic length for the problem, based on the kinematic viscosity ($\nu = \mu/\rho_0$) and time, $\delta = \sqrt{\nu t}$. Then, using $\xi$ and $\delta$ it is possible to recast equations (\ref{eq:NSlam_cont}-\ref{eq:NSlam_ener}) into a self-similar set of equations in which the time dependence is absorbed into the self-similar coordinate $\eta=\xi/\delta$, 

\begin{eqnarray}
\frac{\partial V}{\partial \eta} +  \frac{\eta}{2}\frac{\partial \Theta }{\partial \eta} & =0 & ,  \\
\frac{\partial U}{\partial \eta} \frac{\eta}{2}  + \frac{\partial}{\partial \eta }\left( \frac{1}{\Theta} \frac{\partial U}{\partial \eta} \right) & = 0& , \label{eq:selfsim1}\\
\frac{\partial \Theta}{\partial \eta} \frac{\eta}{2}  + \frac{1}{Pr} \frac{\partial}{\partial \eta }\left(\frac{1}{\Theta} \frac{\partial \Theta}{\partial \eta} \right)  & = 0& , \label{eq:selfsim2}
\end{eqnarray}
where $U = u/\Delta U$, $V = v/\sqrt{\nu/t}$, $\Theta = T/T_0$ and $Pr$ is the Prandtl number. 
The boundary conditions for $U(\eta)$ and $\Theta(\eta)$ are $U(\pm\infty) = \mp0.5$, $\Theta(+\infty) = (1+s)/2$ and $\Theta(-\infty) = (1+1/s)/2$.  
Interestingly, in the self-similar set of equations, $V$ appears only in the continuity equation, allowing 
to solve for $U(\eta)$ and $\Theta(\eta)$ using the momentum and energy equations only. 
 Unfortunately, the equations only admit analytical solution when $s=1$. For other values of $s$, equations (\ref{eq:selfsim1}) and (\ref{eq:selfsim2}) are solved together using Chebychev polynomials \citep{driscoll2008chebop}.

\bibliographystyle{plainnat}
\bibliography{ZMVD_Mixing_Layer}

\end{document}